\documentclass[12pt]{article}
\usepackage[margin=0.95 in]{geometry}
\usepackage{amsmath} 
\usepackage{slashed}
\usepackage{dsfont}
\usepackage{amssymb,amsfonts}
\usepackage{mathtools}
\usepackage[all]{xy}
\usepackage{graphicx}
\usepackage{amsmath}
\usepackage{amssymb}
\usepackage{float}
\usepackage{array}
\usepackage{tikz}
\usepackage{tkz-euclide}
\usetikzlibrary{decorations.markings,arrows}
\tikzset{
    arrowMe/.style={
        postaction=decorate,
        decoration={
            markings,
            mark=at position .5 with {\arrow[thick]{#1}}
        }
    }
}
\pgfarrowsdeclaredouble{<<s}{>>s}{stealth}{stealth}
\pgfarrowsdeclaretriple{<<<s}{>>>s}{stealth}{stealth}
\usepackage{mathtools}
\usepackage{mathrsfs} 
\usepackage[hidelinks]{hyperref}
\usepackage{cite}
\usepackage{tikz}
\usepackage{array, multirow}  
\numberwithin{equation}{section}
\newcommand{\be}{\begin{equation}}
\newcommand{\ee}{\end{equation}}

\def\bea{\begin{eqnarray}}
\def\eea{\end{eqnarray}}
\numberwithin{equation}{section}
\numberwithin{table}{section}

\begin{document}

\hypersetup{pageanchor=false}
\begin{titlepage}
\vbox{\halign{#\hfil    \cr}}  
\vspace*{15mm}
\begin{center}
{\Large \bf The Operator Rings of  Topological Symmetric Orbifolds  \\ \vspace{0.3em}  
and their Large N Limit}

\vspace*{10mm} 

{\large Sujay K. Ashok$^{a,b}$ and Jan Troost$^c$}
\vspace*{8mm}

$^a$The Institute of Mathematical Sciences, \\
		 IV Cross Road, C.I.T. Campus, \\
	 Taramani, Chennai, India 600113

\vspace{.6cm}

$^b$Homi Bhabha National Institute,\\ 
Training School Complex, Anushakti Nagar, \\
Mumbai, India 400094

\vspace{.6cm}

$^c$Laboratoire de Physique de l'\'Ecole Normale Sup\'erieure \\ 
 \hskip -.05cm
 CNRS, ENS, Universit\'e PSL,  Sorbonne Universit\'e, \\
 Universit\'e  Paris Cit\'e 
 \hskip -.05cm F-75005 Paris, France	 

\vspace*{0.8cm}
\end{center}

%\noindent
\begin{abstract} {We  compute the structure constants of topological  symmetric orbifold theories up to third order in the large N expansion. 
The leading order structure constants are dominated by topological metric contractions. The first order interactions are  single cycles joining while at second order we can have double joining as well as splitting. At third order, single cycle joining obtains genus one contributions. 
 We also compute illustrative small N structure constants. Our analysis applies to all second quantized Frobenius algebras, a  large class of algebras that includes the  cohomology ring of the Hilbert scheme of points on K3 among many others. We point out interesting open questions that our results raise.}  
\end{abstract}

\end{titlepage}

\hypersetup{pageanchor=true}

\setcounter{tocdepth}{2}
\tableofcontents

\section{Introduction}
The symmetric group $S_N$ is ubiquitous in physics. It arises as the group that exchanges $N$ identical particles  and from this elementary observation, the group propagates into many aspects of statistical and quantum mechanics. In two-dimensional conformal field theory, the discrete symmetry gauge group $S_N$ has the further consequence of introducing twisted sectors labelled by conjugacy classes of the group  $S_N$  which leads to  interesting  dynamics \cite{Dixon:1985jw}. 
Apart from its universality, a further motivation for understanding these dynamics comes from viewing the $S_N$ discrete gauge group as the Weyl group remnant of a $\mathfrak{su}(N)$ gauge algebra. Indeed, the large $N$ limit in two-dimensional symmetric orbifold conformal field theories has much in common with the 't Hooft large $N$ limit in four-dimensional gauge theories \cite{tHooft:1973alw,El-Showk:2011yvt}. As such, these conformal field theories are natural candidates for holographic duals of string theories in three-dimensional anti-de Sitter space-time \cite{Maldacena:1997re,Belin:2015hwa}. 

In this work, we further our understanding of these theories on several fronts.  We compute a set of  large $N$ symmetric orbifold observables; in particular we study the operator rings of topological symmetric orbifold theories \cite{Li:2020zwo}. The task we set ourselves is to calculate a vast set of large $N$ structure constants. In a previous paper \cite{Ashok:2023mow}, we computed a number of structure constants for the operator ring of the symmetric orbifold of the topologically twisted conformal field theory on $\mathbb{C}^2$. In this paper, we further exploit the mathematical construction of the operator product in second quantized Frobenius algebras \cite{Li:2020zwo,LS2,Kaufmann} to compute large $N$ structure constants at higher order and for a much larger class of theories. These include the symmetric orbifold of $K3$ but also orbifolds of twisted minimal models as well as compact and non-compact Gepner models -- indeed a large class of examples originates in twisted ${\cal N}=(2,2)$ superconformal field theories in two dimensions. 
For selected results on generic correlators in symmetric orbifolds at large N, see e.g. \cite{Burrington:2018upk,Roumpedakis:2018tdb,DeBeer:2019oxm,Dei:2019iym}.

A central result is the master formula (in equation \eqref{MasterFormula}) that exhibits the large $N$ expansion for operator product coefficients in a topological symmetric orbifold theory. Our analysis is applicable to any  seed theory.
New qualitative features of the operator product include contractions with the topological metric and a general genus one contribution to the structure constants -- it appears at third order in the $g_s = {1}/{\sqrt{N}}$ expansion. 

The topological theory captures information about the untwisted symmetric orbifold theories, including all of its extremal correlators \cite{Li:2020zwo}. 
 To illustrate the efficacy of our approach we compute two sets of four point functions (see equation \eqref{doublefusion}). The calculation recovers and extends earlier results obtained in the literature; in particular, we exhibit the dependence of the extremal four-point function on the Euler number of the seed conformal field theory.

Our paper begins with a brief review of the operator ring structure and some illustrative structure constant calculations at small $N$ in section \ref{Review}. In section \ref{TriplesOfPermutations} we identify the triples of permutations that contribute at a given order in the large $N$ expansion. In section \ref{StructureConstants} we compute the  structure constants at zeroth to third order in the large $N$ coupling constant $g_s = 1/\sqrt{N}$. Section \ref{FourPointFunctions} is dedicated to the calculation of a set of four-point functions. We conclude in section \ref{Conclusions} with a summary and the description of a few  open problems. There are four Appendices which contain examples and more technical details.

\section{Review and Exact Examples at Small N}
\label{Review}
 In this section we propose a close relation between topological symmetric orbifolds \cite{Li:2020zwo} and second quantized Frobenius algebras \cite{LS2,Kaufmann} generalizing the observations of \cite{Li:2020zwo}.  
We review the relation between the Ramond-Ramond ground states of the twisted seed conformal field theory and the seed Frobenius algebra \cite{Lerche:1989uy} and propose that this  correspondence  lifts to the symmetric orbifold of both. We review the product in the second quantized Frobenius algebra  \cite{LS2,Kaufmann} and provide small $N$ examples of products in the  algebra. These examples complement the treatment in \cite{Li:2020zwo} in that we explicitly study fully gauge invariant operators. Moreover,   the examples provide an opportunity to illustrate certain properties of  the large $N$ structure constants of the operator ring that already surface at small values of $N$.

\subsection{Topological Quantum Field Theory and Frobenius Algebra}
We study a two-dimensional topological conformal field theory. These are examples of topological quantum field theories. In two dimensions, these theories can be described by (graded) commutative Frobenius algebras \cite{DijkgraafPhD,Dijkgraaf:1997ip,Dubrovin:1994hc}. A Frobenius algebra is  an  associative algebra $A$, with a unit and a linear form $T$. The linear form is such that the bi-linear form $T(ab)$ induced by $T$ is non-degenerate.  We make the connection to topological field theory as follows.
Let $e_i$ be a  basis of the Frobenius algebra $A$. Then, the structure constants of the algebra in this basis are defined by the product
\begin{align} 
e_i e_j &= {c_{ij}}^k\, e_k~.
\end{align}
This captures the product of operators (or the map from two states to one) in the topological  field theory. 
In particular, taking $e_0$ to be the identity element $e_0=1$ we obtain ${c_{0j}}^k = {\delta_j}^k$. 
The linear form $T$ defines the topological metric $\eta$
\be 
T(e_i e_j) = \eta_{ij}~ \, ,
\ee 
which is equivalent to a map from two (incoming) states to $\mathbb{C}$. 
Combining these two relations we have the equations
\be 
c_{ijl}:=T(e_ie_je_l) = T({c_{ij}}^k e_k e_l)={c_{ij}}^k \eta_{kl} \, .
\ee
We also obtain $c_{0jk} = \eta_{jk}$. Since we have a graded Frobenius algebra, we  moreover have the graded symmetry
\be 
{c_{ij}}^k = (-1)^{ij} {c_{ji}}^k~.
\ee
In this paper, we will most often ignore signs arising from the grading of the homogeneous elements of the Frobenius algebra -- they can painstakingly be re-established if necessary. For more information on the relation between topological quantum field theory in two dimensions and Frobenius algebras, see e.g. chapter 4 of \cite{Dijkgraaf:1997ip}.

Examples of topological field theories in two dimensions are provided by 
 ${\cal N}=(2,2)$ superconformal field theories \cite{Lerche:1989uy}.   We  can twist the superconformal field theories such that only the (chiral,chiral) ring survives in the topological  field theory.\footnote{There exists a T-dual or mirror twist.} The (chiral,chiral) ring is a Frobenius algebra with the bi-linear form $T$ provided by the evaluation of the operator in the vacuum state of the twisted conformal field theory. 

For the benefit of the reader, we review a few  facts about these superconformal field theories and their properties \cite{Lerche:1989uy}. Firstly, there is a spectral flow automorphism that relates the NS-NS sector (chiral,chiral) ring to the space of Ramond-Ramond ground states. The (left,right) R-charges of the Ramond-Ramond ground states are shifted by $(-c/6,-c/6)$ compared to the R-charges of the chiral ring elements, where $c$ is the central charge of the  ${\cal N}=(2,2)$ superconformal field theory. The unit operator in the NS-NS sector is mapped to the R-R ground state with minimal charges $(-c/6,-c/6)$.
 The highest charge chiral ring element
 is mapped to the Ramond-Ramond 
 ground state with charges $(c/6,c/6)$. Anomalous R-charge conservation implies that the topological two-point function $T$ has R-charge $-c/3$ i.e. it maps the Ramond-Ramond ground state of charge $(c/6,c/6)$ to a number. 
There is a product of Ramond-Ramond ground states that is inherited from the product of chiral primaries through spectral flow. The reason we review both sectors is that the NS-NS sector gives a direct handle on the ring structure while the R-R perspective is handier to keep track of the (shifted) degrees of operators, as well as the connection to differential forms when such an interpretation is available.
In summary, it is well-known that a topologically twisted ${\cal N}=(2,2)$ superconformal field theory in two dimensions gives rise to a Frobenius algebra. The two structures in the theory that are fundamental are the operator ring (or the structure constants ${c_{ij}}^k$) as well as the bi-linear form $T$ (or topological two-point function $\eta_{ij}$).

Finally, we recall that a standard example of a Frobenius algebra is the cohomology ring of  a compact orientable manifold $M$ equipped with integration. For a compact K\"ahler manifold, the de Rham cohomology is equivalent to the Dolbeault cohomology. For the case of a Ricci flat manifold $M$, the latter in turn  is equivalent to the (chiral,chiral) ring of the ${\cal N}=2$ superconformal field theory with target space $M$ \cite{Lerche:1989uy}.
The cohomology forms a ring with structure constants ${c_{ij}}^k$ while the integration provides the linear form $T$. 
In the geometric case, we have the relation $c=3d$ between the central charge $c$ and the complex dimension $d$ of the manifold $M$. The highest charge chiral ring element corresponds to the volume form on the manifold. However,  it is important to realize that we do not restrict to this setting and that we will be able to generalize considerably statements that are otherwise  tied to geometry.

\subsection{The Topological Symmetric Orbifold}
In the previous subsection, we introduced a seed ${\cal N}=(2,2)$ superconformal field theory that gives rise to a  Frobenius algebra $A$ after topological twisting. 
In this subsection, we review the concept of a second quantized Frobenius algebra \cite{Dijkgraaf:1996xw,LS2,Kaufmann}. From a seed Frobenius algebra $A$, one can rather uniquely\footnote{The construction is unique up to a choice of grading and discrete torsion \cite{Kaufmann}.} determine a second quantized algebra $A^{[n]}$ which is the $N$-th tensor product power $A^{\otimes N}$ of the algebra $A$, supplemented with twisted sectors and
divided by the symmetric group $S_N$ which exchanges the factors of the tensor product. We conjecture that we should interpret this construction as describing the topologically twisted $S_N$ orbifold conformal field theory of a seed ${\cal N}=(2,2)$ model. More broadly still, we propose that this correspondence extends to topologically twisted supersymmetric massive models in two dimensions. Our main motivations for the conjecture are that the second quantized Frobenius algebra construction is tailored on the construction of orbifolds in two-dimensional field theory and that the structure is essentially unique \cite{Kaufmann}. 

The definition of the general second quantized Frobenius algebra can be found in \cite{Kaufmann}.
This reference also establishes its broad scope and uniqueness. It is moreover shown that the algebra coincides with the cohomology ring of the Hilbert schemes of complex surfaces, defined in \cite{LS2} when restricted to the geometric case with $c/3=d=2$. In the latter case,  our conformal field theory seed  is the supersymmetric sigma model on the complex surface and it is well-understood that the twisted seed theory indeed describes the cohomology of the surface \cite{Lerche:1989uy}. Arguments as to why this extends to the symmetric orbifold are reviewed and given in \cite{Li:2020zwo}.  In the following, we  find it  convenient to use the reference \cite{LS2} as the basis of our description of the definition of the operator product in the second quantized Frobenius algebra.
A detailed review of the definition of the multiplication in the second quantized algebra \cite{LS2} can be found in \cite{Li:2020zwo}. Here, we shall provide an executive summary.

\subsubsection{The Second Quantized Frobenius Algebra}
\label{SecondQuantizedFrobeniusAlgebras}
Suppose that we are given a Frobenius algebra $A$. 
The tensor product algebra, denoted $A^{\otimes n}$ is also a Frobenius algebra.\footnote{To improve the typography of our text, we interchangeably use the symbols big $N$ and small $n$.} The product of tensor product elements is the graded tensor product of the various factors:
\be 
(a_1\otimes \ldots \otimes a_n)\cdot  (b_1\otimes \ldots \otimes b_n) =% \epsilon(a,b) 
(a_1b_1)\otimes \ldots \otimes (a_nb_n)~.  
\ee 
Here $\epsilon$  tracks the number of odd elements that are exchanged. On the right hand side we used  the product of elements in the algebra $A$. On the tensor product algebra $A^{\otimes n}$, the bi-linear form  $T$ acts as follows:
\be 
T(a_1\otimes \ldots \otimes a_n)= T(a_1)\ldots T(a_n)~. 
\ee 
The tensor product is a Frobenius algebra once more. 
Secondly, we define the twisted sectors of the orbifold Frobenius algebra.
We introduce the notation $H \setminus B$ for the set of orbits of the space $B$ under the action of the group $H$. We define the set $\{1,2,\dots,n \}=[n]$ and introduce vector spaces:
\begin{equation}
A \{ S_n \} = \oplus_{\pi \in S_n} A^{\otimes \langle \pi \rangle \setminus [n]} \pi \, .
\end{equation}
To each orbit of the group $\langle \pi \rangle$ generated by the permutation $\pi \in S_n$, we have associated a factor of the seed Frobenius algebra $A$ and these multiply the permutation $\pi$. We have temporarily introduced a twisted sector for each permutation $\pi$ in the symmetric group $S_n$. Permutations act naturally on both untwisted and twisted sectors and the final space we are interested in is the set of invariants $A^{[n]}$ under the action of the symmetric group. 
Our task is to construct a product on $A \{ S_n \}$ which restricts nicely to the set of invariants $A^{[n]}$. 

The product is defined as follows. 
Firstly, to be able to multiply two twisted sector operators, we define the subgroup $\langle \pi_1,\pi_2 \rangle$ generated by both the twists $\pi_1$ and $\pi_2$ in operators one and two. This subgroup generates orbits that are larger than those of each respective twist. Each operator  
associated to smaller orbits can be mapped into an operator in a smaller number of tensor products of $A$ through multiplication. These maps are denoted $f^{H,K}$ where $H \subset K$ are subgroups of $S_n$.  This provides a recipe to bring operators in different twisted sectors into a common space where we can multiply them.
When we multiply them though, a correction factor is added that represents loop corrections, weighted by the graph defect or genus:\footnote{The Riemann surface  associated to the permutations is constructed through triangulation in Lemma 2.7 of \cite{LS2}. We will shortly provide an example of such a construction.}
\begin{equation}
g(\pi_1,\pi_2)(B) = \frac{1}{2} (|B|+2 -| \langle \pi_1 \rangle \setminus B |- |\langle \pi_2 \rangle \setminus B |- |\langle \pi_1 \pi_2 \rangle \setminus B|)\, ,
\label{GraphDefect}
\end{equation}
where $B$ is a set of labels of factors in $A^{\otimes n}$ that form a single orbit under $\langle \pi_1,\pi_2 \rangle$.
Finally, the adjoint $f_{K,H}=f_{\langle \pi_1,\pi_2 \rangle, \langle \pi_1 \pi_2 \rangle}$  of the multiplication map $f^{H,K}$ redistributes the  operator resulting from multiplication and loop correction onto the  orbits of the final twist $\pi_1 \pi_2$. The final formula for the multiplication $m_{\pi_1,\pi_2}$ of operators $a$ and $b$ in the twisted sectors $\pi_1$ and $\pi_2$  reads \cite{LS2}:
\begin{equation}
m_{\pi_1,\pi_2}(a \otimes b) = f_{\langle \pi_1,\pi_2 \rangle, \langle \pi_1 \pi_2 \rangle} ( f^{\pi_1,\langle \pi_1,\pi_2\rangle}(a) f^{\pi_2,\langle \pi_1,\pi_2\rangle}(b) e^{g(\pi_1,\pi_2)}   )
\label{Multiplication}
\end{equation}
where $e$ is the Euler class $e(A)$ of the Frobenius algebra $A$. It is defined through the multiplication map $\Delta^{(2)}: A \otimes A \rightarrow A: a \otimes b \mapsto ab$ and its adjoint $\Delta^{(2)}_\ast$ through the formula $\Delta^{(2)} \circ \Delta^{(2)}_\ast (e_0) = e$. 
In more detail, we have that the Euler class equals  $e =\chi \, \text{vol}$ where $\text{vol}$ is the dual of $1$ with respect to the bi-linear form. Indeed, we can write the class $e$ as 
\begin{equation}
\label{deltadeltastarv1}
e=\Delta^{(2)} \Delta_\ast^{(2)}(e_0)={c^{ji}}_0 {c_{ij}}^k e_k = {c^{ji}}_0 {c_{ij0}} \, \text{vol} = \chi \, \text{vol}~,
\end{equation}
where we have used that  the Euler number is  $\chi\coloneqq  \eta^{ji} \eta_{ij} $.

Let us illustrate the  abstract multiplication formula \eqref{Multiplication} in a simple example. Consider the permutations $\pi_1=(1,2,\dots,m)$ and $\pi_2=(m,m-1,\dots,1)$ accompanied with operators $a,b$ and the multiplication:
\begin{equation}
m_{\pi_1,\pi_2}(a \otimes b) \coloneqq a \pi_1 \, b \pi_2 \, . \label{ExampleMultiplication}
\end{equation}
The labels $B=\{1,2,\dots,m-1,m \}$ form a single orbit under the group $\langle \pi_1,\pi_2 \rangle$ and the maps $f^{\pi_i,\langle \pi_1,\pi_2 \rangle}$ map the operators $a$ and $b$ into the same algebra $A$ in which we can multiply them to find $ab$. The graph defect is zero. We then need the adjoint map $f_{\langle \pi_1,\pi_2 \rangle,()}$ where $()$ denotes the identity permutation that results from multiplying the permutations $\pi_1$ and $\pi_2$. There is a multiplication operator $\Delta^{(m)}$ that maps the tensor product $A^{\otimes m}$ into the space $A$ through multiplication of all tensor product factors. The adjoint map  $f_{\langle \pi_1,\pi_2 \rangle,()}$ is the adjoint $\Delta_\ast^{(m)}$ of this multiplication map. It  redistributes the resulting operator over the orbits of the set $B$ with respect to the permutation $\pi_1 \pi_2=()$, i.e. all $m$ tensor product factors. Therefore, the product equals:
\begin{equation}
a \pi_1 \, b \pi_2 =  \Delta_\ast^{(m)} (ab)\, ()\,.
\end{equation}
%will be made explicit in Section \ref{OperatorProducts} when we analyze  the large-$N$ expansion of operator products. 

\subsubsection{Adjoint Maps}
It is useful to develop explicit formulae for these adjoint maps in terms of the structure constants and the topological metric. 
The product map $\Delta^{(m)}$  can be more formally described as:
\begin{equation}
\Delta^{(m)} : A^{\otimes m} \rightarrow A: a_1 \otimes \dots \otimes a_m \mapsto a_1 \dots a_m
\, .
\end{equation}
When we calculate the coefficients of the map in a specific basis $e_i$, we find:
\begin{align}
T(e_l \, \Delta^{(m)} (e_{i_1} \otimes \dots \otimes e_{i_m}))
&=  {c_{i_1 i_2}}^{j_3}
{c_{j_3 i_3}}^{j_3} \dots {c_{j_m i_m}}^{j_{m+1}}  T(e_l  e_{j_{m+1}})
\nonumber \\
&=  {c_{i_1 i_2}}^{j_3}
{c_{j_3 i_3}}^{j_4} \dots {c_{j_m i_m}}^{j_{m+1}}  \eta_{l,j_{m+1}}
\, . 
\label{productformula}
\end{align}
The adjoint $\Delta^{(m)}_\ast$ of the operator $\Delta^{(m)}$  satisfies the equation
\begin{align}
T(e_l \, \Delta^{(m)} (e_{i_1} \otimes \dots \otimes e_{i_m}))&= T( \Delta^{(m)}_\ast (e_l) \, \, e_{i_1} \otimes \dots \otimes e_{i_m}) ~.
\end{align}
For the expression of the adjoint map in a particular basis, we propose the explicit formula -- we ignore an overall sign  --:
\begin{align}
\Delta_\ast^{(m)} (e_l)
&=%(-1)^l 
{c^{k_1 k_2 j_3}} {{c_{j_3}}^{k_3 j_4}}\ldots  {c_{j_{m}}}^{k_m j_{m+1}} \eta_{l,j_{+1}}~ e_{k_1} \otimes e_{k_2} \otimes e_{k_3} \dots \otimes   e_{k_m}  \, .
\label{adjointansatz}
\end{align}
We check: 
\begin{align}
T(\Delta_\ast^{(m)} (e_l) \cdot e_{i_1} \otimes \dots e_{i_m} )
&= %(-1)^l
{c^{k_1 k_2 j_3}} {{c_{j_3}}^{k_3 j_4}}  \ldots  {c_{j_{m}}}^{k_m j_{m+1}} \eta_{l,j_{m+1}} ~ T( e_{k_1} \otimes  \dots   \otimes e_{k_m} \cdot e_{i_1} \otimes  \dots   \otimes e_{i_m} )
\nonumber \\
&= %(-1)^l
{c^{k_1 k_2 j_3}} {{c_{j_3}}^{k_3 j_4}}  \ldots 
 {c_{j_{m}}}^{k_m j_{m+1}} \eta_{l,j_{m+1}} ~ 
\eta_{k_1,i_1} \dots \eta_{k_m,i_m} \nonumber \\
&=  
{c_{i_1 i_2}}^{j_3}
{c_{j_3 i_3}}^{j_4} \dots {c_{j_m i_m}}^{j_{m+1}}  \eta_{l,j_{m+1}}
\, .
\end{align}
%
%This proves that the proposed formula is correct.
Comparing with \eqref{productformula}, 
we find that the proposed formula for the adjoint map in \eqref{adjointansatz} is indeed correct. We will refer to this explicit expression in future sections.

\subsection{Charge Conservation and  Useful Formulae}
It is important to understand how the elements of the second quantized Frobenius algebra match up with the symmetric product ${\cal N}=(2,2)$ conformal field theory (chiral, chiral) ring elements.  A twisted sector characterized by  a cycle of length $k$ has an associated twist operator $\sigma_k$ that creates the vacuum in this twisted sector. The operator $\sigma_k$ has (left) dimension $h_k$:
\begin{equation}
h_k = \frac{c}{24}(k-\frac{1}{k}) \, .
\end{equation}
Since we have a left (and right) ${\cal N}=2$ superconformal algebra, we also have a seed U(1) R-current $J$ which we can describe in terms of a canonically normalized scalar $\phi$:
\begin{equation}
J = \sqrt{\frac{c}{3}}  \partial \phi
\, .
\end{equation}
A twisted sector state in the NSNS sector 
then reads:
\begin{equation}
\tau_k = \sigma_k \exp ( i \sqrt{\frac{c}{3}} \sum_{i=1}^k \frac{k-1}{2k} \phi^i)
\end{equation}
where the sum over the index $i$ is over the $k$ copies involved in the twist.\footnote{While this formula does not seem to be in the literature, inspiration can be drawn from examples e.g. in  \cite{Jevicki:1998bm,Lunin:2001pw}.}
It is straightforward to compute that this state satisfies the chiral ring relation $h=q/2$ with charge:
\begin{equation}
q(\tau_k) = \frac{c}{6} (k-1) \, .
\end{equation}
In the following, we will associate a charge $q$ equal to 
the (left) R-charge, 
\begin{equation}
q([k]) = \frac{c}{6} (k-1)
\,  \label{Rchargeofpermutation}
\end{equation}
to each cycle $(k)=(a_1,a_2,\dots,a_k)$ of length  $k$ in a permutation. 
The permutation carries a charge which is the sum of the charges in each individual cycle. The reader can keep in mind that this charge is the charge of the product of the operators $\tau_k$. Moreover, for each individual cycle, we can multiply in a chiral ring element of the algebra $A$ and remain in the chiral ring. In that case, the total charge of the operator equals the sum of the R-charges of the operators $\tau_{k}$ and the charge of the operators in the algebra $A$ that multiply the individual cycles. That defines the total charge  of any operator in the second quantized algebra. The R-charges grade the algebra.\footnote{In more generality, we can work with the left plus the right R-charge or with a double grading.} 

To understand the intricate formulas to follow, it is once more useful to have a number of special cases of the general formulas in mind.  Firstly, we analyze the R-charges of various operators.
The topological two-point function carries charge $-c/3$.
We therefore have the R-charge counting rules for the metric and its inverse:
\begin{equation}
q(\eta_{ij}) =\frac{c}{3} \,  ,
\qquad \qquad
q(\eta^{ij}) =-\frac{c}{3} \, .
\end{equation}
This has as a consequence that the adjoint operator to multiplication (defined with respect to the topological two-point function $\eta$) has R-charge:
\begin{equation}
q(\Delta_\ast^{(2)}) = \frac{c}{3}
\, . \label{RChargeDeltaStarTwo}
\end{equation}
Consequently, the R-charge of the Euler class $e$ equals:
\begin{align}
q(e) &= \frac{c}{3}  %= q(\text{vol}) 
\, .
\end{align}
Our reasoning generalizes to the R-charge of the adjoint of higher tensor product multiplication:
\begin{align}
\label{RchargeDeltastar}
q(\Delta_\ast^{(l)}) &= \frac{c}{3}(l-1) \, .
\end{align}
We will also need the composition of $\Delta^{(2)} \Delta_\ast^{(2)}$ evaluated on other elements of the algebra $A$ with basis $e_m$.  We calculate: 
\begin{align}
\Delta^{(2)} \Delta_\ast^{(2)}(e_m) &=
 {c^{lk}}_m e_k e_l
\nonumber \\
&=  {c^{lk}}_m {c_{kl}}^n e_n \, .
\end{align}
Applying R-charge conservation to this formula, we find that
 $q(e_n)=q_n=q_k+q_l$ and also $c/3-q_l+c/3-q_k=c/3-q_m$ which implies $c/3=q_n-q_m$. This then implies, since $c/3$ is the maximal difference between left R-charges, that one can only have $q_n=c/3$ and $q_m=0$. We conclude that the result of the consecutive operations $\Delta^{(2)} \Delta_\ast^{(2)}$ is only non-zero for arguments that contain an identity component $e_0$ of charge zero. 
Finally, we record the following analogous simplifications of structure constants:
\be
\label{specialsc}
{c_{i0}}^k = {\delta_{i}}^k~, 
%\quad  {c_{i,j}}^{\text{vol}}\ne 0~,
\quad  {c_{i,\text{vol}}}^{j} = \delta_{i,0} \, \delta^{j,\text{vol}}~,\quad  
{c^{i,0}}_j = \delta^{i,\text{vol}} \, \delta_{j,0} \, ,
\ee
where we recall that $\text{vol}$ is the unique operator dual to the identity. 
These formulas will be useful in section \ref{OPs}.

\subsection{Exact Examples at Small N}
In this subsection, we perform a few exact calculations at small values of $N$ to provide a flavour for the generic problem and to illustrate the formalism. The calculations in the tensor product algebra were performed in \cite{LS2} and \cite{Li:2020zwo} -- here we harvest these results in order to highlight the structure constants for the $S_n$ invariant observables of the discretely gauged theory. 
Firstly, we provide the full set of products in the second quantized Frobenius algebra for the case $n=2$, i.e. for the $\mathbb{Z}_2$ orbifold of the tensor product of two identical topologically twisted conformal field theories.
We moreover provide the  product of the $[2]$ conjugacy class with itself  for generic finite $n$ to illustrate features of the finite $n$ (in-)dependence of the structure constants. Lastly, we study the smallest example (of conjugacy classes $[3]$ times $[3]$) in which we obtain a genus one contribution to the operator product. We illustrate the latter case with the construction of the Riemann surface whose genus is the graph defect. All these examples are precursors of calculations performed in section \ref{StructureConstants}. 

\subsubsection{Example 1: A Complete Description for N=2}
For the example of two tensor product factors ($n=2$ and discrete gauge group $S_2=\mathbb{Z}_2$) and a single twisted sector labelled by the transposition $(12)$, we can list all the products of gauge invariant operators. 
Firstly, the short list of gauge invariant operators is:
\begin{align}
O_{[1,1]} (\alpha_1;\alpha_2) &= \frac{1}{2}(\alpha_1 \otimes \alpha_2+\alpha_2 \otimes \alpha_1) ()
\\
O_{[2]}(\alpha) &= \alpha (12) \, ,
\end{align}
where $\alpha,\alpha_1$ and $\alpha_2$ are operators in the seed Frobenius algebra $A$ and $()$ represents the identity permutation. 
The operator ring is  -- ignoring commutation signs --:
\begin{align}
O_{[1,1]} (\alpha_1;\alpha_2) O_{[1,1]} (\beta_1;\beta_2)
&= \frac{1}{4} (\alpha_1 \beta_1   \otimes \alpha_2 \beta_2 + \text{ three terms })
() \nonumber\\
&= \frac{1}{2} O_{[1,1]}(\alpha_1 \beta_1;\alpha_2 \beta_2 ) + \frac{1}{2} O_{[1,1]}( \alpha_1 \beta_2;\alpha_2 \beta_1)
 \\
O_{[1,1]} (\alpha_1;\alpha_2) O_{[2]}(\beta)&=
\frac{1}{2} (\alpha_1 \alpha_2 \beta + \alpha_2 \alpha_1 \beta) (12) 
\nonumber \\
&=  O_{[2]}\Big(
%\frac{
\alpha_1 \alpha_2 \beta 
%+ \alpha_2 \alpha_1 \beta}{2} 
\Big)
 \\
O_{[2]}(\alpha)O_{[2]}(\beta)
&= \Delta_\ast(\alpha \beta) () 
%\nonumber\\
%&=\Delta_\ast(\alpha \beta) O_{[\phi]}(1)
% &=% (-1)^l 
% \alpha^i \beta^j {c_{ij}}^l {c^{mk}}_l e_k \otimes e_m ~() 
% \nonumber \\
% &= 
% %(-1)^l 
% \alpha^i \beta^j {c_{ij}}^l {c^{mk}}_lO_{[\phi]} ( \{e_k,e_m \})
 \nonumber \\
 &= (\alpha \beta)^m %(-1)^m 
 {c^{ji}}_m \frac{1}{2} (e_i \otimes e_j + e_j \otimes e_i) ()
 \nonumber \\
 &= (\alpha \beta)^m %(-1)^m 
 {c^{ji}}_m  O_{[1,1]} (e_i;e_j)
 \nonumber \\
&
\coloneqq O_{[1,1]}\Big(\Delta_\ast(\alpha \beta)\Big)
\, .
\end{align}
%where $\alpha = \alpha^i e_i$ and $\beta = \beta^j e_j$. 
This fully describes the algebra $A^{[2]}$ of gauge invariant operators. For high values of $n$, an explicit compact description of the algebra $A^{[n]}$ is a very hard combinatorial problem amply out of reach of present knowledge. 

\subsubsection{Example 2: The N-(in)dependence of 2 Times 2}
The next example we wish to describe exactly is the case of the multiplication of two operators associated to transpositions, inside the algebra $A^{[n]}$ for generic $n$. This product  illustrates an interesting property regarding the $n$ dependence of the structure constants. 
Firstly, we recall the result for the multiplication of sums of permutations in the conjugacy classes for generic $n$:
\begin{equation}
C_{[2]} C_{[2]} = 2  C_{[2,2]}
+ 3 C_{[3]} + \frac{n(n-1)}{2} C_{[1^n]} \, .
\label{AFirstExampleWithoutOperators}
\end{equation}
The calculation with operators associated to the orbit of the two-cycles yields on the other hand:
\begin{align}
 O_{[2]}(\alpha) \ast  O_{[2]}(\beta)
% =&\phantom{+} \alpha (12) \ast \beta (12) + \dots \nonumber \\
%  &+ \alpha(12) \ast \beta (13) + \dots
%  \nonumber \\
%  & + \alpha(12) \ast \beta (34) + \dots
%  \nonumber \\
 &=  2 O_{[2,2]}(\alpha,\beta) 
  +3  O_{[3]}( \alpha \beta) 
  + %\frac{n(n-1)}{2} 
  O_{[1^2,1^{n-2}]}(\Delta_\ast^{(2)}(\alpha \beta))~. \label{AFirstExample}
\end{align}
The calculation can be understood by observing that the first term corresponds to disjoint transpositions which can occur in one order or the opposite order giving rise to a factor of two -- the Frobenius algebra elements remain associated to separate cycles. The second term arises from transpositions that have one element in common. The combinatorics works out as in the case of conjugacy classes while the operators in the Frobenius algebra multiply because they belong to a common orbit under the group generated by the transpositions. The last term arises as in  example (\ref{ExampleMultiplication}). 
The first term is defined in such a way that $(\alpha,\beta)$ are distributed over the two two-cycles in one order, or the other order, with weight one half.  In the third term, the $\Delta_\ast^{(2)}$ operator distributes terms over the two marked inactive $1$ entries evenly. We then symmetrize in all choices of two out of $n$ inactive entries with weight one (for a total of $n(n-1)/2$ terms). 
 The third term preserves R-charge because $\Delta_\ast^{(2)}$ carries R-charge $c/3$ (by equation (\ref{RChargeDeltaStarTwo})).
We note that written in the manner of equation (\ref{AFirstExample}) all the structure constants are $n$-independent. Of course, it is now implicit in the choice of marked inactive colours. In other words, when we set the operators $\alpha$ and $\beta$ equal to the unit operator, we recuperate equation (\ref{AFirstExampleWithoutOperators}) because of the symmetrization procedure of the entries in equation (\ref{AFirstExample}).\footnote{This paragraph discusses the N-dependence of the structure constants of conjugacy classes and operators which have a normalization independent of $N$. For a comparison to the N-dependence of the structure constants when normalizing the operators using the untwisted two-point functions in the traditional manner, see \cite{Ashok:2023mow}.} We will see many further illustrations of this type of  combinatorics of marked inactive entries in the rest of the paper.\footnote{See also \cite{IvanovKerov} for further mathematics background.}

\subsubsection{Example 3: The Simplest Genus One Diagram}
\label{GenusOneExample}
Finally, we treat the simplest example in which we have a graph defect one contribution to the operator product. This example will also feature the  adjoint  $\Delta_\ast^{(3)}$ of a higher multiplication operator. It foreshadows a generic genus one contribution computed in section \ref{StructureConstants}.

The multiplication of two conjugacy classes of length three equals \cite{IvanovKerov}:
\begin{equation}
C_{[3]} C_{[3]}=2C_{[3,3]} + 5C_{[5]} + 8 C_{[2,2]} + (3n -8)C_{[3]} +
\frac{n(n-1)(n-2)}{3}
C_{[1^n]} \, . \label{CCThreeTimesThree}
\end{equation}
The corresponding product in the second quantized Frobenius algebra $A^{[n]}$ equals:
\begin{align}
 O_{[3]}(\alpha) \ast  O_{[3]}(\beta) %&= \alpha (123) \beta (321)
&= 
2 O_{[3,3]}( \alpha, \beta )
+ 5 O_{[5]}( \alpha \beta)
+ 8 O_{[2,2]}(\Delta_\ast(\alpha \beta) )
\nonumber \\
& \phantom{=} \,
+ 3 %(n-3)
O_{[3,1]}(\Delta_\ast(\alpha\beta) )
+ 2 O_{[1^3, 1^{n-3}]}(\Delta_\ast^{(3)} (\alpha \beta))
+ O_{[3]}(\alpha \beta e)  \, .
\label{ThreeTimesThree}
\end{align}
%  
%This should be compared to the result when the operators are the identity, namely, the product of conjugacy class sums $C_{[\lambda]}$ in the center of the group algebra of $S_N$: 
The first term is again symmetrized
(with weight one half) and the third term as well. The fourth term in equation (\ref{ThreeTimesThree}) is interesting in that we need to pick one entry out of $n-3$ to associate to the second factor in $\Delta_\ast(\alpha \beta)$. This symmetrization is what leads to a coefficient $3n-9$ when $\alpha=1=\beta$. The previous to last term has a similar choice of $3$ out of $n$. All in all this agrees with the combinatorics of equation (\ref{CCThreeTimesThree}). The last term in \eqref{ThreeTimesThree} is what we are interested in: the first genus one correction proportional to the Euler class $e$. 
%
%R-charge conservation can be checked to work out using the rules previously specified. 
%
%

\begin{figure}[H]
\begin{center}
\begin{tikzpicture}[scale=1.2]

% Define the coordinates of the hexagon nodes
\coordinate (A) at (0,2);
\coordinate (B) at (1.73,1);
\coordinate (C) at (1.73,-1);
\coordinate (D) at (0,-2);
\coordinate (E) at (-1.73,-1);
\coordinate (F) at (-1.73,1);
\coordinate (CTR) at (0,0);

%\tkzDrawSegments[>=stealth](A,B);
%\tkzDrawSegments[>=stealth](D,E,F);

% Draw the hexagon
%\draw[red] (A) -- (B);

%-- (C) -- (D) -- (E) -- (F) -- cycle;

\tkzDrawSegments[arrowMe=stealth](E,F C,B);
\tkzDrawSegments[>=stealth,arrowMe=>>](F,A D,C);
\tkzDrawSegments[>=stealth,arrowMe=>>>](A,B E,D);

% Draw the long diagonals.

\draw (A) -- (D);
\draw (B) -- (E);
\draw (C) -- (F);

% Draw the node labels
% \node[draw,circle,scale=0.7] (N1) at (A) {};
% \node[draw] (N2) at (B) {};
% \node[draw,circle] (N3) at (C) {};
% \node[draw] (N4) at (D) {};
% \node[draw,circle] (N5) at (E) {};
% \node[draw] (N6) at (F) {};
% \node[draw,circle,dashed] (N7) at (CTR) {};

% Draw the edges
\draw (A) -- (B);
\draw (B) -- (C);
\draw (C) -- (D);
\draw (D) -- (E);
\draw (E) -- (F);
\draw (F) -- (A);

\end{tikzpicture}

\end{center}
\caption{The genus one hexagon}
\label{GenusOneHexagon}
\end{figure}
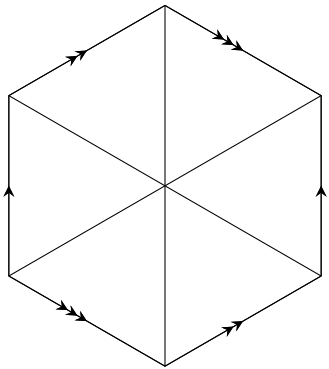

We want to draw the Riemann surface with genus equal to the graph defect for the last term on the right hand side of equation (\ref{ThreeTimesThree}). We follow Lemma 2.7 in \cite{LS2} for the manner in which the Riemann surface is constructed.  We are instructed to draw three vertices corresponding to the $[3]$ conjugacy class in the first operator as well as the second and the third -- in the term at hand, all the permutations permute the same three vertices.\footnote{The reader is welcome to keep  the example product of permutations $(123)(123)=(132)$ in mind.} For ease of visualization, it is convenient to draw two vertices thrice and to put them in alternating order on the edges of a hexagon, as in Figure
\ref{GenusOneHexagon}. The third vertex is placed in the middle. The six faces of the Riemann surface are  the triangles that fill in the hexagon. There are nine edges. 
 The resulting Riemann surface has Euler number $\chi=3-9+6=0$, i.e. it is a torus.\footnote{In fact, the drawing is known as the $K_{3,3}$ or the Thomsen graph or the three-rung M\"obius ladder. It is useful in the three utility problem which is closely related to its appearance here.} The genus can also be obtained from the Riemann-Hurwitz formula. For a surface with $a$ active colours and $s$ insertions of single cycles of length $n_j$, the genus equals:
 \begin{equation}
 g = \frac{1}{2} \sum_{j=1}^s (n_j-1) -a+1 \, .
 \end{equation}
 In the case at hand we have $s=3$, $n_j=3$ and $a=3$ giving genus one. Using this formula, it is possible to see that the case we treated is indeed the simplest available. Note that it has three colours which are active in all three permutations.  
 The fact that this triple of permutations gives rise to a genus one Riemann surface implies (via the multiplication formula (\ref{Multiplication})) that we have an Euler class insertion for this one loop structure constant. 

% 
% Lemma 2.7 in [Lehn-Soerger, K3]. The first example with a genus one diagram, three vertices. Dual graph. We have two incoming strings with colours $(123)$ in that order on each incoming string. The outgoing string has colours $(132)$, in that order. We want to draw a string diagram such that we can draw a line from the two initial colours $a$ to the final colour $a$, and this for each colour $a \in \{ 1,2,3 \}$. This is impossible on the sphere (with three holes) but it is possible on the torus (with three holes). 
%See http://www.weddslist.com/groups/genus/1/hex.php. 
% The dual graph is six triangles forming a hexagon with opposite sides identified.  We have $(V,E,F)=(6,9,3)$ and for the dual $(V,E,F)=(3,9,6)$. 
% Other forms of the graph are at the commented out URL (about the three utilities problem). (Also: four utilities)  %https://en.wikipedia.org/wiki/Three_utilities_problem  
%

That finishes our introduction to the operator product in the second quantized Frobenius algebras and some properties of the product at finite $N$. The examples served to ease into the harder combinatorics of the calculation of the generic large $N$ structure constants of the operator product at multiple orders in perturbation theory.

\section{The Large N Order and Triples of Permutations}
\label{TripleOverlapExpansion}
\label{LargeN}
\label{LargeNExpansion}
\label{TriplesOfPermutations}

Our goal is to compute structure constants for the topological orbifold of the symmetric product of a Frobenius algebra order by order in the $1/\sqrt{N}$ expansion. The operators involved in the  product are labelled by conjugacy classes and they are  given by a sum over permutations that belong to that conjugacy class. The product of such operators will therefore involve sums of  contributions that each arise from the product of permutations (and we suppress the role of the Frobenius algebra for the moment). Not all of these contributions have the same large $N$ behaviour. From the work of \cite{Belin:2015hwa}, it is known that if we normalize these operators such that the two point functions in the untwisted theory have unit coefficient, the large $N$ scaling of each of the terms appearing in the operator product is determined by the triple overlap $|T|$ between the three permutations involved. By triple overlap here, we mean the number of active colours that appear in all three of the permutations.  Thus, the large $N$ order of each term is given by 
\be 
N^{-|T|/2}~.
\ee
%We have already seen glimpses of this for the small-N examples worked out in the previous section. 
It therefore makes sense for us to classify and list products of two permutations based on the number of triple overlap colours. In this section we undertake an exhaustive study of products of permutations, classified by the order of the triple overlap number, all the way up to $|T|=3$. We perform this task up to order three as this is the first order (in the large $N$ expansion) at which one finds a genus one correction. Our classification will then be put to good use in  section \ref{StructureConstants} in which we shall study products of operators labelled by conjugacy classes and extract the zeroth, first, second and third order contributions in a systematic manner. 

% 
%Also, we can define the order as:
% \begin{equation}
% \text{Ord} = -2+2g+p \, .
% \end{equation}
% These are the same, when properly counted (including increasing $p$ when separate cycles are involved in a given permutation).

\subsection{Preliminaries}
We consider three permutations $\pi_i$ that satisfy the relation $\pi_1 \pi_2=\pi_3$.  
We work orbit by orbit with respect to the subgroup $\langle \pi_1 , \pi_2 \rangle \subset S_n$. We can refer to each orbit as a connected part of a diagram. We define the set of colours in one orbit to be $O$. 
Let's define some further sets of colours. The sets $A_i$ are the colours active under permutation $i$ while the sets $I_i$ contain the colours inactive under permutation $i$ in orbit $O$. We therefore have the equations: 
\begin{align}
O &= A_i \sqcup I_i \, ,
\end{align}
where $\sqcup$ denotes a disjoint union of sets. 
We moreover always have the equations:
\begin{align}
\nonumber \\
A_{i} &= (A_{i} \cap I_{i+1}) \sqcup (A_{i} \cap I_{i+2}) \sqcup 
T \, 
\end{align}
for $i=1,2,3$ and where the indices are defined modulo $3$. The set $T$ is the set of colours which are active in all three permutations. The reason for this equation is that a colour $i$ in the set $A_i$ is active under the permutation $\pi_i$. If it is also active under the other two permutations, it belongs to $T$. If it is inactive under one of the other two permutations, it must be active under the other. Hence, the disjoint union is proven. 

Furthermore, if we suppose that we have a non-trivial orbit (and we do), we have that the set $I_1$ of colours in the orbit inactive under $\pi_1$ must be active in $\pi_2$ (and therefore in $\pi_3$) since it is in the non-trivial orbit $O$. 
Thus, we can introduce a unique set $C$ such that
\begin{equation}
A_{2} \cap I_1 = C = A_{3} \cap I_1 \, .
\end{equation}
Similarly, we can name the sets:
\begin{equation}
A_2 \cap I_3 = B  = A_1 \cap I_3 \, , \qquad 
A_3 \cap I_2 = A = A_1 \cap I_2  \, .
\end{equation}
Clearly, the full (non-trivial) orbit $O$ is made of colours that are either triply active, or doubly active. 
All four sets are mutually exclusive and they fill out the above equations entirely:
\begin{align}
O &= T \sqcup A \sqcup B \sqcup C
\nonumber \\
A_1 &= T \sqcup A \sqcup B
\nonumber \\
A_2 &= T \sqcup B \sqcup C
\nonumber \\
A_3 &= T \sqcup C \sqcup A \, .
\end{align}
Another way to put the same information is that $A$ are the colours active under permutations $\pi_{1,3}$ only, $B$ the colours active under $\pi_{1,2}$ only and $C$ the colours active under $\pi_{2,3}$ only. For future reference we denote the colours in the triple overlap by $T=\{t_1, t_2, \ldots \}$, the colours in $A$ by $\{a_1, a_2, \ldots \}$, those in $B$ by $\{b_1, b_2, \ldots \}$ and those in $C$ by $\{c_1, c_2, \ldots \}$.

It is instructive to state which of these sets are mapped into which by the various permutations. Consider an element $b \in B$. We prove that by the permutation $\pi_1$, this element cannot be mapped into an element $a \in A$. Suppose that:
\begin{equation}
\pi_1(b) = a \, .
\end{equation}
We then act by $\pi_2$ and find:
\begin{equation}
\pi_2(\pi_1(b))=\pi_3(b)=b=\pi_2(a)=a
\end{equation}
by the definitions of the permutation $\pi_3$ and the sets $A$ and $B$. This contradicts the fact that $A$ and $B$ have zero intersection and therefore the assumption is wrong. No element of $B$ is mapped into the set $A$ by the permutation $\pi_1$. Similarly, we have that no element of $C$ can be mapped into $B$ by $\pi_2$ 
and no element of $C$ can be mapped into $A$ by $\pi_3$. 
Consider then a cycle in the permutation $\pi_1$. It must necessarily be of the symbolic form $(tabtab\dots)$ since colours $b$ must be mapped to triple overlap colours $t$. Similarly, cycles in $\pi_2$ have the form $(tbctbc\dots)$ and in permutation $\pi_3$ they are $(tactac\dots)$.\footnote{The analysis will have a more manifest cyclic symmetry if one performs it in terms of the permutations $\pi_{1,2}$ and $\pi_3^{-1}$.}

There is a simple but useful corollary to this result. When a cycle does not contain a single triply active colour $t$, then it consists of single type of double active colour. For instance, in permutation $\pi_1$, one can then only have cycles $(a \dots)$ or $(b \dots)$. In such an instance, these cycles form separate orbits under the group  $\langle \pi_1 , \pi_2 \rangle $ generated by permutations $\pi_1$ and $\pi_2$.

\subsection{The Leading Order}
\label{sec3leadingO}
We have that the triple overlap number equals
%\begin{equation}
 %\# T = 
the cardinal number 
$ |T|$
of the set $T$.
%\, .
%\end{equation}
At leading order we have the constraint  $|T|=0$.  The  set $T$ is empty. By the previous reasoning, and in particular, the lack of elements $t$,  this implies that the cycles of $\pi_1$ are of the form $(a\dots)$ or $(b \dots)$, and similarly for the other permutations. 
Since the colours $b$ are not active in permutation $\pi_3$, any such cycle of elements in $B$ in $\pi_1$ has an inverse cycle in $\pi_2$. Thus, the permutations must have the form:
\begin{align}
\pi_1 &= (a_1 \dots a_{j_1})(a_{j_1+1} \dots a_{j_2}) \dots (b_1 \dots b_{l_1}) (b_{l_1+1} \dots b_{l_2}) \dots
\nonumber \\
\pi_2 &= (b_{l_1} \dots b_1) (b_{l_2} \dots b_{l_1+1}) \dots (c_1 \dots c_{m_1}) (c_{m_1+1} \dots c_{m_2}) \dots
\nonumber \\
\pi_3 &= (a_1 \dots a_{j_1})(a_{j_1+1} \dots a_{j_2}) \dots (c_1 \dots c_{m_1}) (c_{m_1+1} \dots c_{m_2}) \dots
\end{align}
At leading order, there is no joining nor splitting but only the possible annihilation of cycles. 
We will compute the structure constants for twisted sector operators associated to   these triples of permutations in section \ref{Products}. In this section, we continue the classification of triples of permutations by their triple overlap number.

\subsection{The Subleading Triple Overlap of One}
\label{sec3subleadingoverlap}

The total triple overlap is the sum of the triple overlaps in each orbit. Thus, in the subleading case where $|T|=1$, we have that all orbits have zero triple overlap except one. All  other orbits are analyzed as in the leading order analysis. We will ignore these auxiliary combinatorics from now on -- see e.g. \cite{Ashok:2023mow} -- and concentrate on the non-trivial orbit with $|T|=1$ only.

In this orbit, we have a single colour that is active under all three permutations. It therefore belongs to a non-trivial cycle of the permutation $\pi_1$. All other cycles of $\pi_1$ lack a triply active colour $t$ and therefore behave as in the leading order analysis. As a consequence, they necessarily belong to other orbits. 
Thus, we can concentrate on a single cycle in $\pi_1$. A similar reasoning holds for the other permutations and we can therefore concentrate on a single cycle in all permutations $\pi_i$. From the general analysis, we immediately conclude that these cycles take the form:
\begin{align}
    \pi_1 &= (t a_1\ldots a_j b_1\ldots b_l) \nonumber \\
    \pi_2 &= (t b_l\ldots b_1 c_1\dots c_m)  \nonumber \\
    \pi_3 &= (t a_1 \ldots a_j c_1 \ldots c_m) ~.
\end{align}
We introduced a convention in which we concentrate on the cycles in the orbit under discussion and leave out extra information on potential other orbits which would clutter our expressions. 

\subsection{Triple Overlap Two}
\label{doubleoverlap}
\label{sec3doubleoverlap}

In this subsection, we consider the case of two triply overlapping colours, $|T|=2$. 
If these two colours are in two separate cycles in both the first and the second permutation, then we have two separate orbits of overlap $|T|=1$ and the discussion largely reduces to that of the previous subsection. We have two disconnected processes. The more interesting cases are where at least for one of the two permutations, the two triply overlapping colours are in one and the same cycle. By symmetry (including inversion), we can suppose that at least in permutation $\pi_1$, the two colours are in the same cycle. We are left with two cases for permutation $\pi_2$: the triply overlapping colours $T=\{t_1,t_2 \}$ are in one cycle in $\pi_2$, or in two different cycles.
We discuss the two possibilities separately.

\subsubsection{(1,1;2)}
In the first case, we have a single cycle for $\pi_1$ and $\pi_2$. We can then compute the permutation $\pi_3$ uniquely:
\begin{align}
\pi_1 &= (t_1 a_1 \dots a_{j_1}  b_{1} \dots b_{l_1} t_2 a_{j_1+1} \dots a_{j_2} b_{l_1+1} \dots b_{l_2} )  \\
\pi_2 &=(t_2b_{l_1}\ldots b_1c_1\ldots c_{m_1} t_1 b_{l_2}\ldots b_{l_1+1}c_{m_1+1}\ldots c_{m_2})~, 
\end{align} 
leading to 
\be 
\pi_3= (t_1 a_1, \ldots a_{j_1} c_1 \ldots c_{m_1})(t_2a_{j_1+1}\ldots a_{j_2}c_{m_1+1}\ldots c_{m_2})~. 
\ee
We ensured that the colours $b$ remain inactive in $\pi_3$, leading to a unique ansatz for permutation $\pi_2$ and a unique result for $\pi_3$ which is a double cycle. Hence the title of this subsection,  summarizing the pertinent cycle structure.

\subsubsection{(1,2;1)}

The case in which the two colours $t_1$ and $t_2$ are in different cycles in $\pi_2$ leads to the triple of permutations:
\begin{align}
\pi_1 &= (t_1 a_1 \dots a_{j_1}  b_{1} \dots b_{l_1} t_2 a_{j_1+1} \dots a_{j_2} b_{l_1+1} \dots b_{l_2} ) \, 
\nonumber \\
\pi_2 &=  (  t_2 b_{l_1}  \dots b_1 c_1 \dots c_{m_1})(
t_1 b_{l_2}  \dots b_{l_1+1} c_{m_1+1} \dots c_{m_2}) \nonumber \\
\pi_3 &= (t_2 a_{j_1+1} \ldots a_{j_2} c_{m_1+1} \dots  c_{m_2} t_1a_1\ldots a_{j_1} c_1 \ldots c_{m_1}) \, .
\end{align}
The cycle structure is $(1,2;1)$. 
 Left and right multiplication by inverse group elements on the defining relation $\pi_3 = \pi_1 \pi_2$ relates this case to the previous one. 

\subsection{The Triple Triple Overlap}
\label{sec3tripletripleoverlap}

In this subsection, we consider the case where we have a set  $\{t_1, t_2, t_3\}$ of three triply active colours. 
We wish to treat only those cases that do not reduce to a product of previous sub-cases. Clearly, each case with a permutation which contains a cycle with the three triply active colours is new. The only other case that could potentially be new is when two triply active colours belong to one cycle in $\pi_1$ and two other triply active colours belong to one cycle in $\pi_2$. For this case, the structure of these permutations, without loss of generality, is:
\begin{align}
\pi_1 &= (t_1 a_{1} \dots a_{j_1}  b_1 \dots b_{l_1} t_2 a_{j_1+1} \dots a_{j_2}  b_{l_1+1} \dots b_{l_2})(t_3 a_{j_2+1} \dots a_{j_3}  b_{l_2+1} \dots b_{l_3})
\nonumber \\
\pi_2 &= (t_2 b_{l_1} \dots b_1 c_{m_1+1} \dots c_{m_2} t_3 b_{l_3} \dots b_{l_2+1} c_{m_2+1} \dots c_{m_3})(t_1 b_{l_2} \dots b_{l_1+1} c_1 \dots c_{m_1}) \, .
\end{align}
Given this structure, the permutation $\pi_3$ is of the form:
\begin{align}
\pi_3 &= (t_1 a_1 \dots a_{j_1} c_{m_1+1} \dots c_{m_2} t_3 a_{j_2+1} \dots a_{j_3} c_{m_2+1} \dots c_{m_3} t_2 a_{j_1+1} \dots a_{j_2} c_1 \dots c_{m_1} ) \, .
\end{align}
Therefore, all new cases involve a permutation that incorporates all three triply active colours in one cycle. For simplicity, we choose this permutation to be the permutation $\pi_1$ in the systematic treatment below. 
Thus, our starting point is the specification of the permutation $\pi_1$: 
\begin{align} 
\pi_1 &= (t_1 a_{1} \dots a_{j_1}  b_1 \dots b_{l_1} t_2 a_{j_1+1} \dots a_{j_2}  b_{l_1+1} \dots b_{l_2}t_3 a_{j_2+1} \dots a_{j_3}  b_{l_2+1} \dots b_{l_3}) \, .
\end{align}

\subsubsection{(1,3;1)}
In the first sub-case, 
we choose the permutation $\pi_2$ to consist of three cycles containing a triply active colour each; we then find the three permutations:
\begin{align}
\pi_1 &= (t_1 a_{1} \dots a_{j_1}  b_1 \dots b_{l_1} t_2 a_{j_1+1} \dots a_{j_2}  b_{l_1+1} \dots b_{l_2}t_3 a_{j_2+1} \dots a_{j_3}  b_{l_2+1} \dots b_{l_3}) \\
\pi_2 &=(t_1b_{l_3}\ldots b_{l_2+1} c_1 \dots c_{m_1})(t_2 b_{l_1} \dots b_1 c_{m_1+1} \dots c_{m_2})( 
t_3 b_{l_2} \dots b_{l_1+1}
c_{m_2+1} \dots c_{m_3})\\
    \pi_3&= (t_1 a_1\ldots a_{j_1} c_{m_1+1} \dots c_{m_2} t_2a_{j_1+1}\ldots a_{j_2} c_{m_2+1} \dots c_{m_3} t_3a_{j_2+1}\ldots a_{j_3}dc_1\ldots c_{m_1})~.
\end{align}

\subsubsection{(1,2;2)}
If we instead distribute the triply active colours in the permutation $\pi_2$ between two cycles, we find the triple of permutations 
 \begin{align}
\pi_1 &= (t_1 a_{1} \dots a_{j_1}  b_1 \dots b_{l_1} t_2 a_{j_1+1} \dots a_{j_2}  b_{l_1+1} \dots b_{l_2}t_3 a_{j_2+1} \dots a_{j_3}  b_{l_2+1} \dots b_{l_3}) \\
\pi_2 &=(t_1b_{l_3}\ldots b_{l_2+1} c_{m_2+1} \dots c_{m_3}
t_2 b_{l_1} \dots b_1
c_{m_1+1} \dots c_{m_2}
)( 
t_3 b_{l_2} \dots b_{l_1+1}c_1\ldots c_{m_1})\\
    \pi_3 &=(t_1 a_1\ldots a_{j_1} c_{m_1+1} \dots c_{m_2})(t_2 a_{j_1+1}\ldots a_{j_2}c_1\ldots c_{m_1}t_3a_{j_2+1}\ldots a_{j_3} c_{m_2+1}\ldots c_{m_3})~.
\end{align}

\subsubsection{(1,1;1)}
We now consider the cases where the three triply active colours are in a single cycle in both permutations $\pi_1$ and $\pi_2$. For the permutation $\pi_2$, we can choose two ways in which we order  the triply active colours $t_{1,2,3}$; either they occur in the same order as in $\pi_1$ or they occur in the opposite order.
When they are in the same order, we find that the third permutation is also a single cycle:
\begin{align}
\pi_1 &= (t_1 a_{1} \dots a_{j_1}  b_1 \dots b_{l_1} t_2 a_{j_1+1} \dots a_{j_2}  b_{l_1+1} \dots b_{l_2}t_3 a_{j_2+1} \dots a_{j_3}  b_{l_2+1} \dots b_{l_3}) \\
\pi_2&= (t_1 b_{l_3} \ldots b_{l_2+1}c_{1}\ldots c_{m_1}t_2 b_{l_1} \ldots b_1 c_{m_1+1}\ldots c_{m_2} t_3 b_{l_2}\ldots b_{l_1+1} c_{m_2+1}\ldots c_{m_3}) \\
%\end{equation}
%\begin{equation}
\pi_3&=
(t_1a_1\ldots a_{j_1}c_{m_1+1}\ldots c_{m_2}t_3 a_{j_2+1}\ldots a_{j_3}
c_1\ldots c_{m_1}
t_2a_{j_1+1}\ldots a_{j_2}c_{m_2+1}\ldots c_{m_3})
\, .
\end{align}
%The orbits 
%
\subsubsection{(1,1;3)}
For the opposite order, we find that the third permutation splits into a triple cycle permutation:
\begin{align}
\pi_1 &= (t_1 a_{1} \dots a_{j_1}  b_1 \dots b_{l_1} t_2 a_{j_1+1} \dots a_{j_2}  b_{l_1+1} \dots b_{l_2}t_3 a_{j_2+1} \dots a_{j_3}  b_{l_2+1} \dots b_{l_3}) \\
\pi_2&= (t_1 b_{l_3} \ldots b_{l_2+1}c_{1}\ldots c_{m_1} t_3 b_{l_2}\ldots b_{l_1+1} c_{m_1+1}\ldots c_{m_2} t_2 b_{l_1} \ldots b_1 c_{m_2+1}\ldots c_{m_3})  \\
%\end{equation}
%We compute $\pi_3$ in this case:
%\begin{equation}
\pi_3 &= (t_1 a_1 \dots a_{j_1} c_{m_2+1}\ldots c_{m_3} )
(t_2 a_{j_1+1}+ \dots a_{j_2} c_{m_1+1} \dots c_{m_2} ) 
(t_3 a_{j_2+1} \dots a_{j_3} c_{1} \dots c_{m_1} )
\, .
\end{align}
% % \item We consider the case when all three triply common colours are adjacent:
% % \begin{align} 
% % \pi_1 &= (c_1  c_2 c_3 a_{1} \dots a_{j_1}  b_1 \dots b_{l_1}) \, .
% % \end{align}
% % Then we have the following possibility:
% % %
% % \begin{align}
% %     \pi_2 &=(c_2c_3c_1b_{l_1}\ldots b_1)\\
% %     \pi_3 &=(c_2c_1 c_3a_1 \ldots a_{j_1})
% % \end{align}
% %
% The graph defect $g$ equals
%
% %
% \be 
% g= \frac12(j_1+l_1+3+2-(1 +(j_1+1)+(l_1+1)) = 1~.
% \ee 
% %
We have systematically identified  all processes which contain zero, one, two or three triple overlaps. It would certainly be interesting to  analyze the systematics of the permutation group combinatorics and how it interacts with the large $N$ expansion further.  We stop at the third order since it is the order at which we will see the first genus one contribution to the structure constants of the chiral ring -- see subsection \ref{GenusOneExample}.

%

% \subsection{Higher Overlap (1,1;1)}
% Can we generally argue what $\pi_2$ looks like if we want the end result to be a single cycle, from two single cycles ? Let's see. We start with a generic single cycle $\pi_1$:
% \begin{align}
% \pi_1 &= (t_1 a_{1} \dots a_{j_1}  b_1 \dots b_{l_1} t_2 a_{j_1+1} \dots a_{j_2}  b_{l_1+1} \dots b_{l_2}
% %t_3 a_{j_2+1} \dots a_{j_3}  b_{l_2+1} \dots b_{l_3} t_4 a_{j_3+1} \dots
% \ldots  t_r
% a_{j_{r-1}+1} \dots a_{j_r} b_{l_{r-1}+1} \dots b_{l_{r}} ) \, .
% \end{align}
% We consider a single cycle $\pi_2$:
% \begin{align}
% \pi_2&= (t_{1} b_{l_r} \ldots b_{l_{r-1}+1}c_{1}\ldots c_{m_1} t_{x_1} b_{l_{x_1}}\ldots b_{l_{x_1-1}+1}
% t_{x_2} \dots) 
% \end{align}
% but only such that $\pi_3$ becomes a single cycle. What are the possibilities ?
%\end{equation}
%\begin{equation}

% We find:
% \begin{align}
% \pi_3&=
% (t_1a_1\ldots a_{j_1}
% c (following b_1) (rest of that list of c's) 
% %\ldots c_{m_2}t_3 a_{j_2+1}\ldots a_{j_3}
% %c_1\ldots c_{m_1}
% %t_2a_{j_1+1}\ldots a_{j_2}c_{m_2+1}\ldots c_{m_3})
% \end{align}

\section{The Operator Products}
\label{OPs}
\label{Products}
\label{OperatorProducts}
\label{StructureConstants}
In the previous section, we calculated the structure of the triples of permutations that contribute at a given value $|T|$ of the triple overlap of the permutations. The latter parameter governs the leading large $N$ behaviour of the associated structure constant. Notably, a given conjugacy class structure constant may acquire contributions from differing triple overlaps. In that case, higher order terms in the large $N$ expansion of a lower triple overlap structure constant can mingle with the leading large $N$ term of a higher triple overlap interaction. Thus, it is useful to split the contributions to the product of conjugacy class sums in terms of contributions from various values of triple overlaps. We thus define:
\be 
O_{\rho_1} \ast O_{\rho_2} =\sum_{\rho_3}\left[ \left(\sum_{|T|=0}^\infty {{\cal C}_{\rho_1 \rho_2}}^{\rho_3}({|T|}) \right) O_{\rho_3} \right]~. 
\ee 
Here, we have suppressed the Frobenius algebra elements associated to each of the operators $O_{\rho_i}$ and have specified only the permutation labels. 
Note that the structure constant at fixed triple overlap number $|T|$ is invariant under simultaneous conjugation of $\pi_{1,2}$ and $\pi_3$. Thus, even at fixed triple overlap number, we indeed have a conjugacy class coefficient on the right hand side.   

For each triple of conjugacy classes, we can determine the lowest triple overlap of triples of permutations that contributes to their structure constant. This will determine the leading large $N$ behaviour of that structure constant. Schematically, we can ask for the leading order structure constant ${{\cal C}_{\rho_1 \rho_2}}^{\rho_3}$
\begin{equation}
O_{\rho_1} \ast O_{\rho_2} = ({{\cal C}_{\rho_1 \rho_2}}^{\rho_3} (|T|_{\text{min}}) +\dots) ~ O_{\rho_3} + \dots \, .
\end{equation}
If one wants to compute a subleading order structure constant, one has to take into account a non-minimal triple overlap $|T|$ as well as possible subleading corrections from a leading order (minimal $|T|$) structure constant. With these structural remarks in mind, we start computing the perturbation theory.

\subsection{The Leading Order}
In this subsection, we compute the leading order contribution to the product of two operators labelled by arbitrary permutations. The leading order contributions have zero triple overlap. From the discussion in Section \ref{sec3leadingO} we know that we must have either disjoint multiplication of the permutations or the full annihilation of cycles. For simplicity only, we focus on operators associated to permutations which have distinct cycle lengths (within one given permutation).  This makes for an easier calculation of the symmetry factor.\footnote{When some cycle lengths are equal, the calculations are similar to computations performed in  \cite{Ashok:2023mow}.}
We would like to calculate the product 
\begin{equation}
O_{[n_1,n_2,\dots]}
( \alpha_1,\alpha_2,\dots ) 
\ast 
O_{[p_1,p_2,\dots]}
(\beta_1, \beta_2,\dots)
\end{equation}
at leading order $|T|=0$, and under the assumption that the integers $n_i$ are mutually distinct as are the integers $p_j$. The operators $\alpha_i$ are associated to the cycles of length $n_i$. Once more, for simplicity only, we assume that the operators associated to untwisted tensor product factors are the identity.  Each term in the gauge invariant operators is a familiar permutation multiplied by a set of operators associated to the (non-trivial) orbits of the permutations. In a first contribution on the right hand side, the permutations $\pi_{1,2}$ are entirely disjoint and the conjugacy class of the composite permutation corresponds to the union of the partitions $n_i$ and $p_i$ (and an appropriate number of ones). A second contribution arises from permutations $\pi_2$ that have a single cycle which is the inverse of a single cycle in $\pi_1$. And so on. Whenever a cycle is annihilated, the corresponding operators $\alpha_i$ and $\beta_j$ multiply into $\alpha_i \beta_j$ and then must be redistributed over the tensor product of spaces corresponding to that cycle with the operator $\Delta_\ast^{(n_i)}$ -- see example (\ref{ExampleMultiplication}).
The remaining task is to determine the symmmetry factor for each term. This is a matter of counting the number of permutations $\pi_1$ and $\pi_2$ in the conjugacy classes that can give rise to a permutation $\pi_3$ in the conjugacy class on the right hand side. Before we announce the final result for the product, let us perform this calculation separately. 

We concentrate on a permutation $\pi_1$ in the given conjugacy class, and have:
\begin{equation}
\frac{n!}{\prod_i n_i (n-\sum_i n_i)!} 
\end{equation}
choices. Suppose we wish to determine the number of permutations $\pi_2$ that give rise to the conjugacy class in which we have annihilated the cycles labelled by $n_s$ where $s$ runs over some subset $S$ of indices of the set of $n_i$. Of course, the subset $n_s$ is then also a subset of the $p_j$. In this instance, we have 
\begin{equation}
\frac{(n-\sum_i n_i)!}{\prod' p_j  \dots (n- \sum_i n_i -\sum'_j p_j)!}
\end{equation}
choices for the permutation $\pi_2$ where the primed sum is over all indices that are not in the subset $S$. Indeed, the part of $\pi_2$ which corresponds to the set $S$ is already fixed (since it is inverse to that part in $\pi_1$). The other colours $p_j$ can be chosen among those colours not active in $\pi_1$. That gives rise to the counting above. Finally, the set of permutations $\pi_3$ is combined into all elements of a conjugacy class and we must therefore divide the total number of terms  by the number of elements in the final conjugacy class. It is important to note that this is influenced by the number of $n_i$ equal to $p_j$.  We must also keep in mind that there are marked entries in the final permutation corresponding to the operators $\Delta_\ast^{(n_s)}$.  Among the tensor product factors that are not permuted by permutation $\pi_3$, there are special entries associated to the set of integers $n_s$. We also need to symmetrize over all possible choices for those. There are $n-\sum' n_i - \sum' p_j$ invariant factors under the  permutation $\pi_3$. We choose $n_{s_1}, n_{s_2}$ et cetera among these factors to associate the $\Delta_\ast$ operators to. These are another:
\begin{equation}
\frac{(n-\sum' n_i - \sum' p_j)!}{\prod_s n_s ! (n-\sum n_i - \sum'p_j)!}
\end{equation}
choices. We have to divide by this factor as well. We conclude that we need to multiply by: 
\begin{equation}
%\prod_{i}' \left( \begin{array}{c} m_{n_i}([n_i]) + m_{n_i}([p_i]) \\ m_{n_i}([n_i]) \end{array} \right)
2^{ | \{ n_i \} \cap \{p_j \} |-|S|}
\frac{\prod_i' n_i \prod_j' p_j (n-\sum' n_i - \sum' p_j)!}{n!}
\times 
\frac{\prod_s n_s ! (n-\sum n_i - \sum'p_j)!}{(n-\sum' n_i - \sum' p_j)!}
\, .
\end{equation}
%$ | \{ n_i \} \cap \{p_j \} |$
The first factor takes care of the factors of two associated to left-over $n_i$ equal to $p_j$. See \cite{Ashok:2023mow}.
%\footnote{We have written it in a fastidious fashion since in this manner it coincides with a factor determined in more general settings in \cite{Ashok:2023mow}.}
The final symmetry factor is the product of all these  numbers which equals:
\begin{align}
% &\frac{n!}{\prod_i n_i (n-\sum_i n_i)!} 
% \frac{(n-\sum_i n_i)!}{\prod' p_j  \dots (n- \sum_i n_i -\sum'_j p_j)!}
% \times \frac{\prod_i' n_i \prod_j' p_j (n-\sum' n_i - \sum' p_j)!}{n!}
\text{Symmetry Factor} &= 
2^{ | \{ n_i \} \cap \{p_j \} |-|S|} \prod_{s \in S} (n_s-1)! \, .
\end{align}
We therefore find the dominant term in the multiplication of operators associated to arbitrary conjugacy classes: 
\begin{align}
& O_{[n_1,n_2,\dots]}
( \alpha_1,\alpha_2,\dots) 
\ast  
O_{[p_1,p_2,\dots]}
( \beta_1, \beta_2,\dots )\Big|_{|T|=0} = \nonumber 
\\
&
2^{ | \{ n_i \} \cap \{p_j \} |}
O_{[n_1,n_2,\dots p_1,p_2,\dots]}
(  \alpha_1,\alpha_2,\dots  \beta_1, \beta_2,\dots ) \nonumber \\
&+ 2^{ | \{ n_i \} \cap \{p_j \} |-1} \sum_{i,j} \delta_{n_i,p_j}  
%\frac{(n-\sum n_j -\sum p_k + 2 n_i )!}{n_i (n- \sum_j n_j -\sum_k p_k + n_i)!}
(n_i-1)!
O_{[n_1, \dots \slashed{n}_{i}\dots p_1 \dots \slashed{p}_{j}\dots 1^{n_i}]} (  
\dots \slashed{\alpha}_i, \dots 
\slashed{\beta}_j,\dots 
;\Delta_\ast^{(n_i)}(\alpha_{i} \beta_{j} ) )\nonumber \\
&+
2^{ | \{ n_i \} \cap \{p_j \} |-2} 
\sum_{i_1,i_2,j_1,j_2} \delta_{n_{i_1},p_{j_1}}\delta_{n_{i_2},p_{j_2}} (n_{i_1}-1)!(n_{i_2}-1)!
%\frac{(n-\sum n_j -\sum p_k + 2 n_{i_1}+2 n_{i_2} )!}{n_{i_1} n_{i_2} (n- \sum_j n_j -\sum_k p_k + n_{i_1}+n_{i_2})!}
\nonumber \\
&\times 
O_{[ \ldots, \slashed{n}_{i_1},\ldots ,\slashed{n}_{i_2}, \ldots  %\slashed{p}_{j_1}\ldots\slashed{p}_{j_2} 
%,\ldots 
,1^{n_{i_1}}, 1^{n_{i_2}} ]} 
( \dots ,\slashed{\alpha}_{i_1},\dots, \slashed{\alpha}_{i_2}, \dots  % 
;\Delta_\ast^{(n_{i_1})}(\alpha_{{i_1}} \beta_{{j_1}}) ;\Delta_\ast^{(n_{i_2})}(\alpha_{{i_2}} \beta_{{j_2}}))+\ldots   
\end{align}
This is the generic leading order interaction in second quantized Frobenius algebras. Compared to the leading order result for non-compact models  \cite{Ashok:2023mow}, there is the novel possibility of annihilating cycles, in turn opened up by the existence of a non-trivial topological metric $\eta$ which carries non-zero degree.  One can check that the first term agrees with a special case of the formulas in \cite{Ashok:2023mow}.
% (i)
% \begin{equation}
% \frac{n!}{\prod_i n_i (n-\sum_i n_i)!}
% \frac{(n-\sum_i n_i)!}{p_1  p_2  \dots (n- \sum_i n_i -\sum_j p_j)!} \times \frac{\prod_i n_i \prod_j p_j (n-\sum n_i - \sum p_j)!}{n!} = 1
% \end{equation}
% (ii)
% \begin{align}
% &\frac{n!}{\prod_i n_i (n-\sum_i n_i)!} 
% \frac{(n-\sum_i n_i)!}{p_2  \dots (n- \sum_i n_i -\sum'_j p_j)!}
% \times \frac{\prod_i' n_i \prod_j' p_j (n-\sum' n_i - \sum' p_j)!}{n!}\nonumber \\
% &= \frac{ (n-\sum n_i - \sum p_j + 2n_1)!}{n_1(n- \sum_i n_i -\sum_j p_j +n_1)!}
% \end{align}

%Namely, we have topological free field behaviour. 
%The N-dependence is trivial. 
%If cycles are not all distinct, symmetry factors have to be taken into account (much as in our previous paper, it seems to us). 

\subsection{The First Order}

We now consider the  product of operators for which there is a subleading contribution to the structure constant. We study a product in which there is a single colour that is common to all three operators, and that corresponds to the triple overlap $|T|=1$. From our analysis in Section \ref{sec3subleadingoverlap} it is clear that the elementary process corresponds to the  product of two single cycle operators giving rise to a third single cycle operator.    
\be 
O_{[n_1]}(\alpha) \ast O_{[p_1]}(\beta) \Big|_{|T|=1} = \sum_{q_1} {\cal C}_{|T|=1}(q_1)\, O_{[q_1]}(\gamma)~.
%= \sum_{|T|=0}^\infty {\cal C}_{|T|}  C_{[q_1(|T|)]}~. 
\ee 
% \be
% {\cal C} = \sum_{|T|=0}^\infty {\cal C}_{|T|} \, .
% \ee
%For each right hand side conjugacy class, we can determine the lowest order $|T|$ for which ${\cal C}_{|T|}$ is non-zero. 
%Statement: if the left hand side conjugacy classes are single cycle, and the right hand side as well, then $|T|=1$ gives the only non-zero structure constant, i.e. ${\cal C}={\cal C}_{1}$. Proof ?
As we shall see, the sum on the right hand side  arises from a sum over all possible double overlaps between the two operators on the left hand side.

Let us consider the product of a pair of permutations appearing in the conjugacy classes on the left hand side. Following our notation for the colours, we have the individual term
\begin{equation}
\label{LOpermutations}
\alpha (t a_1 a_2 \dots a_j b_1 b_2 \dots b_l ) \ast \beta (t b_l b_{l-1} \dots b_2 b_1 c_1\ldots c_m ) = 
%O(\alpha,\beta;j,l,m)~
\gamma (t a_1 \dots a_j c_1 \dots c_m) ~.
\end{equation}
Since the lengths of the cycles are fixed, we have the relations
\be
\label{cyclelengths}
n_1 = 1+j+l~, \quad p_1 = 1+l+m~, \quad\text{and}\quad q_1 = 1+j+m~.
\ee 
It is important to note that the cycle length of the permutation $\pi_3= \pi_1\pi_2$ can be expressed in terms of the  lengths of the single cycle permutations and the number $l$ of (extra) double overlaps between the permutations $\pi_1$ and $\pi_2$: 
\be 
\label{q1fixed}
q_1 = n_1+p_1 - 2l-1~.
\ee 
Therefore,  the final result will be a sum over conjugacy classes whose length is determined by the  double overlap between the colours of the fusing operators. Finally, from the form of the permutation $\pi_3$, it is clear that the inequality $q_1\ge 2$ must hold for the triply active colour $t$ to remain part of a non-trivial cycle.  

The total number of colours involved in the orbit is given by $1+j+l+m$.
%=(n_1+p_1+q_1-1)/2. 
We have the orbits in the space $O$ of active colours, $O=\{t, a_i, b_r, c_s\}$:
\begin{align}
\langle \pi_1 \rangle \setminus O
&= \{ \{ t, a_i, b_r \}, c_1,\dots, c_m  \}
\nonumber \\
\langle \pi_2 \rangle \setminus O
&=  \{ \{ t, b_r, c_s \}, a_1, \ldots a_j \}
\nonumber \\
\langle \pi_1 \pi_2 \rangle \setminus O
&=  \{ \{ t, a_i, c_s \}, b_1, \ldots b_l \}
\nonumber \\
\langle \pi_1 , \pi_2 \rangle \setminus O
&= \{ t,a_i ,b_r ,c_s \} \, .
\end{align}
We have omitted the curly brackets on sets with one element. The graph defect (\ref{GraphDefect}) can be computed to be
\begin{align}  
g &= \frac12(1+j+m+l+2 -(m+1)-(j+1)-(l+1)) = 0~.
\end{align}
We then apply the multiplication maps $f^{\cdot}$ discussed in subsection \ref{SecondQuantizedFrobeniusAlgebras}.
We need the maps $f^{\pi_1,\langle \pi_1,\pi_2 \rangle}$, $f^{\pi_2,\langle \pi_1,\pi_2 \rangle}$ as well $f_{\langle \pi_1,\pi_2 \rangle, \pi_3}$. The first map associates the operator $\alpha \in A$ to the full set of active colours $H$. The second map does the same for the operator $\beta$. We then multiply them.
Thus, at graph defect zero, we obtain the  product $\alpha \beta$ at this preliminary stage.
The third map then co-multiplies the resulting operator over the orbits of permutation $\pi_3$. 
%These maps reduce the number of tensor product factors. 
%In the case at hand, the resulting number of factors is one. 
 We thus need to sprinkle the product over the $l+1$ orbits of $\pi_3$ using the adjoint map
$\Delta_\ast^{(l+1)}$. 
%Although one of the orbits is "big", I gather that each orbit is treated on an equal footing in this distribution. 
We thus find 
\begin{equation}
\alpha \pi_1\ast \beta \pi_2 = \Delta_\ast^{(l+1)}(\alpha \beta) \pi_1 \pi_2 \, ,
\end{equation}
and identify the Frobenius algebra element $\gamma$ in equation (\ref{LOpermutations}):
\be 
\gamma = \Delta_{\ast}^{(l+1)}(\alpha \beta)~.
\ee 
%
% Using the formulae for the R-charges in We can check that this is consistent with R-charge conservation. We have that
% %
% \begin{align}
% Q(\text{l.h.s})=Q(\alpha) + Q(\beta) + (n_1-1)+(p_1-1)~,\\ 
% Q(\text{r.h.s})=Q(\alpha) + Q(\beta) + Q(\Delta_{\ast}^{(l+1)}) + (q_1-1)~.
% \end{align}
% %
Using the expressions \eqref{cyclelengths}  of the cycle lengths and the formula \eqref{Rchargeofpermutation} for the R-charges associated to the permutation as well as the  R-charge \eqref{RchargeDeltastar} of the operator
$\Delta_{\ast}^{(l+1)}$,
we confirm that the product is consistent with the conservation of R-charge.

We are now ready to extend this result for an individual term to the full product of conjugacy classes by adding the appropriate symmetry factors to the calculation. In other words we compute the numerical coefficient that appears in the product of conjugacy classes, assuming the non-degeneracy of cycle numbers as before. We compute the structure constant at each given value of  the double overlap $l$. Given that the cycle lengths are fixed and using equation \eqref{cyclelengths}, this also fixes both the numbers $j$ and $m$ that characterize the permutations  \eqref{LOpermutations}. 
\paragraph{i.} We begin with a count of the number of elements in the first conjugacy class:
\begin{equation}
|C_{n_1;n}|=\frac{n!}{n_1 (n-n_1)! } \, .
\end{equation}

% \paragraph{ii.} The number of ways in which we can pick the triple overlap  colour $t$ is given by 
% \begin{equation}
% n_1~.
% \end{equation}

\paragraph{ii.} We next identify the colour $b_1$ in $\pi_1$, to carry it over into the permutation $\pi_2$. We pick a colour $b_1$ in 
\begin{equation}
n_1
\end{equation}
ways. This automatically fixes the triple overlap colour $t$ (since it follows $b_l$) as well as the rest of the colours $a_i$ in the first permutation.

\paragraph{iii.} 
%For any choice of $a_1\ldots a_j$ and $b_1, \dots, b_l$ in the first permutation, the ordering of these colours in the second and resulting permutation is fixed.  However the colours $c_1,\dots, c_m$ in the second permutation can be chosen among $n-n_1$ colours. Since the ordering among the cokours is irrelevant, this gives rise to a factor:
We then pick $m$ c-colours out of $n-n_1$ colours and order them, for a total factor:
\begin{equation}
\frac{(n-n_1)!}{ (n-n_1-m)!}= \frac{(n-n_1)!}{ (n-q_1-l)!} \, ,
\end{equation}
where we have rewritten $m$ in terms of $q_1$ and $l$. This completes the identification of the permutation $\pi_2$. 
% They can be ordered in 
% \begin{equation}
% m!
% \end{equation}
% different ways. 

\paragraph{iv.} We now divide by the number of elements in the last conjugacy class. We therefore multiply by:
\begin{equation}
\frac{q_1 (n-q_1)!}{n!} \, .
\end{equation}
\paragraph{v.} Finally we still need to pick $l$ colours out of the remaining $n-q_1$ colours (over which we distribute the result of the $\Delta_\ast$-operation). This contributes a factor:
\begin{equation}
\frac{l!(n-q_1-l)! }{(n-q_1)!} \, .
\end{equation}
The complete symmetry factor is a product of all these factors, and we find  
\begin{equation}
{\cal C}_{|T|=1} = l!\, q_1~.
%\frac{ (n-n_1-p_1+2l+1)!}{(n-n_1-p_1+l+1)!} \, .
\end{equation}
% Example check. $n=8$. Classes $j=1,l=2,m=3$ $n_1=4$, $n_2=6$. Classes: $(C_{[1^4,4]} \times C_{[1^2,6]})_{n_3=1} = 
% 30 C_{[1^3,5]}$. The coefficient is 120 IF one does not restrict to $n_3=1$. 
% % C11 * C19 = 120 C16. 
% Second example: $n=7$ $j=1,l=1,m=3,n_1=3,n_2=5$ and $n_3=5$. 
% $(C_{[1^4,3]} \times C_{[1^2,5]})_{n_3=1} = 5 \times 2 C_{[1^2,5]}$. It is 20 with no restriction. 
% % We must hit Kerber.
% % C5 * C12 = 20 C12
% $n=4$. Classes $j=1,l=1,m=1$. $C_{[1,3]} \times C_{[1,3]} = 4 C_{[1,3]}$. We get: $3$.
% % C4 times C4 = 4 C4 [Kerber]

% R-charges on the left: $q(\alpha) + n_1-1$ and $q(\beta) + n_2-1$ and on the right: $q(\alpha)+q(\beta) + q(\Delta_\ast^{(l+1)})+n_1+n_2-2l-2$. They are equal iff  $q(\Delta_\ast^{(l+1)})= 2l$. What is $\Delta_\ast^{(l+1)} (\mathds{1})$ ?
%
Combining these results, we have the following operator product expansion in the symmetric orbifold theory: 
\begin{equation}
 O_{[n_1]}(\alpha) \ast  O_{[p_1]}(\beta) \Big
|_{|T|=1}= \sum_{l=0}^{\text{min} (n_1,p_1)-1} \sum_{ q_1\ge 2}  l!~q_1~ \delta_{q_1+2l+1-n_1-p_1,0} ~
O_{[q_1, 1^l]}(\Delta_\ast^{(l+1)} (\alpha \beta)) ~.
\end{equation}
In the upper limit of the sum, we have taken into account that $l$ has to be smaller or equal to $n_1-1$ and $p_1-1$ by construction. 
For $l=0, 1$, this reproduces results previously obtained in \cite{Li:2020zwo,  Ashok:2023mow}. 
As stressed in \cite{Li:2020zwo}, this result agrees with and already goes beyond the classic results obtained in \cite{Jevicki:1998bm, Lunin:2000yv, Lunin:2001pw} since the operator $\Delta_\ast^{(l+1)}$ includes terms computed in those references, and more.  The present calculation is a broad generalization both in terms of models to which it applies and in terms of double overlap number. 

\subsection{The Second Order}

We next consider the contribution to the product of operators that occur due to a triple overlap $|T|=2$. We omit the contribution that arises from a product of two $|T|=1$ processes and consider only novel contributions. From our analysis in Section \ref{doubleoverlap}. we know that there are two 
essentially new possibilities, in which the relevant orbit involves a pair of one-cycles and a two-cycle\footnote{As we have seen in the leading order calculation, the operator $\gamma$ on the right hand side is not necessarily associated to a single cycle but can associate operators to multiple cycles. We denote it by $\gamma$ in order to unclutter the initial formulae.}:
\begin{align}
O_{[n_1]}(\alpha) \ast O_{[p_1,p_2]}(\beta_1, \beta_2)\Big|_{|T|=2} & = \sum_{q_1} {\cal C}_{|T|=2}(q_1)\, O_{[q_1]}(\gamma)\\
O_{[n_1]}(\alpha) \ast O_{[p_1]} (\beta) \Big|_{|T|=2} & = \sum_{q_1, q_2} {\cal C}_{|T|=2}(q_i)\, O_{[q_1, q_2]}(\gamma)~.
\end{align}
We discuss these cases in turn. 

\subsubsection{(1,2;1)}

Let us consider the operator product of any two permutations that occur in the first case. We have 
%The first diagram is:
% \begin{align}
% % \alpha C_{(a_1 \dots a_{j'} b_1 \dots b_j c a_{j'+1} \dots a_k b_{j+1} \dots b_l c')}
% % &\ast  C_{(c b_j b_{j-1} \dots b_1)(c' b_l b_{l-1} \dots b_{j+1})} ( \{ \beta_1, \beta_2 \})\nonumber \\
% \alpha C_{( a_1 \dots a_{j_1}  b_{1} \dots b_{l_1} c_1 a_{j_1+1} \dots a_{j_2} b_{l_1+1} \dots b_{l_2} c_2) }&\ast C_{(  c_1 b_{l_1}  \dots b_1 e_1 \dots e_n)( 
% c_2 b_{l_2}  \dots b_{l_1+1} d_1 \dots d_m)}( \{ \beta_1, \beta_2 \}) \nonumber \\
% &= (...) C_{(c_1 a_{j_1+1} \ldots a_{j_2} d_1 \dots  d_m c_2a_1\ldots a_{j_1} e_1 \ldots e_n)}
% \end{align}
% We need:
\begin{multline}
\alpha ( t_2 a_1 \dots a_{j_1}  b_{1} \dots b_{l_1} t_1 a_{j_1+1} \dots a_{j} b_{l_1+1} \dots b_{l} )   \ast \beta_1 (  t_1 b_{l_1}  \dots b_1 c_1 \dots c_{m_1}) 
\\
\hspace{8cm} \beta_2 ( 
t_2 b_{l}  \dots b_{l_1+1} c_{m_1+1} \dots c_m)  \\
= \gamma (t_1 a_{j_1+1} \ldots a_{j} c_{m_1+1} \dots  c_m t_2a_1\ldots a_{j_1} c_1 \ldots c_{m_1})
\label{Permutations121}
\end{multline}
We have split the cycle lengths as follows:
\begin{align}
\label{n1p1p2defn}
    n_1= 2+j+l~, \quad p_1 &= 1+l_1+m_1~, \quad p_2 = 1+l+ m-l_1 -m_1 \\
    q_1 &= 2+j+m~.
\end{align}
As in the previous case, the cycle length of the permutation $\pi_3= \pi_1\pi_2$ can be expressed in terms of the other lengths and the number of double overlaps between $\pi_1$ and $\pi_2$: 
\be 
q_1 = n_1+p_1+p_2 - 2l-2~.
\label{Expressionq1}
\ee 
Thus, on the right hand side we will have a sum of terms depending on the number of double overlaps. 
We have the orbits in the space $O=\{ t, a, b, c\}$ of dimension $j+m+l+2$.
\begin{align}
\langle \pi_1 \rangle \setminus O
&= \{ \{t_1, t_2, a_i, b_r\}, c_s \}
\nonumber \\
\langle \pi_2 \rangle \setminus O
&= \{ \{t_1, b_r, c_s\}, \{t_2, b_r, c_s\}, a_i \}
\nonumber \\
\langle \pi_1 \pi_2 \rangle \setminus O
&= \{\{t, a, c\}, b_r \}
\nonumber \\
\langle \pi_1 , \pi_2 \rangle \setminus O
&= \{ t, a,b,c \} \, .
\end{align}
The graph defect is zero.
%\begin{align}
%    g= \frac12\big(2+j+m+l+2 -(1+m)-(2+j)-%(1+l) \big) = 0~.
%\end{align}
This implies that we have a product $\alpha \beta_1 \beta_2$ %(in that order), 
that we distribute back over $l+1$ entries, i.e. the orbits of the permutation $\pi_3=\pi_1 \pi_2$. 
%We use that the genus is zero.  
Thus, we have:
\begin{multline}
\alpha (t_2 a_1 \dots a_{j_1}  b_{1} \dots b_{l_1} t_1 a_{j_1+1} \dots a_{j} b_{l_1+1} \dots b_{l})   \ast \beta_1 (  t_1 b_{l_1}  \dots b_1 c_1 \dots c_{m_1}) 
\\ 
\hspace{8cm} \beta_2 ( 
t_2 b_{l}  \dots b_{l_1+1} c_{m_1+1} \dots c_m)  \\
= \Delta_\ast^{(l+1)} (\alpha \beta_1 \beta_2) (t_1 a_{j_1+1} \ldots a_{j} c_{m_1+1} \dots  c_m t_2a_1\ldots a_{j_1} c_1 \ldots c_{m_1}) \, .
\label{121permutationsmaintext}
\end{multline}
Thus we have identified $\gamma = \Delta_{\ast}^{(l+1)} (\alpha \beta_1 \beta_2)$, which one can check is consistent with R-charge conservation. One has to distribute the resultant operator over the $q_1$-cycle and $l$ inactive colours.  
% We study conjugacy class products with double overlap and structure:
% \begin{equation}
% C_{[p]} \ast C_{[n_1,n_2]} = {\cal C}_{|2|} C_{[n_1+n_2+p-2l_2-2]}
% \end{equation}
%
% We can also use the lengths of the cycles 
% \begin{align}
%     p&=l_2+j_2+2~,\quad n_1=l_1+k+1~,\quad n_2=l_2-l_1+m+1~,\quad \text{and}\nonumber \\
%     n_3&=j_2+m+k+2=n_1+n_2+p-2-2l_2~.
% \end{align}
% We note that $(p,n_1,n_2,n_3)$ fix the cycle structure. These are four numbers. Note that the splitting numbers $l_1,j_1$ leave the cycle structure invariant. To get the conjugacy class coefficients (at this order in $N$), we sum over $(l_1,j_1)$ ??
It only remains to compute the symmetry factor.

\paragraph{i.} The number of elements in the first conjugacy class is 
\begin{equation}
|C_{n_1;n}| = \frac{n!}{n_1 (n-n_1)!} \, .
\end{equation}

\paragraph{ii.} We pick a number $l_1$ between $0$ and $l$ for which we have 
\begin{equation}
l+1
\end{equation}
choices. The number $l_1$ determines the number of $b$-colours in the first cycle of $\pi_2$ as well as  the number of $b$-colours in the second cycle of the permutation $\pi_2$. It then also determines the numbers $m_1$ and $m$ through the lengths of the two cycles.  All numbers are then fixed except $j_1$. 

\paragraph{ii.}  Next, we must identify the colour $b_1$ in the fixed permutation $\pi_1$, to carry it over into the permutation $\pi_2$. We pick a colour $b_1$ in 
\begin{equation}
n_1
\end{equation}
ways.  Since we know $l_1$, this fixes the colours $t_1$ and $b_1,\dots, b_{l_1}$.

\paragraph{iii.} 
To fix the colours $a_i$, we need to pick a number $j_1$ which we can choose in 
 \begin{equation}
 j+1 = n_1 -l-1
 \end{equation}
ways. We have used equation \eqref{n1p1p2defn}  in writing this equality. 
This completes the identification of all colours in the permutations $\pi_1$ and $\pi_2$.

% \paragraph{ii.} The number of ways in which we pick two triple overlap colours is given by the binomial coefficient 
% %$\begin{pmatrix} n_1\\2\end{pmatrix}$.
% \be 
% \begin{pmatrix}
% n_1\\ 2
% \end{pmatrix}~.
% \ee
% % \begin{equation}
% % \frac12 n_1 (n_1-1)~.
% % \end{equation} 
% \paragraph{ii'.} We associate one of the two colours to the cycle of length $p_1$ and call it $t_1$ and the other in the cycle of length $p_2$ we name $t_2$. There are $2$ choices for this and so  combining this with the factor from ii. above,  we obtain the factor
% \begin{equation}
% n_1(n_1-1)~.
% \end{equation}
% %choices for this. 
% %This then fixes the $b$ parts of $\pi_2$.

\paragraph{iv.} We then pick $m_1$ $c$-colours out of $n-n_1$ and order them,  and $m-m_1$ c-colours out of the remaining $n-n_1-m_1$ and order them for a total factor of:
\begin{equation}
\frac{(n-n_1)!}{(n-n_1-m_1)!} \frac{(n-n_1-m_1)!}{(n-n_1-m)!} = \frac{(n-n_1)!}{(n-n_1-m)!} \, .
\end{equation}
%(If $n_1$ and $n_2$ are equal, we may have overcounted possibilities by a factor of $2$.)

\paragraph{v.} We distribute all products over the elements of the right hand side conjugacy class and therefore multiply by the inverse of the cardinal number:
\begin{equation}
\frac{1}{|C_{q_1;n}|} = \frac{q_1 (n-q_1)!}{n!} \, .
\end{equation}
\paragraph{vi.} Lastly we pick out $l$ colours from the $n-q_1$ entries leading to the (inverse) factor:
\begin{align}
\frac{l!(n-q_1-l)!}{(n-q_1)!}~.
\end{align}
The product of all these factors gives the structure constant. Substituting for $q_1=n_1+p_1+p_2 - 2l-2$, and $m=n_1-q_1-l$, we find the structure constant:
\begin{align}
{\cal C}_{|T|=2} &= (j+1)(l+1)\, l!\, q_1 \nonumber \\
&=  (n_1-l-1)(l+1)!  (n_1+p_1+p_2-2l-2) ~.
\end{align}

\noindent
Thus we find, at the $|T|=2$ level, the operator product:
\begin{multline}
O_{[n_1]}(\alpha) \ast O_{[p_1,p_2]}(\beta_1, \beta_2)\Big|_{|T|=2}  = \sum_{l=0}^{n_1-2}\sum_{q_1\ge 2} (n_1-l-1)~(l+1)!~
%\frac{(n-n_1-p_1-p_2+2l+2)!}{(n-n_1-p_1-p_2+l+2)!} \\  
 q_1~\delta_{q_1+2l+2-n_1-p_1-p_2,0} \\ 
 \times O_{[q_1,1^l]}\big(\Delta_{\ast}^{(l+1)} (\alpha \beta_1 \beta_2)\big)~.
\label{121maintextanswer}
\end{multline}
We end this subsection by stressing implicit assumptions about the cycle lengths that we have made in the course of the calculation. The form of the permutations in \eqref{121permutationsmaintext} implies that the number $l$ satisfies the bound $l \le \text{min} (p_1-1,p_2-1,n_1-2)$. Combining this with the expression for $q_1 = n_1+ p_1+p_2 - 2l-2$ in \eqref{Expressionq1},  we conclude that $q_1$ will necessarily be  
%(OLD and incorrect ? : smaller or equal to $n_1$; replace by : ?)
greater or equal to $\text{max}(n_1+p_2-p_1,n_1+p_1-p_2,p_1+p_2-n_1+2)$. Similar assumptions underlie all the following calculations and we will specify them in due course. If one considers quantum numbers that do not satisfy these requirements, a more refined calculation is necessary. The applicability range remains very broad.

\subsubsection{(1,1;2)}

We now focus on the second case
\begin{align}
    O_{[n_1]}(\alpha) \ast O_{[p_1]} (\beta)\Big|_{|T|=2} & = \sum_{q_1, q_2} {\cal C}_{|T|=2}(q_i) C_{[q_1, q_2]}(\gamma)~.
\end{align}
The product of any two permutations within the conjugacy classes is given by
\begin{multline}
 \alpha (a_1 \dots a_{j_1} b_1 \dots b_{l_1} t_1 a_{j_1+1} \dots a_{j} b_{l_1+1} \dots b_{l} t_2)
\ast \beta (t_1 b_{l_1} \dots b_1 c_1\ldots c_{m_1} t_2 b_{l} \dots b_{l_1+1}c_{m_1+1} \ldots c_{m} )   
\\
 = \gamma\,   
(t_1 a_{j_1+1} \dots a_{j} c_{m_1+1} \ldots c_{m})\, ( t_2 a_1 \dots a_{j_1} c_1\ldots c_{m_1} )~.
\end{multline}
The cycle lengths have been split as follows:
\begin{align}
    n_1 &= 2+j+l~, \hspace{2.5cm} p_1 = 2+l+m~, \\
    q_1 &= 1+j-j_1+m-m_1~,\quad q_2 = 1+j_1+m_1
    \label{qiin112}
\end{align}
The length of the permutation $\pi_3$ can again be expressed in terms of the lengths $n_1, p_1$ and the double overlap $l$:
\be 
q_1+q_2 = 2+j+m = n_1+p_1 - 2l-2~.
\ee 
Once again, it is important to note that both $q_1$ and $q_2$ have to have length at least $2$ so that the triply active colours $t_i$ are part of non-trivial cycles. 

We have the orbits in the space $O=\{ t, a, b, c\}$ of dimension $j+m+l+2$.
\begin{align}
\langle \pi_1 \rangle \setminus O
&= \{ \{t_1, t_2, a_i, b_r\}, c_s \}
\nonumber \\
\langle \pi_2 \rangle \setminus O
&=\{\{t, b, c\}, a_i \}
\nonumber \\
\langle \pi_1 \pi_2 \rangle \setminus O
&=  \{ \{t_1, a_{i_1}, c_{s_1}\}, \{t_2, a_{i_2}, c_{s_1}\}, a_i \}
\nonumber \\
\langle \pi_1 , \pi_2 \rangle \setminus O
&= \{ t, a,b,c \} \, .
\end{align}
The graph defect is again zero: 
\begin{align}
    g= \frac12\big(2+j+m+l+2 -(1+m)-(1+j)-(2+l) \big) = 0~.
\end{align}
This implies that we multiply the operators $\alpha$ and $\beta$, and then distribute the product back over the cycles $q_1$, $q_2$ and over $l$ inactive colours, leading to the right hand side operator:
%We use that the genus is zero.  
%
\be 
\gamma = \Delta_{\ast}^{(l+2)}(\alpha \beta)~.
\ee 
%
%This is also consistent with the R-charge conservation.
% ; equating the R-charges on either side of the operator product, we find the relation:
% % 
% \be 
% Q(\Delta^{(n)}_\ast) = n_1-1+p_1-1 -(q_1+q_2-2) = 2l+2~.
% \ee 
% %
%Using \eqref{RchargeDeltastar} this fixes $n=l+2$.
To find the numerical structure constant for the product of conjugacy classes, we follow the usual procedure. 
% R-charge comparison:
% \begin{align}
%     q(LHS)&= q(\alpha)+q(\beta)+j_2+ l_2+1 + q(\beta)+l_2+m_2+1~,\\
%     q(RHS)&= q(\alpha)+q(\beta)+q(\Delta_\ast)+ 1+j_2-j_1+m_2-m_1+1 + 1+j_1+m_1-1 \nonumber\\ 
% &=q(\alpha)+q(\beta)+q(\Delta_\ast)+j_2+m_2 
% \end{align}
% So we have $q(\Delta_\ast) = 2l_2+2$, which fixes it to be $\Delta_{\ast}^{2l_2+2}$. 
% Compare to [LT] for this case. 
% \begin{align} 
% n_1&=2+j_2+l_2~,
% n_2=2+l_2+m_2~,\\
% n_{31}&=1+j_2-j_1+m_2-m_1 ~,
% n_{32}=1+j_1+m_1~. 
% \end{align} 

%Without loss of generality we assume that $n_1 < p_1$ and we begin by fixing the form of the smaller permutation. 
% \noindent
% Steps ${\bf i.}$ and ${\bf ii.}$ are identical to the previous case. 
Most of the steps, up  to dealing with the third permutation, are identical to those carried out for the previous case. Step {\bf i.} fixes the permutation $\pi_1$. We split the double overlap number $l$ into two parts in step {\bf ii.}. In step {\bf iii.} we identify the colour $b_1$ and in step {\bf iv.} we pick $j_1$.
A crucial subtlety appears now. Fixing $j_1$ will fix $m_1$. Therefore, the maximal number of ways in which we can pick $j_1$ equals $j+1$ or $m+1$ -- whichever one is smaller. For simplicity of notation, we continue with the first possibility but we keep the subtlety in mind. 
The $c$-colours in $\pi_2$ are chosen in step {\bf v.}. The contribution from all these steps can be calculated to be 
\be 
\frac{n!}{(n-n_1-m)!}\, (l+1)\, (j+1)\,,
\ee
with the relations $j= n_1-l-2$ and $n_1+m = q_1+q_2+l$, that can be inferred from the cycle lengths.

\paragraph{vi.} It now remains to  divide by the number of elements in the third conjugacy class:
\begin{equation}
\frac{1}{|C_{q_1,q_2;n}|} = \frac{q_1 q_2 (n-q_1-q_2)!}{n!}
\end{equation}

\paragraph{vii.} We choose $l$ factors from among the $n-q_1-q_2$ colours, leading to the factor:
\begin{equation}
\frac{l!(n-q_1-q_2-l)! }{(n-q_1-q_2)!}~.
\end{equation}
The structure constant is the product of all these factors:
\begin{align}
%(l_2+1)(n_1-1) (p_1-1) n_{31} n_{32}\frac{ (n-n_{31}-n_{32})!}{(n-n_1-n_2+l_2+2)!}
{\cal C}_{|T|=2} %&=(j+1)(l+1)\, l! q_1 q_2 \nonumber \\
&= (n_1-l-1)(l+1)! q_1 q_2~.
\end{align}
Substituting $q_1+q_2 = n_1+p_1-2l-2$, we obtain, for the product of conjugacy classes:
\begin{multline}
    O_{[n_1]}(\alpha) \ast O_{[p_1]}(\beta)\Big|_{|T|=2}  = %\frac{1}{2} 
    \sum_{l=0}^{n_1-2}(n_1-l-1)\, 
\sum_{\substack{q_1,q_2 \ge 2 
          \\
               \{ q_1,q_2 \} }
      }
%^{n_1+p_1-4}
(l+1)! q_1 q_2\, \delta_{q_1+q_2, n_1+p_1-2l-2} \\
    \times 
    %\frac{ (n-n_1-p_1+2l+2)!}{(n-n_1-p_1+l+2)!} 
    O_{[q_1,q_2,1^l ]}(\Delta_{\ast}^{(l+2)}(\alpha \beta))~.
\end{multline}
We sum over all possible sets $\{ q_1,q_2 \}$.\footnote{To be more precise, when $q_1 \neq q_2$, we only sum over the set $\{ q_1,q_2 \}$ once. The possibility $q_1=q_2$ is also allowed.}
%The factor of $1/2$ is present because we must not over count the possibility of exchanging $q_1$ and $q_2$. 
We recall from the subtlety above that when $n_1>p_1$, the factor within the $l$-sum will be $(p_1-l-1)$. Thus we may write the more general result
\begin{multline}
    O_{[n_1]}(\alpha) \ast O_{[p_1]}(\beta)\Big|_{|T|=2}  = %\frac{1}{2} 
    \sum_{l=0}^{n_1-2} (\text{min}(n_1,p_1)-l-1)
\sum_{\substack{q_1,q_2 \ge 2 
          \\
               \{ q_1,q_2 \}}}
%^{n_1+p_1-4}
(l+1)! q_1 q_2\, \delta_{q_1+q_2, n_1+p_1-2l-2} \\
    \times 
    %\frac{ (n-n_1-p_1+2l+2)!}{(n-n_1-p_1+l+2)!} 
    O_{[q_1,q_2,1^l ]}(\Delta_{\ast}^{(l+2)}(\alpha \beta))~.
\end{multline}
%This result is derived from an alternative perspective in the appendix. 
In our derivation of the structure constant, we have implicitly assumed that  $m_1$ satisfies the constraint $0 \le m_1 \le m$, with $m = p_1-l-2$. These impose non-trivial constraints on the cycle lengths of the permutation $\pi_3$. Consider the expressions for the lengths $q_i$ in equation \eqref{qiin112}. To find the extreme bounds, we set $j_1=0$ in the expression for $q_1$ and $j_1=j$ in the expression for $q_2$ and find that the $0 \le  m_1 \le m$ bound and a similar  bound for $m-m_1$ is satisfied if and only if we have
\be 
q_i \ge n_1-l-1 ~,
%\le n_1+p_1~, 
\quad\text{for}\quad i=1,2~.
\ee 
Thus, the structure constant we derived above for each $l$ is valid only for those values of $q_i$ that satisfy this inequality. 
%
% \subsubsection*{Check}
%
% Let's re-attempt this calculation, now by fixing a right (!) hand side as early as possible such that all cycles are necessarily non-trivial. 
%
% % i. We fix $q_1,q_2, l$. We pick a definite permutation in this conjugacy class. We have (after inversion):
% % \begin{equation}
% % \frac{q_1 q_2 l!(n-q_1-q_2-l)! }{n!}~.
% % \end{equation} 
% % possibilities. 
%
% ii. We pick triple overlap colours for a factor of 
% \begin{equation}
% q_1 q_2 \, .
% \end{equation}
%
% iii. To determine $\pi_1$, we pick a $j_1$ which goes from $0$ to $q_2-1$. This is 
% \begin{equation}
% q_2 
% \end{equation}
% possibilities. 
%
% iii'. To further determine $\pi_1$, we pick a $l_1$ between $n_1-2-j_1$ and $0$, for 
% \begin{equation}
% n_1-1-j_1
% \end{equation}
% choices. 
%
% iii''. For given $l_1$, we pick $l_1$ and $l-l_1$ colours and order them for a factor of 
% \begin{equation}
%  \frac{(n-q_1-q_2)!}{(n-q_1-q_2-l_1)!} \frac{(n-q_1-q_2-l_1)!}{(n-q_1-q_2-l)!} \, .
% \end{equation}
%
% % The number of elememts in the $\pi_2$ conjugacy class equals:
% % \begin{equation}
% % \frac{n!}{p_1 (n-p_1)!} \, .
% % \end{equation}
%
% We have 
% \begin{equation}
% q_1+q_2+l = n_1+p_1-l-2 \, .
% \end{equation}
%
% If we randomly combine, we find:
% \begin{equation}
% l! q_1 q_2 q_2 \sum_{j_1} (n_1-1-j_1)
% \end{equation}

We can check this structure constant by comparing the cycle distributions $(1,2;1)$ and $(1,1;2)$. 
We can think of them as being related through 
the exchange of the second and third conjugacy classes. The relative factor between the two calculations should then be the ratio of the number of elements in the generalized conjugacy classes which equals $q_1 q_2/p_1$ (by which we mean the product of the lengths of the two-cycles divided by the length of the longest single cycle). This is the case.  This works out because the expression here is $\text{min}(p_1,n_1)$ which matches the fact that $\text{min}(n_1,q_1)=n_1$ in the previous calculation and we must identify $q_1=p_1$. Both formulas have the limited range of validity that we have indicated in their calculation. There is an overlapping range of validity which explains why they coincide.\footnote{Since the range of validity is different in both calculations, as can be checked, one suspects that both calculations can be argued to be valid even beyond the range of values that we have used in our proof. We leave this further combinatorial task for the future.}

%Also, we determined that $(1,3;1)$ should be calculable, $(1,1;3)$ as calculable as this last one, and $(1,2;2)$ more difficult at $|T|=3$. 

\subsection{The Large N Expansion of the Operator Product}

% Example calculation of single cycle permutation with generic permutation.

% Firstly, by large $n$ order.

% One-and-a-half-ly, by number of 'deletions' $l$.  These correct a given joining at the same order in $n$. 

% Secondly, by number of joinings (i.e. the number of cycles in the end result). These have contributions at different orders in $n$. 

% \begin{align}
% C_{[n_1]}(\alpha) &\ast C_{[p_1, p_2, \ldots]}(\{ \beta_j\})  = 
% %\#  C_{[n_1,p_1,\ldots]} + \# \delta_{n_1,p_i} \Delta_\ast^{(n_1)} C_{\hat{p}_i}
% \nonumber  \\
%  & \sum_{p_j} \sum_{l=0}^{\text{min}(n_1,p_j)}  l!~(n_1+p_j-2l-1)~
% C_{[p_1, \ldots, \slashed{p}_j, \ldots n_1+p_j-2l-1, 1^{l}]}\big(\beta_1,\ldots \slashed{\beta}_j \ldots \Delta_\ast^{(l+1)} (\alpha \beta_j)\big)\nonumber  \\
% & +(n_1-1)\sum_{p_j} (p_j-1)\sum_{q_j, l} (l+1)! q_j(n_1+p_j-q_j-2-2l)\nonumber \\ 
% & \hspace{4.5cm}\times C_{[p_1, \ldots \slashed{p}_j\ldots q_j, n_1+p_j-q_j-2l_j-2, 1^{l}]}\big(\beta_1, \ldots \slashed{\beta}_j, \ldots \Delta_\ast^{(l+2)}(\alpha\beta_j )\big)
% \nonumber \\
% &+ (n_1-1)\sum_{p_j, p_k} \sum_{l} (l+1)!(n_1+p_j+p_k-2l-2) \nonumber\\
% &\hspace{3cm}\times C_{[p_1, \ldots \slashed{p}_j\ldots \slashed{p}_k \ldots n_1+p_j+p_k-2l-2, 1^l]}
% \big(\beta_1, \ldots \slashed{\beta}_j\ldots \slashed{\beta}_k, \ldots \Delta_\ast^{(l+1)}(\alpha\beta_j\beta_k) \big)\nonumber\\
%  + \dots
% \end{align}
% The first term is at first order, the second and third terms are second order, and represent joining and recombination respectively, the fifth and ... lines are third order, and the remaining dots are higher order. 
% 
%
We also computed the structure constants at third order in the $1/\sqrt{N}$ expansion -- the details are described in Appendix \ref{T3structureconstants}. Our results can be collected together in a master formula for the operator product expansion:
\begin{align}
\label{MasterFormula}
O_{[n_1]}(\alpha) &\ast O_{[p_1, p_2, \ldots]}( \beta_1, \beta_2, \ldots)  = 
{O_{[n_1,p_1,p_2,\dots]}( \alpha,\beta_1, \beta_2, \ldots )}
\nonumber \\
 & %\textcolor{blue}
 +  \sum_{l=0}^{n_1-1}l!  \sum_{\{ p_j \}} \sum_{q_j\ge 2}q_j \delta_{q_j, n_1+p_j-2l-1}~
O_{[p_1, \ldots, \slashed{p}_j, \ldots q_j, 1^{l}]}\big(\beta_1,\ldots \slashed{\beta}_j \ldots \Delta_\ast^{(l+1)} (\alpha \beta_j)\big) \nonumber  \\
& + 
\Big\{\sum_{l=0}^{n_1-2}(n_1-l-1)(l+1)! \sum_{q_j\ge 2} q_j 
 \sum_{\{ p_j, p_k \}}\delta_{q_j, n_1+p_j+p_k-2l-2}  \nonumber\\
&\hspace{2cm}\times  O_{[p_1, \ldots \slashed{p}_j\ldots \slashed{p}_k \ldots q_j, 1^l]}
\big(\beta_1, \ldots \slashed{\beta}_j\ldots \slashed{\beta}_k, \ldots \Delta_\ast^{(l+1)}(\alpha\beta_j\beta_k) \big)\nonumber\\
& + \sum_{l=0}^{n_1-2}(n_1-l-1)(l+1)!~
%\frac12
\sum_{\substack{ q_j \ge 2 \\
     \{ q_j,q_k\} }}
 q_j q_k
  \sum_{\{ p_j \}} \delta_{q_j+q_k, n_1+p_j-2l-2}\nonumber \\ 
& \hspace{2cm}\times O_{[p_1, \ldots \slashed{p}_j\ldots q_j, q_k, 1^{l}]}\big(\beta_1, \ldots \slashed{\beta}_j, \ldots \Delta_\ast^{(l+2)}(\alpha\beta_j )\big) \Big\}
\nonumber \\
&+\Big\{
%\textcolor{red}
{\frac12}
\sum_{l=0}^{n_1-3} (n_1-l-1)(n_1-l-2)(l+2)!\sum_{q_m\ge 3} q_m 
\sum_{\{ p_i, p_j, p_k \}} \delta_{q_m, n_1+p_i+p_j+p_k -2l -3} \nonumber \\
&\hspace{2cm}\times  O_{[p_1, \ldots \slashed{p}_i\ldots \slashed{p}_j\ldots \slashed{p}_k, \ldots q_m, 1^l ]}(\beta_1, \ldots \slashed{\beta}_i, \ldots \slashed{\beta}_j\ldots \slashed{\beta}_k\ldots \Delta_\ast^{(l+1)}(\alpha\beta_i\beta_j\beta_k)) \nonumber \\
&+\frac{1}{4}\sum_{l=0}^{n_1-3}(n_1-l-1)(n_1-l-2)(l+2)! 
\sum_{q_j\ge 3}  q_j\sum_{\{ p_i \}}\delta_{q_j, n_1+p_i-2l-3}\nonumber \\
&\hspace{2cm}\times  O_{[p_1, \ldots, \slashed{p}_i, \ldots q_j, 1^l]}(\Delta_\ast^{(l+1)}(\alpha\beta_i e))\nonumber\\
&+\frac12\sum_{l=0}^{n_1-3}(n_1-l-1)(n_1-l-2)(l+2)!
%\frac{1}{2!}
\sum_{ \substack{ q_k \ge 2 \\
     \{ q_k,q_m\} }} ~q_k\, q_m \nonumber\\
 & \hspace{2cm}\times    \sum_{\{p_i, p_j\}}\delta_{q_k+q_m, n_1+p_i+p_j-2l-3 } O_{[p_1, \ldots \slashed{p}_i \ldots \slashed{p}_j, \ldots q_k, q_m, 1^l]}(\Delta_\ast^{(l+2)}(\alpha\beta_i\beta_j))\nonumber \\
 &+\frac{1}{2}\sum_{l=0}^{n_1-3}(n_1-l-1)(n_1-l-2)(l+2)! 
 %\frac{1}{3!}
 \sum_{ \substack{ q_j \ge 2 \\
     \{ q_j,q_k,q_m\} }}  ~q_j\, q_k\, q_m\nonumber \\
&\hspace{2cm}\times \, \sum_{\{ p_i\}}\delta_{q_j+q_k+q_m, n_1+p_i-2l-3}\,  O_{[p_1, \ldots \slashed{p}_i\ldots q_j, q_k, q_m, 1^l]}(\Delta_\ast^{(l+3)}(\alpha\beta_i)) \Big\}+ \ldots 
\end{align}
Here we have assumed that the cycle length $n_1$ is minimal among cycle lengths and that all of the necessary assumptions of genericity for the cycle lengths and the double overlap parameter $l$ are satisfied. E.g. the length $n_1$ differs from all lengths $p_i$, simplifying the leading order term. We have collected the terms in increasing order of triple overlap $|T|$. The second and third order terms have been collected within curly brackets. Within the set of terms with the same overlap $|T|$, we have arranged the terms so that the number of non-trivial cycles on the right hand side is in ascending order. We recall that if we normalize operators such that they have unit two-point function in the untwisted theory, the triple overlap $|T|$ determines the large $N$ order of the structure constant to be $N^{-\frac{|T|}{2}}$. 
We note the genus one correction involving the Euler class $e$ that appears in the third from last term. It has triple overlap $|T|=3$ and order $N^{-\frac{3}{2}}$.  
Finally, we observe that all the structure constants are $N$-independent in the natural group theory normalization.

As a check of the above formula, we isolate the terms determined by the Farahat-Higman formula \cite{FarahatHigman}; these are the operator products in the topologically twisted conformal field theory on $\mathbb{C}^2$ \cite{Li:2020zwo,Ashok:2023mow}. These are terms in which no topological metric intervenes: this essentially implies that we only consider the $l=0$ terms on the right hand side. Furthermore, we only allow for joining operations among the cycles; as a consequence, we omit all terms with more than one $q_i$ in the final permutation. Finally, we set all operators to the identity operator and the Euler class $e$ to zero. We are then left with the simplified operator product:
%Let's write out these terms separately:
\begin{multline}
\label{MasterFormulaFH}
O_{[n_1]}(1) \ast O_{[p_1, p_2, \ldots]}( 1,  \ldots)  = 
{O_{[n_1,p_1,p_2,\dots]}( 1, \ldots )}
 \\
 +   \sum_{\{ p_j \}} (n_1+p_j-1)~
O_{[p_1, \ldots, \slashed{p}_j, \ldots n_1+p_j-1]}\big(1,\dots \big)  \\
 + (n_1-1) 
 \sum_{\{ p_j, p_k \}} (n_1+p_j+p_k-2)    
O_{[p_1, \ldots \slashed{p}_j\ldots \slashed{p}_k \ldots n_1+p_j+p_k-2]}
\big(1,\dots \big)
\\
+(n_1-1)(n_1-2) 
\sum_{\{ p_i, p_j, p_k \}} (n_1+p_i+p_j+p_k  -3)
O_{[p_1, \ldots \slashed{p}_i\ldots \slashed{p}_j\ldots \slashed{p}_k, \ldots n_1+p_i+p_j+p_k  -3 ]}(1,\ldots)  + \dots
\end{multline}
This perfectly matches the Farahat-Higman formula \cite{FarahatHigman, Ashok:2023mow} for the case in which we assume that all the $n_1, p_i, q_i$ are distinct, as we do.    

\section{A Few Four-point Functions}
\label{FourPointFunctions}

Given the structure constants of the topological theory, extremal correlators (of chiral operators with a single anti-chiral operator) in the untwisted theory can be calculated by repeated multiplication in the ring. For the symmetric orbifold of $\mathbb{C}^2$, this technique was applied  in \cite{Li:2020zwo,Ashok:2023mow} to compute extremal correlators efficiently and to prove a general conjecture \cite{Pakman:2009ab} on their combinatorial nature. In this section we compute a sample four point function in the symmetric orbifold $A^{[n]}$ of a general Frobenius algebra $A$. We study the (untwisted) extremal correlator of single cycle operators:
\begin{equation}
\langle ( O_{[n_4]}(\alpha_4))^\dagger~ O_{[n_3]}(\alpha_3) ~O_{[n_2]}(\alpha_2)  ~ O_{[n_1]}(\alpha_1)  \rangle%_{\text{phys}}
\end{equation}
for the special case in which we have $n_2=n_3=2$, namely the four-point function 
\begin{equation}
F = \langle ( O_{[n_4]}(\alpha_4))^\dagger~ O_{[2]}(\alpha_3)   ~ O_{[2]}(\alpha_2)~ O_{[n_1]}(\alpha_1) \rangle%_{\text{phys}} 
\, .
\end{equation}
We compute the successive operator products of the first three operators and then read off the coefficient of single cycle operators in the resulting fusion:
\be 
 O_{[n_3]}(\alpha_3)\ast  O_{[n_2]}(\alpha_2) \ast   O_{[n_1]}(\alpha_1) = {{\cal C}_{123}}^{4}\,  O_{[n_4]}(\alpha_4) + \ldots  
\ee
where we fuse from right to left. 
As indicated, on the right hand side, we also
restrict to operators with only identity operators associated to the inactive colours. 
We will find that for $n_2=n_3=2$, the structure constant is non-vanishing  only for $n_4 \in \{n_1, n_1+2\} $. This is a consequence of charge conservation and the upper bound on the charge in the seed theory.

The first step in the calculation is an application of our summary formula \eqref{MasterFormula}: 
\begin{align}
O_{[2]}(\alpha_2)  \ast  O_{[n_1]}(\alpha_1 )
=& O_{[n_1,2]} (\alpha_1,\alpha_2)
 +(n_1+1) O_{[n_1+1]}(\alpha_1\alpha_2)
  \nonumber \\
& +\frac{1}{2} \sum_{q_1=2}^{n_1-2} q_1 q_2 \delta_{q_1+q_2-n_1} O_{[q_1,q_2]}(\Delta_\ast^{(2)} (\alpha_1 \alpha_2))
\nonumber \\
& + (n_1-1)       O_{[n_1-1,1]} ( \Delta_\ast^{(2)}(\alpha_1 \alpha_2)) \, .
%\nonumber \\
%& + \dots 
\label{firststep}
\end{align}
In order to write the second line we have assumed that the integer $n_1$ is odd (in order to avoid dealing with the case where $q_1$ can be equal to $q_2$ separately).  
% The first term is propagation, the second term is joining and the third is (a small) pinching and the fourth term is splitting. 
At the next step, we multiply  the operator $ O_{[2]}(\alpha_3)$ with each of the four terms on the right hand side of equation \eqref{firststep}. This process is straightforward for the first three sets of terms as it involves repeated application of our master formula. However, the product of the operator  $O_{[2]}(\alpha_3 )$ with the last term in equation \eqref{firststep} is of a new type as it involves operator products in which non-trivial elements of the Frobenius algebra are associated to the inactive colours. One therefore has to derive this novel operator product from first principles, and this is done in Appendix \ref{novelOP}. 

Next, we  list the single cycle operators (with trivial operators associated to the inactive colours) that we obtain by fusing the operator $ O_{[2]}(\alpha_3)$ with each of the terms on the right hand side of equation \eqref{firststep}. We denote the successive terms by $X_i$. For the first two terms, we find 
\begin{align}
\label{secondstep1v1}
X_1:=& O_{[2]}(\alpha_3) \ast O_{[n_1,2]} (\alpha_1,\alpha_2) =
(n_1+2) \, O_{[n_1+2]}(\alpha_3 \alpha_1 \alpha_2)
+  O_{[n_1,1,1]} (\alpha_1 ; \Delta_\ast^{(2)} (\alpha_3 \alpha_2) ) +\ldots  \\
%\end{align}
% On the left hand side, we count the number of terms in $O_{[n_1,2]}$:
% \begin{equation}
%  \frac{n!}{2 n_1 (n-n_1-2)!} \, .
% \end{equation}
% We count the terms in which $[2]$ agrees with $,2$. This is one term. Next, we count the number of terms in which $[2]$ overlaps in one with each cycle. There are $n_1 \times 2$ such terms. The total number of terms is:
% \begin{equation}
%  (2n_1+1) \times \frac{n!}{2 n_1 (n-n_1-2)!} \, .
% \end{equation}
% On the right hand side, we have:
% \begin{equation}
% (n_1+2) \frac{n!}{(n_1+2) (n-n_1-2)!}
% + \frac{n!}{n_1 (n-n_1)!} \times \frac{(n-n_1)!}{2 (n-n_1-2)!}
% = lhs
% \end{equation}
%
% % Total (impossible) count:
% % Number of terms on lhs: 
% % \begin{equation}
% % \frac{n!}{2 (n-2)!} \times \frac{n!}{2 n_1 (n-n_1-2)!}
% % \, .
% % \end{equation}
% % On rhs:
% % \begin{equation}
% % (n_1+2) \frac{n!}{(n_1+2) (n-n_1-2)!}
% % + \frac{n!}{n_1 (n-n_1)!} \times \frac{(n-n_1)!}{2 (n-n_1-2)!}
% % \end{equation}
%\begin{align}
X_2:=&(n_1+1) O_{[2]}(\alpha_3) \ast O_{[n_1+1]} (\alpha_1 \alpha_2) = (n_1+1)(n_1+2)\,  O_{[n_1+2]}(\alpha_3 \alpha_1 \alpha_2) \nonumber\\
&\hspace{7cm}+ n_1(n_1+1) O_{[n_1,1]} (\Delta_\ast^{(2)}(\alpha_3 \alpha_1 \alpha_2)) + \ldots 
\label{secondstep2v1}
\end{align}
% Check. Number of (relevant) terms on lhs:
% \begin{equation}
% \frac{n!}{(n_1+1) (n-n_1-1)!} \times \Big( (n_1+1)(n-n_1-1)+ (n_1+1) \Big)  = \frac{n! (n-n_1)} {
% (n-n_1-1)!}\, .
% \end{equation}
% The number of terms on the rhs:
% \begin{equation}
% \frac{n!}{(n-n_1-2)!} + \frac{n!}{(n-n_1)!}
% (n-n_1)
% = \frac{n!}{(n-n_1-1)!} ( n-n_1-1 + 1)
% %= \frac{n!}{ (n-n_1-1)!} \left( (n-n_1-1) + (n-1)  \right) 
% \end{equation}
The ellipses indicate single and multi-cycle terms that are of no consequence to the right hand side operators to which we limit ourselves. Before we proceed further it is useful to recall properties of the co-adjoint multiplication in the Frobenius algebra. 
%This would also help in writing the terms in \eqref{firststep} more explicitly. 
We recall the expression for the co-multiplication in terms of basis elements $e_k$ and structure constants:
\be 
\label{deltastar}
\Delta_{\star}^{(2)}(\alpha_1\alpha_2) = (\alpha_1\alpha_2)^m \, {c^{lk}}_m \, e_l\otimes e_k~,
\ee
where we denote by $(\alpha_1\alpha_2)^m$, the component of the element of the Frobenius algebra along the $m$th basis element $e_m$, with $m=0$ denoting the identity element. We can then rewrite the equations \eqref{secondstep1v1} and \eqref{secondstep2v1} in the following way:
\begin{align}
\label{secondstep1v2}
X_1
%\alpha_3 O_{[2]} \ast O_{[n_1,2]} (\alpha_1,\alpha_2) 
&=
(n_1+2) \, O_{[n_1+2]}(\alpha_3 \alpha_1 \alpha_2)
+ \, (\alpha_3\alpha_2)^m\, {c^{lk}}_m\,   O_{[n_1,1,1]} (\alpha_1 ; e_l; e_k ) +\ldots\nonumber \\
&= (n_1+2)\, O_{[n_1+2]} (\alpha_3 \alpha_1 \alpha_2)
+ \ldots\\
X_2
%\alpha_3 O_{[2]} \ast O_{[n_1+1]} (\alpha_1 \alpha_2) 
&= (n_1+1)(n_1+2)\,  O_{[n_1+2]} (\alpha_3 \alpha_1 \alpha_2)
+ n_1(n_1+1)\, (\alpha_3\alpha_2 \alpha_1)^m\, {c^{lk}}_m\, O_{[n_1,1]} (e_l; e_k) + \ldots \nonumber\\
&= (n_1+1)(n_1+2)\,  O_{[n_1+2]} (\alpha_3 \alpha_1 \alpha_2)
+ n_1(n_1+1)\, (\alpha_3\alpha_2 \alpha_1)^0 \, \, O_{[n_1]}(\text{vol})+\ldots
\label{secondstep2v2}
\end{align}
In the second equality for the term $X_1$ we have used the fact that both $e_l$ and $e_k$ have to be the identity operator, since these are the extremal correlators we are interested in; the corresponding structure constant vanishes because of the upper bound on the charge of operators in the Frobenius algebra.\footnote{Throughout this section, we assume the Frobenius algebra arises from the topological twist of a  ${\cal N}=(2,2)$ superconformal field theory.} For the term $X_2$ we use the fact that when $e_k$ is the identity operator, we have ${c^{l0}}_m = \delta_{m,0} \delta^{l, \text{vol}}$.  See equation \eqref{specialsc}.  
 For the next term in the double fusion, we can again write down the single cycle terms that appear in the product:
\begin{align}
 O_{[2]}(\alpha_3) \ast O_{[q_1,q_2]} (\Delta^{(2)}_\ast (\alpha_1 \alpha_2)) 
=& n_1
O_{[q_1+q_2=n_1]} \Big(\alpha_3 \Delta^{(2)}\Big( \Delta^{(2)}_\ast (\alpha_1 \alpha_2) \Big)\Big)
\nonumber \\
&
+ \delta_{q_1,2} 
(\alpha_1 \alpha_2)^m {c^{lk}}_m
%\Delta_\ast^{(2)}(\alpha_3,e_l)
O_{[n_1-2,1,1]} (e_k,\Delta_\ast^{(2)}(\alpha_3 e_l))
\nonumber \\
&+ \delta_{q_2,2} 
(\alpha_1 \alpha_2)^m {c^{lk}}_m
%\Delta_\ast^{(2)}(
O_{[n_1-2,1,1]} (e_l;\Delta_\ast^{(2)}(\alpha_3 e_k))+\ldots
\end{align}
The first term is from the usual master formula; the next two terms arise when the operator  $O_{[2]}$ annihilates the two-cycle in the operator $O_{[2, n_1-2]}$, and we have also made use of the co-product formula \eqref{deltastar}.
% Check on number of terms. Lhs:
% \begin{equation}
% \frac{n!}{q_1 q_2 (n-q_1-q_2)!} \times (q_1 q_2 + \delta_{q_1,2} + \delta_{q_2,2})
% \end{equation}
% Rhs: 
% \begin{equation}
% \frac{n!}{(n-q_1-q_2)} + (\delta_{q_1,2} +\delta_{q_2,2}) \frac{n! (n-n_1+2)(n-n_1+1)}{(n_1-2) (n-n_1+2)!  2} \, .
% \end{equation}
% We  note that the first term is also there for both the values $q_1=2$ and $q_2=2$.
To write this expression more explicitly, we once again make use of the definition of the co-adjoint product in equation \eqref{deltastar} and also recall the formula (see equation \eqref{deltadeltastarv1})
 \be 
\label{deltadeltastarv2}
\Delta^{(2)} \Delta_\ast^{(2)}(e_{\ell})={c^{ji}}_0 {c_{ij}}^k e_k \delta_{\ell,0} = {c^{ji}}_0 {c_{ij0}} \, \text{vol}\,  \delta_{\ell,0} = \chi \, \text{vol}\, \delta_{\ell,0} = e\, \delta_{\ell,0}~,
 \ee 
where $\chi$ is the Euler character and $e$ is the Euler class. We then have
\begin{align}
 O_{[2]}(\alpha_3) \ast O_{[q_1,q_2]} (\Delta^{(2)}_\ast (\alpha_1 \alpha_2)) =&\,  n_1\,\chi\, (\alpha_1\alpha_2)^0 \,  
O_{[n_1]} (\alpha_3 \text{vol})\,
 \\
&
+ \delta_{q_1,2}\, 
(\alpha_1 \alpha_2)^m (\alpha_3 e_{l})^n {c^{lk}}_m\, {c^{pq}}_n \, 
O_{[n_1-2,1,1]} (e_k;e_p;e_q)
\nonumber\\
&+ \delta_{q_2,2}\, 
(\alpha_1 \alpha_2)^m (\alpha_3 e_{k})^n {c^{lk}}_m\, {c^{pq}}_n \, 
O_{[n_1-2,1,1]} (e_l;e_p;e_q)
+\ldots  \nonumber
\end{align}
%Can we summarize the following equations for $q_1 \ge 1$ ?
Once again for the extremal correlators of interest we set $e_p$ and $e_q$ to be the identity operator in the last two terms, and this sets the structure constants to zero. Thus, only the first term contributes, and only for the component of $\alpha_3$ along the identity $e_0$.   
Performing the sum over $q_1, q_2$, we obtain 
\begin{align}
X_3:=& O_{[2]}(\alpha_3) \ast \frac12\sum_{q_1,q_2=2}^{n_1-2}q_1q_2\, \delta_{q_1+q_2,n_1} O_{[q_1,q_2]} (\Delta^{(2)}_\ast (\alpha_1 \alpha_2))\nonumber\\
=&  \frac{\chi}{6}n_1(n_1-1)(n_1-2)(n_1+3)(\alpha_1\alpha_2\alpha_3)^0 \,  
O_{[n_1]} (\text{vol})\, + \ldots 
% \nonumber \\
% &+ 2(n_1-2)
% (\alpha_1 \alpha_2)^m (\alpha_3 e_{l})^n {c^{lk}}_m\, {c^{pq}}_n \, 
% O_{[n_1-2,1,1]} (e_k;e_p;e_q)~.
\end{align}
For the contribution of the last term to the double fusion, we turn to the formula \eqref{extraOP} derived in the appendix. Using the co-product \eqref{deltastar}, and focusing only on the terms with the single cycles on the right hand side, we obtain
% \begin{align}
% \alpha O_{[2]} \ast O_{[n_1-1,1]} (\beta_1;\beta_2) &=
% O_{[n_1-1,2,1]} (\beta_1;\alpha;\beta_2)
% +2 O_{[n_1-1,2]} (\beta_1,\alpha \beta_2)\nonumber\\
% & + \frac{1}{2}
% \sum_{q_1=1}^{n_1-2} q_1 q_2 \delta_{q_1+q_2,n_1-1} O_{[q_1,q_2,1]} (\Delta_\ast^{(2)} (\alpha \beta_1); \beta_2) \nonumber\\
% &+ n_1 (O_{[n_1,1]} (\alpha \beta_1;\beta_2)+\alpha \beta_1 \beta_2 O_{[n_1]})~.
% \end{align}
\begin{align}
X_4:=& (n_1-1) O_{[2]} (\alpha_3) \ast O_{[n_1-1,1]} (\Delta^{(2)}_\ast (\alpha_1 \alpha_2)) \nonumber\\ 
=& (n_1-1)
(\alpha_1 \alpha_2)^m  {c^{kl}}_m\Big\{ 
n_1 O_{[n_1,1]}(\alpha_3 e_l; e_k)
+ n_1 \, {c_{lk}}^n  \, 
 O_{[n_1]}(\alpha_3\,  e_n) \,
\nonumber \\
&\hspace{3cm} +(n_1-2)\, (\alpha_3 e_l)^n\, {c^{pq}}_n\, O_{[n_1-2,1,1]}(e_p;e_q;e_k) 
\Big\}+\ldots 
\end{align}
The last term  arises from the $q_1=1, q_2=n_1-1$ and $q_1=n_1-1, q_2=1$ cases in the sum in equation \eqref{extraOP}. 
Let us focus on each term in the $X_4$ term in turn and impose the requirement that  there is no non-trivial operator associated to the inactive colours. 
For the first term, we set $k=0$. This automatically sets $e_l=\text{vol}$ and $m=0$ on account of the structure constant ${c^{l0}}_m = \delta^{l,\text{vol}}\delta_{m,0}$. This in turn implies that only the component of $\alpha_3$ along $e_0$ contributes (as $\alpha_3$ multiplies $e_l$ in the operator and the volume operator has maximal degree). 
In the second term, we use the identity
\be 
{c^{kl}}_m\, {c_{lk}}^n=\chi\, \delta_{m,0}\, \delta^{n,\text{vol}}~,
\ee 
which once again implies that only the component of $\alpha_3$ along $e_0$ matters. 
In the last term, the constraint sets $q=k=0$. We now use the special structure constants ${c^{p0}}_n = \delta^{p,\text{vol}}\delta_{n,0}$ and ${c^{l0}}_m = \delta^{l,\text{vol}}\delta_{m,0}$ and see that this term vanishes on account of the term $(\alpha_3 e_l)^n$; this vanishes since $e_l$ is  the generalized volume form and has zero projection along the identity element. We finally have 
\be 
X_4 = (\chi+1)n_1 (n_1-1) \, (\alpha_1\alpha_2\alpha_3)^0 \, O_{[n_1]}(\text{vol}) + \ldots 
\ee 
%
%For $q_1,q_2$ not two:
%
% \begin{align}
% \alpha_3 O_{[2]} \ast (\alpha_1 O_{[n_1]} \ast \alpha_2 O_{[2]} ) 
% &= \alpha_3 \alpha_1 \alpha_2 O_{[n_1+2]}
% \nonumber \\
% & + \dots + \dots
% \nonumber \\
% & + (n_1-1) \alpha_3 O_{[2]} \ast  O_{[n_1-1,1]} ( \Delta_\ast^{(2)}(\alpha_1 \alpha_2))
% \nonumber \\
% & + 
% \end{align}
% For the last line, we must understand the product of each term in $\Delta_\ast^{(2)}$. Is it just equal to the argument ?
%We split the rest of the calculation into three cases, depending on the value of $n_4$ compared to $n_1$.
%When we summarize the results incorporated in the four terms, 
Summing the four contributions to the double fusion, we find:
\begin{multline}
\label{doublefusion}
 O_{[2]}(\alpha_3) \ast  O_{[2]}(\alpha_2) \ast  O_{[n_1]}(\alpha_1) =
 (n_1+2)^2 O_{[n_1+2]} (\alpha_1 \alpha_2 \alpha_3) \\
  + 2 n_1^2\left(1+\frac{\chi}{12}(n_1^2-1)\right) (\alpha_1 \alpha_2 \alpha_3)^0 O_{[n_1]} (\text{vol})
 % + \frac{\chi}{6} n_1^2 (n_1^2-1) (\alpha_1 \alpha_2 \alpha_3)^0 O_{[n_1]} (\text{vol}) 
 + \dots
 \end{multline}
In order to compare our results with some extremal correlators computed previously in the literature it is useful to recall that the double fusion coefficients ${{\cal C}_{123}}^4$ calculated in this section are related to the extremal correlators ${\cal C}_{1234^{\dagger}}$ by the relation ${\cal C}_{1234^{\dagger}} = {{\cal C}_{123}}^4\, g_{44^\dagger} $, where $g$ is the Zamolodchikov metric, which  in the normalization of \cite{Pakman:2009ab} is inversely proportional to the length of the cycle. 
The first term is then a  generalization of the Hurwitz number computed in \cite{Pakman:2009ab, Ashok:2023mow} for the trivial Frobenius algebra. % (with non-zero central charge $c$). 
As in \cite{Ashok:2023mow}, we see that this coefficient obtains contributions from both single and double cycle intermediate states. 
For the term proportional to $O_{[n_1]}$, after taking into account the normalization factor of $n_1$, the coefficient of $2n_1$  matches the result in \cite{Pakman:2009zz} for the number of Feynman diagrams, or equivalently, the number of equivalence classes of maps that contribute to the extremal correlator. Our calculation strongly suggests that each Feynman diagram of \cite{Pakman:2009zz} contributes one to the final correlator --- in any case, we have determined the end result. Furthermore, we see that for a generic seed Frobenius algebra, there is an additional contribution to the coefficient of this operator due to  
the double contraction of structure constants, that represents a loop in the seed theory. It gives rise to a  new contribution proportional to the Euler number $\chi$. 

The calculation of the four-point function, which is an application of a small subset of our formulas for the operator products in the chiral ring, demonstrates once more the  calculational power of our approach.

\section{Conclusions}
\label{Conclusions}

Our main result, summarized in equation \eqref{MasterFormula}, is the large $N$ expansion of the operator ring of topological symmetric orbifolds, up to third order in perturbation theory. 
%At third order, we found a novel genus one contribution to the operator product. 
%Subsequently, by repeated use of the operator product, we showed how to compute simple extremal four point functions (in the physical theory) involving single cycle operators. 
We argued that the formalism used in this work is appropriate for all second quantized Frobenius algebras, regardless of their geometric interpretation. Our analysis  then applies not only to the theories with seed target $K3$ or $T^4$ but also $\mathbb{C}^2/\Gamma$, asymptotically linear dilaton theories of central charge six, as well as compact and non-compact Gepner models of any central charge and Calabi-Yau targets of any dimension. This is a non-trivial consequence of the algebraic construction of second quantized Frobenius algebras \cite{Kaufmann} which broadly generalizes the geometric construction of the cohomology ring of the Hilbert schemes of surfaces \cite{LS2}. 

Our concrete calculations represent a sizeable addition to the set of known three-point functions in these models and allow us to compute a large class of extremal correlators in the untwisted theory in large $N$ perturbation theory. We provided an illustration of this fact by computing a new class of extremal four-point functions including corrections proportional to the Euler number of the Frobenius algebra. 

It would be interesting to understand the $N$-independence of our structure constants better. For the trivial Frobenius algebra (with only the identity operator but non-zero central charge), the N-independence follows from properties of the symmetric group \cite{FarahatHigman}. These may be best understood in a semi-group generalization of the group of permutations \cite{IvanovKerov}. It is an interesting  challenge to extend the proof of $N$-independence of the structure constants for properly normalized operators to the general case. A very particular subcase was proven in \cite{LQW}. Perhaps a combination of the semi-group approach with the conservation of R-charge can yield progress. 

Our genus expansion of three-point functions begs for a string theoretic interpretation. In the case of the trivial Frobenius algebra, Hurwitz numbers provide a rather direct map between the symmetric group combinatorics and the counting of Riemann surfaces. In our general case, a world sheet interpretation would need to incorporate the Frobenius algebra structure constants into the string theory data. It would be wonderful to be able to draw dressed string diagrams that represent the genus expansion of structure constants faithfully. If we accept that single cycle operators represent single string states, we need to promote first quantized string diagrams to second quantized amplitudes in order to incorporate the multi-cycle states that form an integral and leading part of the operator ring. In any case, the string interpretation of the genus expansion is an interesting open problem (akin to the string interpretation of the correlators of two-dimensional Yang-Mills theory \cite{Gross:1993hu,Cordes:1994fc}). 

We believe symmetric orbifold theories form an interesting and universal subset of the set of conformal field theories and believe that the in-depth analysis of the operator algebra of their simplest, topological members in the 't Hooft large $N$ limit is a similarly universal and interesting result. We may be hopeful that it contributes to our understanding of the genus expansion of gauge theories and from there, to holographic quantum gravity.

\appendix

\section{Frobenius Algebras: Simple Examples
}
\label{FrobeniusAlgebraExamples}

We have formally defined Frobenius algebras and given several abstract formulae in the main text for multiplication maps and their adjoints. It may be useful to briefly go over a few simple models to see how these are realized in practice. 
%\subsubsection*{A Few Illustrations}

\subsection{The Chiral Ring of the Minimal Model}

Our first  example of a Frobenius algebra arises from the chiral ring of an ${\cal N}=2$ minimal model seed conformal field theory with central charge $c=3-6/k$ for a strictly positive integer $k$. The corresponding Frobenius algebra or chiral ring \be 
A_{MM}=\frac{\mathbb{C}[x]}{\langle x^{k-1} \rangle}
\ee
is the ring of polynomials in one variable $x$, modded out by the ideal generated by $x^{k-1}$. The  bi-linear form $T$ is 
$ 
T(x^i x^j)=\eta_{ij}=\delta_{i+j,k-2}~
$
and the structure constants are given by
\be 
{c_{ij}}^{\ell} = \delta_{\ell, i+j}~.
\ee 
By using the topological metric, one finds the symmetric tensor $c_{ij\ell} = \delta_{i+j+\ell, k-2}$.  
In this case the adjoint $\Delta^{(2)}_\ast$ of the multiplication reads:
\begin{align}
\Delta^{(2)}_\ast (x^{\ell}) &= \sum_{k_1,k_2=0}^{k-2} {c^{k_1k_2}}_{\ell}\, x^{k_1}\otimes x^{k_2}  \\
&=\sum_{k_1,k_2=0}^{k-2}\, \delta_{k_1+k_2, (k-2) +\ell} ~ x^{k_1} \otimes x^{k_2} \, .
\end{align}
By composing this with the multiplication map, we obtain
\begin{align} 
\Delta^{(2)}\Big( \Delta^{(2)}_\ast (x^{\ell})\Big) &=  \sum_{k_1,k_2=0}^{k-2}\,\delta_{ k_1+k_2, k-2+\ell} ~ x^{k_1+k_2} 
=\sum_{k_1=0}^{k-2}\, x^{k-2+\ell} = (k-2)x^{k-2+\ell}\nonumber \\
&=  (k-2) x^{k-2} \, \delta_{\ell,0}~,
\end{align}
which allows one to identify the Euler class $e$ as $e(A_{MM}) =(k-2) x^{k-2}$.
The higher co-products can be similarly computed using the structure constants. For $n=3$ we have
\begin{align}
    \Delta_\ast^{(3)} (x^{\ell}) &=\sum_{j_i, k_i=0}^{k-2}  {c^{k_1k_2j_3 }}{c_{j_3}}^{k_3j_4} \eta_{\ell j_4}~ x^{k_1}\otimes x^{k_2}\otimes x^{k_3}\nonumber \\ 
    &=\sum_{j_i, k_i=0}^{k-2}   {c^{k_1k_2j_3 }}\delta_{j_3, k_3-\ell}~ x^{k_1}\otimes x^{k_2}\otimes x^{k_3}\nonumber \\ 
    &=\sum_{k_i=0}^{k-2} \delta_{k_1+k_2+k_3, 2(k-2)+\ell} ~ x^{k_1}\otimes x^{k_2}\otimes x^{k_3}~.
\end{align} 
The action of the higher $n$-product $\Delta_\ast^{(n)} $ on $x^{\ell}$ can similarly be shown to give a non-vanishing result with unit coefficient for $x^{k_1}\otimes \ldots x^{k_n}$, provided the constraint $\sum_i k_i = \ell+ (n-1)(k-2)$ is satisfied. 

\subsection{The Cohomology Ring of the Four-Torus }

A second example Frobenius algebra is the cohomology ring $A_{T^4}$ of the four-torus.  We have a central charge $c=6$ seed conformal field theory. The two-point function is the integration of the forms over the four-torus. There are a total of sixteen basis elements in the cohomology.  In a choice of complex structure, they are given by the one-forms $dz_{1,2}, d\bar{z}_{1,2}$ to the power zero or one, multiplied together. 
The Euler class is zero since the Euler characteristic is. We illustrate this model by exhibiting explicitly the co-product $\Delta_\ast^{(2)}$:
\begin{equation}
\Delta_\ast^{(2)} (\omega_l) = \sum_{k_i} {c^{k_1 k_2}}_l \omega_{k_1} \otimes \omega_{k_2} =
\sum_{k_i^\vee} {c_{k^\vee_1 k^\vee_2}}^{l^\vee} \omega_{k_1} \otimes \omega_{k_2}\, .
\end{equation}
Each form $k$ has a dual with respect to the volume form, which we denoted by $k^\vee$. For example, if $\omega_l$ is the volume form, then we can only have  the identity for $k_i^\vee$ and the tensor product of two volume forms on the right hand side -- the reader can work out various other special cases. The general expression is a combination of contracted epsilon symbols. 
For the four-manifold $K3$,  there is a similar story. The Euler class is non-zero (and equal to twenty-four times the volume form) and the topological two-point function is of type $(4,20)$ and corresponds to a $E_8 \oplus E_8 \oplus 4  H$ lattice (where $H$ is a hyperbolic lattice of signature $(1,1)$). 
The list of examples is much longer and includes for instance compact and non-compact Gepner models. See e.g. \cite{Belin:2020nmp} for yet another class of interesting examples.

\section{The Structure Constants at Third Order}
\label{T3structureconstants}
\label{ThirdOrderStructureConstants}
In this Apppendix we  consider the contribution to the product of operators that occur due to a triple overlap $|T|=3$. 
%\subsection{The Enumeration of Various Subcases}
As before we only consider novel contributions, and from our analysis in Section \ref{sec3tripletripleoverlap} we know that the possible cycle structures are: 
%where the numbers refer to the cycle structures in $\pi_1, \pi_2$ and $\pi_3$: $(1,3;1),(1,2;2),(1,1;1), (1,1;3)$.  
\begin{align}
    O_{[n_1]}(\alpha)\ast O_{[p_1, p_2, p_3]}(\beta_1,\beta_2,\beta_3)\Big|_{|T|=3} &= \sum_{q_1} {\cal C}_{|T|=3}(q_1) O_{[q_1]}(\gamma)\\
     O_{[n_1]}(\alpha)\ast O_{[p_1, p_2]}(
     \beta_1,\beta_2)\Big|_{|T|=3} &= \sum_{q_i} {\cal C}_{|T|=3}(q_i) O_{[q_1,q_2]}(\gamma)\\
      O_{[n_1]}(\alpha)\ast O_{[p_1]}(\beta) \Big|_{|T|=3} &= \sum_{q_i} {\cal C}_{|T|=3}(q_i) O_{[q_1,q_2,q_3]}(\gamma)\\
      O_{[n_1]}(\alpha)\ast O_{[p_1]}(\beta)\Big|_{|T|=3} &= \sum_{q_1} {\cal C}_{|T|=3}(q_1) O_{[q_1]}(\gamma)
\end{align}
We discuss each of these cases in turn. 

\subsubsection{(1,3;1)}

Consider a particular set of three permutations involved in the operator product:
\begin{align} 
\pi_1 &= \alpha  (t_1 a_{1} \dots a_{j_1}  b_1 \dots b_{l_1} t_2 a_{j_1+1} \dots a_{j_2}  b_{l_1+1} \dots b_{l_2}t_3 a_{j_2+1} \dots a_{j}  b_{l_2+1} \dots b_{l}) \\
\pi_2 &=\beta_1(t_1b_{l}\ldots b_{l_2+1} c_1 \dots c_{m_1})\beta_2(t_2 b_{l_1} \dots b_1 c_{m_1+1} \dots c_{m_2})
\beta_3( t_3 b_{l_2} \dots b_{l_1+1}
c_{m_2+1} \dots c_{m}) \\
\pi_3&=\gamma (t_1 a_1\ldots a_{j_1} c_{m_1+1} \dots c_{m_2} t_2a_{j_1+1}\ldots a_{j_2} c_{m_2+1} \dots c_{m} t_3a_{j_2+1}\ldots a_{j}c_1\ldots c_{m_1}) ~.
\end{align}
The cycle lengths are split as follows: 
\begin{align}
    n_1 &= 3+j + l~,\hspace{1.6cm} q_1 = 3+j+m~, \\
    p_1 &= 1+l-l_2 + m_1~, \quad p_2 = 1+l_1+m_2-m_1~,\quad p_3 = 1+l_2-l_1 + m-m_2 ~.
\end{align}
The cycle length of the permutation $\pi_3$ can be determined in terms of the double-overlap $l$:
\be 
q_1 = n_1+p_1+p_2+p_3 - 2l-3~.
\ee 
We should be aware of a small complication at this stage. If we fix the cycle lengths of the permutation $\pi_2$ to be $(p_1,p_2,p_3)$ with triple overlap colours $(t_1,t_2,t_3)$ then there are alternative permutations $\pi_1$ and $\pi_3$ with the same cycle lengths that also satisfy $\pi_1 \pi_2 = \pi_3$, namely:
\begin{align} 
\pi_1 &= \alpha  (t_1 a_{1} \dots a_{j_1} 
 b_{l_1+1} \dots b_{l_2}t_3  a_{j_1+1} \dots a_{j_2} b_1 \dots b_{l_1} t_2 a_{j_2+1} \dots a_{j}  b_{l_2+1} \dots b_{l}) \\
\pi_2 &=\beta_1(t_1b_{l}\ldots b_{l_2+1} c_1 \dots c_{m_1})\beta_2(t_2 b_{l_1} \dots b_1 c_{m_1+1} \dots c_{m_2})
\beta_3( t_3 b_{l_2} \dots b_{l_1+1}
c_{m_2+1} \dots c_{m}) \\
\pi_3&=\gamma (t_1 a_1\ldots a_{j_1} c_{m_2+1} \dots c_{m} t_3 a_{j_1+1}\ldots a_{j_2} c_{m_1+1} \dots c_{m_2} t_2 a_{j_2+1}\ldots a_{j} c_1\ldots c_{m_1}) ~.
\end{align}
These have the effect of doubling the number of possible contributions in the reasoning below.

Consider the orbits of the various permutations in the space $O=\{ t, a, b, c\}$ of dimension $j+m+l+3$.
\begin{align}
\langle \pi_1 \rangle \setminus O
&= \{ \{t_1, t_2, t_3, a_i, b_r\}, c_s \}
\nonumber \\
\langle \pi_2 \rangle \setminus O
&= \{ \{t_1, b_{r_1}, c_{s_1}\}, \{t_2, b_{r_2}, c_{s_2}\}, \{t_3, b_{r_3}, c_{s_3}\}, a_i \}
\nonumber \\
\langle \pi_1 \pi_2 \rangle \setminus O
&= \{\{t_1, t_2, t_3, a_i, c_s\}, b_r \}
\nonumber \\
\langle \pi_1 , \pi_2 \rangle \setminus O
&= \{ t, a,b,c \} \, .
\end{align}
The graph defect is zero.
%can be computed as follows: 
%\begin{align}
%    g= \frac12\big(3+j+m+l +2 -(1+m)-%(3+j)-(1+l) \big) = 0~.
%\end{align}
Thus we find the operator  
\be 
\gamma = \Delta_\ast^{(l+1)}(\alpha\beta_1\beta_2\beta_3)~.
\ee 
We proceed to compute the numerical coefficient that would appear in the product of conjugacy classes. 

\paragraph{i.} The number of elements in the first conjugacy class is 
\begin{equation}
|C_{n_1;n}|=\frac{n!}{n_1 (n-n_1)! } \, .
\end{equation}

% \paragraph{ii.} We pick three triple overlap colours in the following number of ways:
% \begin{equation}
% \begin{pmatrix}n_1 \\ 3
% \end{pmatrix}\,. 
% \end{equation}

% \paragraph{ii'.} The number of ways in which the three colours $t_i$ can be associated to the cycles of lengths $p_i$ is given by
% \begin{equation}
% 3!~.
% \end{equation}

\paragraph{ii.} We split the double overlap number $l$ into three parts:
\begin{equation}
\frac{(l+2)(l+1)}{2}
\, .
\end{equation}
This fixes the number of $b$-colours in both the permutations $\pi_1$ and $\pi_2$.

\paragraph{iii.} We identify the colour $b_1$ in the first permutation $\pi_1$. This can be done in 
\be 
n_1
\ee
ways. Given the number $l_1$, this fixes the colour $t_2$ and the colours $b_1$ to $b_{l_1}$. 

\paragraph{iv.} In order to fix the $a$-colours, we need to pick numbers $j_1$ and $j_2$, which can be done in 
\be 
\frac12(j+1)(j+2)=\frac12(n_1-l-1)(n_1-l-2)
\ee 
ways. This fixes all the colours in the first permutation. 

\paragraph{v.} The $m$ colours of type $c$ can be chosen in 
\begin{equation}
% \frac{(n-n_1)!}{(n-n_1-m_1)!} \frac{(n-n_1-m_1)!}{(n-n_1-m_2)!} \frac{(n-n_1-m_2)!}{(n-n_1-m_3)!} =  
\frac{(n-n_1)!}{(n-n_1-m)!} \, 
\end{equation}
ways.

\paragraph{vi.} We divide by the number of elements in the last conjugacy class:
\begin{equation}
\frac{1}{|C_{q_1;n}|}=\frac{q_1 (n-q_1)! }{n!} \, .
\end{equation}
\paragraph{vii.} We  choose $l$ colours from $n-q_1$ colours:
\begin{equation}
\frac{l! (n-q_1-l)!}{(n-q_1)!}~.
\end{equation}

\paragraph{viii.} Finally, we take into account the doubling feature mentioned at the  beginning, which adds a factor of
\begin{equation}
2 \, .
\end{equation}
The structure constant is the product of all these factors:
\begin{align}
% &\frac{n!}{n_1 (n-n_1)! }
% n_1 (n_1-1)(n_1-2)
% \frac{(l+2)(l+1)}{2}
% \frac{(n-n_1)!}{(n-q_1-l)!}\frac{q_1 (n-q_1)! }{n!} \nonumber \\
% &= 
{\mathcal C}_{|T|=3}= \frac12(n_1-l-1)(n_1-l-2)
(l+2)! q_1\, .
\end{align}
We summarize the operator product:
\begin{multline}
    O_{[n_1]}(\alpha)\ast O_{[p_1, p_2, p_3]}(\beta_1,\beta_2,\beta_3)\Big|_{|T|=3} = \frac12\sum_{l=0}^{n_1-3}(n_1-l-1)(n_1-l-2)(l+2)!\\
\times\sum_{q_1\ge 3} ~q_1 ~\delta_{q_1+2l+3 -n_1-p_1-p_2-p_3, 0} ~ O_{[q_1, 1^l ]}(\Delta_\ast^{(l+1)}(\alpha\beta_1\beta_2\beta_3))~.
\end{multline} 
We conclude with listing the assumptions made about the cycle lengths. These arise from imposing non-negativity of the integers $m_1$, $m_2-m_1$ and $m-m_2$. They lead to the constraints
$l \le \text{min}(p_i-1, n_1-3) $. Along with the expression for $q_1$, we find that $q_1$  is necessarily greater than or equal to $\text{max}(n_1-p_1+p_2+p_3-1, n_1+p_1-p_2+p_3-1,n_1+p_1+p_2-p_3-1, p_1+p_2+p_3-n_1+3)$ for our reasoning to be valid.

\subsubsection{(1,1;3)}

We turn to the next operator product and, as usual, we begin with a set of permutations involved in the operator product: 
\begin{align} 
\pi_1 &= \alpha (t_1 a_{1} \dots a_{j_1}  b_1 \dots b_{l_1} t_2 a_{j_1+1} \dots a_{j_2}  b_{l_1+1} \dots b_{l_2}t_3 a_{j_2+1} \dots a_{j}  b_{l_2+1} \dots b_{l}) \\
\pi_2&= \beta (t_1 b_{l} \ldots b_{l_2+1}c_{1}\ldots c_{m_1} t_3 b_{l_2}\ldots b_{l_1+1} c_{m_2+1}\ldots c_{m} t_2 b_{l_1} \ldots b_1 c_{m_1+1}\ldots c_{m_2})  \\
\pi_3 &= \gamma\, (t_1 a_1 \dots a_{j_1} c_{m_1+1}\ldots c_{m_2} )
(t_3 a_{j_2+1} \dots a_{j} c_{1} \dots c_{m_1} )
(t_2 a_{j_1+1}+ \dots a_{j_2} c_{m_2+1} \dots c_{m} ) \, .
\end{align}
The cycle lengths are:
\begin{align}
    n_1&=3+j+l~, \quad p_1 = 3+l+m \\
    q_1 &= 1+j_1+m_2-m_1~, \quad q_2= 1+j-j_2+m_1~,\quad q_3 = 1+j_2-j_1+m-m_2~.
\end{align}
They satisfy the relation:
\be 
q_1+q_2+q_3=n_1+p_1 - 2l - 3~
\ee 
and we have the condition that $q_i \ge 2$. 
As in the $(1,3;1)$ case, we find that if we fix the permutation $\pi_3$ to be as shown, there exist alternative permutations $\pi_1$ and $\pi_2$ such that the relation $\pi_1\pi_2 = \pi_3$  is satisfied: 
\begin{align} 
\pi_1 &= \alpha (t_1 a_{1} \dots a_{j_1}
b_{l_1+1} \dots b_{l_2}t_3  a_{j_2+1} \dots a_{j} 
b_1 \dots b_{l_1} t_2a_{j_1+1} \dots a_{j_2}  b_{l_2+1} \dots b_{l}) \\
\pi_2&= \beta (t_2 b_{l_1} \ldots b_{1}c_{1}\ldots c_{m_1} t_3 b_{l_2}\ldots b_{l_1+1} c_{m_1+1}\ldots c_{m_2} t_1 b_{l} \ldots b_{l_2+1} c_{m_2+1}\ldots c_{m})  \\
\pi_3 &= \gamma\, (t_1 a_1 \dots a_{j_1} c_{m_1+1}\ldots c_{m_2} )
(t_3 a_{j_2+1} \dots a_{j} c_{1} \dots c_{m_1} )
(t_2 a_{j_1+1}+ \dots a_{j_2} c_{m_2+1} \dots c_{m} ) \, .
\end{align}
Once again this leads to a doubling of the contributions to the final structure constant. 

Let us consider the orbits in the space $O=\{ t, a, b, c\}$ of dimension $3+j+m+l$.
\begin{align}
\langle \pi_1 \rangle \setminus O
&= \{ \{t_1, t_2, t_3, a_i, b_r\}, c_s \}
\nonumber \\
\langle \pi_2 \rangle \setminus O
&= \{ \{t_1, t_2, t_3, b_r, c_s\}, a_i \}
\nonumber \\
\langle \pi_1 \pi_2 \rangle \setminus O
&= \{\{t_1, a_{i_1}, c_{s_1}\},\{t_2, a_{i_2}, c_{s_2}\},\{t_3, a_{i_3}, c_{s_3}\}, b_r \}
\nonumber \\
\langle \pi_1 , \pi_2 \rangle \setminus O
&= \{ t_1, t_2, t_3, a_i, b_r, c_s \} \, .
\end{align}
The graph defect is zero. 
%
%\begin{align}
%    g= \frac12\big(3+j+m+l+2 -(1+m)-(1+j)-(3+l) \big) = 0~.
%\end{align}
% 
We identify the right hand side operator $\gamma$:
\be 
\gamma= \Delta_\ast^{(l+3)} (\alpha \beta )~.
\ee 
% We once again check R-charge conservation; matching the R-charges on both sides leads to 
% %
% \be 
% Q(\Delta_\ast^{(l+3)}) = n_1+p_1-(q_1+q_2+q_3) +1 = 2l+4~,
% \ee 
% %
% which checks out. 
We now turn to the numerical evaluation of the structure constant.  
\noindent
Step {\bf i.} fixes permutation $\pi_1$ as before. We again split the double overlap number $l$ in three parts in step  {\bf ii.}. In step {\bf iii.} we identify the colour $b_1$. In step {\bf iv.} we pick the numbers $j_1$ and $j_2$. We pick the $c$-colours in the second permutation $\pi_2$ in step {\bf v.} All of these steps together provide a factor:
\be 
\frac{n!}{(n-n_1-m)!}\, \frac{1}{4}(j+1)(j+2)(l+1)(l+2)\, ,
\ee 
where we identify $j=n_1-l-3$ and $n_1+m=q_1+q_2+q_3+l$ from the expressions for the cycle lengths. 
% \noindent
% Steps {\bf i.} and {\bf ii.} are identical to the previous case. 
% % i. We count the number of elements in the first conjugacy class:
% % \begin{equation}
% % |C_{n_1;n}|=\frac{n!}{n_1 (n-n_1)! } \, .
% % \end{equation}

% % ii. We pick the triple overlap colours for a factor of
% % \begin{equation}
% % n_1 (n_1-1) (n_1-2)/3! \, .
% % \end{equation}

% \paragraph{ii'.} We pick the placement of the triply active colours in $\pi_2$, correlated with the ordering in $\pi_1$ and find
% \begin{equation}
% \frac{(p_1-1)(p_1-2)}{2}
% \end{equation}
% ways of placing them. 

% \noindent
% Steps {\bf iii.} and {\bf iv.} are identical to the previous case. 

% % iii. We split the number $l$ into three parts. We have two out of $l+2$ choices for a factor of:
% % \begin{equation}
% % \frac{(l+2)(l+1)}{2}
% % \end{equation}

% % iv. We pick $m_1$ colours out of $n-n_1$ and order them , then $m_2-m_1$ colours out of $n-n_1-m_1$ and order them, and then $m_3-m_2$ out of $n-n_1-m_2$ and order them. This gives a factor of:
% % \begin{equation}
% % % \frac{(n-n_1)!}{(n-n_1-m_1)!} \frac{(n-n_1-m_1)!}{(n-n_1-m_2)!} \frac{(n-n_1-m_2)!}{(n-n_1-m_3)!} =  
% % \frac{(n-n_1)!}{(n-n_1-m)!} \, .
% % \end{equation}

\paragraph{vi.} We divide by the number of elements in the  conjugacy class of $\pi_3$:
\begin{align}
\frac{1}{|C_{q_1,q_2,q_3;n}|} &= \frac{q_1 q_2 q_3 (n-q_1-q_2-q_3) !}{ n!} ~.
\end{align}

\paragraph{vii.} We choose $l$ colours out of $n-(q_1+q_2+q_3)$ to give a factor
\be 
\frac{l!(n-q_1-q_2-q_3-l)}{(n-q_1-q_2-q_3)!}~.
\ee 

\paragraph{viii.} Lastly we take into account the alternative set of permutations that gives the same $\pi_3$: This leads to a factor of 
\be 
2~.
\ee

We multiply all the factors to find the structure constant:
% \begin{align}
% %k_{31} k_{32} k_{33}(n_1-1)(n_1-2)\frac{(n-(k_{31}+k_{32}+k_{33}))!}{(n-n_1-m_3)! }=
% \frac{(l_3+2)(l_3+1)}{2}
% k_{31} k_{32} k_{33}(n_1-1)(n_1-2)\frac{(n-(k_{31}+k_{32}+k_{33}))!}{(n-n_1-n_2+3+l_3)! } \, .
% \end{align}
% We use 
% \begin{equation}
% n_1+n_2-3 = k_{31}+k_{32}+k_{33}+2l_3 \, .
% \end{equation}
%
% \begin{align}
% \frac{(l_3+2)(l_3+1)}{2}
% k_{31} k_{32} k_{33}(n_1-1)(n_1-2)\frac{(n-n_1-n_2+2l_3+3)!}{(n-n_1-n_2+l_3+3)! } \, .
% \end{align}
\begin{align}
    {\cal C}_{|T|=3} &=\frac12 (j+1)(j+2)\, (l+1)(l+2)\,  l!\, q_1q_2q_3 \nonumber\\
    &= \frac12(n_1-l-1)(n_1-l-2)(l+2)!q_1q_2q_3~.
\end{align}
We thus have the operator product
\begin{multline}
O_{[n_1]}(\alpha)\ast O_{[p_1]}(\beta)\Big|_{|T|=3} = \frac{1}{2}\sum_{l=0}^{n_1-3} (n_1-l-1)(n_1-l-2) (l+2)!
 \\
%\times\frac{1}{3!}
\sum_{\substack{q_i \ge 2 \\
                \{ q_i, q_j, q_k \} }
    } ~q_1\, q_2\, q_3~ \delta_{q_1+q_2+q_3, n_1+p_1-2l-3}\, O_{[q_1, q_2, q_3, 1^l]}(\Delta_\ast^{(l+3)}(\alpha\beta))~.
\end{multline}
%In the final expression we have an additional factor of $\frac{1}{3!}$, which is applicable to the case in which all the $q_i$ are distinct. For particular cases in which we all $q_i$ are identical, for instance, this factor will be omitted. 
As before, there are constraints on the length of the cycles in $\pi_3$ that are obtained by imposing the non-negativity of the numbers $m_1$, $m_2-m_1$ and $m-m_2$.  They are given by  
\be 
q_i \ge n_1-l-2 ~, \quad\text{for}\quad i=1,2,3~.
\ee 

\subsubsection{(1,2;2)}

We consider a set of permutations involved in the operator product:
\begin{align} 
\label{perms122l1}
\pi_1 &= \alpha (t_1 a_{1} \dots a_{j_1}  b_1 \dots b_{l_1} t_2 a_{j_1+1} \dots a_{j_2}  b_{l_1+1} \dots b_{l_2}t_3 a_{j_2+1} \dots a_{j}  b_{l_2+1} \dots b_{l}) \\
\pi_2 &=\beta_1 (t_1 b_{l}\ldots b_{l_2+1} c_{m_2+1} \dots c_{m}
t_2 b_{l_1} \dots b_1
c_{m_1+1} \dots c_{m_2})
\beta_2(t_3 b_{l_2} \dots b_{l_1+1}c_1\ldots c_{m_1})\\
    \pi_3 &=\gamma\, (t_1 a_1\ldots a_{j_1} c_{m_1+1} \dots c_{m_2}) (t_2 a_{j_1+1}\ldots a_{j_2}c_1\ldots c_{m_1}t_3a_{j_2+1}\ldots a_{j} c_{m_2+1}\ldots c_{m})~.
\label{perms122l3}
\end{align}
The cycle lengths are given by
\begin{align}
    n_1&= 3+j+l~,\\
    p_1 &= 2+l+l_1-l_2+m-m_1~,\quad p_2 = 1+l_2-l_1+m_1\\
    q_1 &= 1+j_1+m_2-m_1~,\quad q_2 = 2+j-j_1 + m+m_1-m_2~.
\end{align}
The requirement that the triply active colours be part of a non-trivial cycle imposes $q_i \ge 2$. 
The cycle lengths of $\pi_3$ can be written:
\be 
q_1+q_2 = n_1+p_1+p_2-2l-3~.
\ee
%
% 
% \begin{align} 
% \pi_1 &= \alpha (t_1 a_{1} \dots a_{j_1} 
% b_{l_1+1} \dots b_{l_2}t_3 a_{j_1+1} \dots a_{j_2} 
% b_1 \dots b_{l_1} t_2a_{j_2+1} \dots a_{j}  b_{l_2+1} \dots b_{l}) \\
% \pi_2 &=\beta_1 (t_1 b_{l}\ldots b_{l_2+1} c_{m_2+1} \dots c_{m}
% t_2 b_{l_1} \dots b_1
% c_{m_1+1} \dots c_{m_2})
% \beta_2(t_3 b_{l_2} \dots b_{l_1+1}c_{m_1+1}\ldots c_{m_2})\\
%     \pi_3 &=\gamma\, (t_1 a_1\ldots a_{j_1} c_{m_1+1} \dots c_{m_2}) (t_2 a_{j_2+1} \ldots a_j c_{m_1+1}\ldots c_{m_2}t_3a_{j_1+1}\ldots a_{j_2} c_{m_2+1}\ldots c_{m})~.
% \end{align}
% Is there another combination of three permutations where $t_1$ is the colour associated to $q_1$, $t_3$ is the colour associated to $p_2$ and $t_2$ is the third triple active colour, but in which $t_1,t_3,t_2$ appear in thAt order in $\pi_1$ ? 
%
In the permutations shown in \eqref{perms122l1}-\eqref{perms122l3}, we have assumed that two of the triply active colours appear in the $p_1$-cycle; we have to take into account the other possibility as well, in which two of the triply active colours appear in the $p_2$-cycle. This essentially amounts to exchanging $p_1$ and $p_2$. This leads to a doubling of the contribution to the structure constant. 
%Another possibility (?) is to exchange $(q_1,q_2)$. (Yes, but that is already summed over.)

Consider the orbits of the various permutations in the space $O=\{ t, a, b, c\}$ of dimension $j+m+l+3$.
\begin{align}
\langle \pi_1 \rangle \setminus O
&= \{ \{t_1, t_2, t_3, a_i, b_r\}, c_s \}
\nonumber \\
\langle \pi_2 \rangle \setminus O
&= \{ \{t_1, t_2, b_{r_1}, c_{s_1}\}, \{t_3, b_{r_2}, c_{s_2}\}, a_i \}
\nonumber \\
\langle \pi_1 \pi_2 \rangle \setminus O
&= \{\{t_1, a_{i_1}, c_{s_1}\}, \{t_2, t_3, a_{i_2}, c_{s_2}\}, b_r \}
\nonumber \\
\langle \pi_1 , \pi_2 \rangle \setminus O
&= \{ t, a,b,c \} \, .
\end{align}
The graph defect is zero.
%\begin{align}
%    g= \frac12\big(3+j+m+l +2 -(1+m)-%(2+j)-(2+l) \big) = 0~.
%\end{align}
One identifies the operator
\be 
\gamma = \Delta_\ast^{(l+2)}(\alpha\beta_1\beta_2)~.
\ee
%
% This can be confirmed by equating the R-charges on both sides of the operator product:
% \begin{align} 
% Q(\Delta_\ast^{(n)})&=n_1-1+p_1+p_2-2 - (q_1+q_2-2) \cr
% &= n_1+p_1+p_2-q_1-q_2 -1 = 2l+2 ~,
% \end{align}
% %
% which sets $n=l+2$. 
We now compute the numerical coefficient of the product of conjugacy classes. 
\noindent
% Step {\bf i.} fixes permutation $\pi_1$ as before. We split the double overlap number $l$ in three parts in step  {\bf ii.}, as before. In step {\bf iii.}, we identify the colour $b_1$. In step {\bf iv.} we pick $j_1$ and $j_2$. All of these steps together provide a factor:
% \begin{equation}
% |C_{n_1;n}|
% \frac{(l+1)(l+2)}{2}
% n_1
% \frac12(n_1-l-1)(n_1-l-2) \, .
% \end{equation}
% \paragraph{v.} We pick the $c$-colours to fill out $\pi_2$.
% They can be chosen in 
% \begin{equation}
% \frac{(n-n_1)!}{(n-n_1-m)!}
% \end{equation}
% ways.
Steps {\bf i.} to {\bf v.} are identical to the previous two cases, leading to a factor 
\be 
\frac{n!}{(n-n_1-m)!}\, \frac{1}{4}(j+1)(j+2)(l+1)(l+2)\, ,
\ee 
where we identify $j=n_1-l-3$ and $n_1+m=q_1+q_2+l$ from the expressions for the cycle lengths. 

\paragraph{vi.} We divide by the cardinal number of the third conjugacy class:
\begin{equation}
\frac{1}{|C_{q_1,q_2;n}|}
= \frac{q_1  q_2 (n-q_1-q_2)!}{n!}
\end{equation}

\paragraph{vii.} Finally we have the factor associated to choosing $l$ colours from $n-q_1-q_2$ colours:
\begin{equation}
\frac{l! (n-q_1-q_2-l)!}{(n-q_1-q_2)!}
\end{equation}

\paragraph{viii.} Finally, we have a factor of
\be 
2
\ee 
from the exchange of $p_1$ and $p_2$, as discussed above. 
The product of all these factors gives the structure constant:
\begin{align}
{\mathcal C}_{|T|=3}
%&= \frac12(j+1)(j+2)(l+1)(l+2)\, l!\, q_1 q_2\nonumber \\ 
&= \frac12 (n_1-l-1)(n_1-l-2) (l+2)!
q_1  q_2 ~.
\end{align}
Substituting $q_1+q_2$ in terms of the cycle lengths  and $l$, we find 
\begin{multline}
     O_{[n_1]}(\alpha)\ast O_{[p_1, p_2]}(
     \beta_1,\beta_2)\Big|_{|T|=3} = \frac12\sum_{l=0}^{n_1-3}(n_1-l-1)(n_1-l-2)(l+2)!\\
     \times
     %\frac{1}{2!}
     \sum_{\substack{ q_i \ge 2 \\
     \{ q_1,q_2\} }
          } ~q_1\, q_2~\delta_{q_1+q_2,n_1+p_1+p_2-2l-3} O_{[q_1,q_2,1^l]}(\Delta_\ast^{(l+2)}(\alpha\beta_1\beta_2))~.
\end{multline}
The constraints on the validity of the derivation follow from the non-negativity of $m_1$ and $m_2-m_1$ and $m-m_2$. We have that $l \le \text{min}(p_1-1, p_2-1, n_1-3)$, and that $q_1 \ge n_1-l-2$, $q_2 \ge n_1-l-1$ as well as $q_2 \ge n_1+p_2-2$. 

\subsubsection{(1,1;1)}

We  turn to the final case with triple overlap three.  Consider a set of three single cycle permutations involved in the operator product:
\begin{align}
\pi_1 &= \alpha (t_1 a_{1} \dots a_{j_1}  b_1 \dots b_{l_1} t_2 a_{j_1+1} \dots a_{j_2}  b_{l_1+1} \dots b_{l_2}t_3 a_{j_2+1} \dots a_{j}  b_{l_2+1} \dots b_{l}) \\
\pi_2&= \beta (t_1 b_{l_3} \ldots b_{l_2+1}c_{1}\ldots c_{m_1}t_2 b_{l_1} \ldots b_1 c_{m_1+1}\ldots c_{m_2} t_3 b_{l_2}\ldots b_{l_1+1} c_{m_2+1}\ldots c_{m}) \\
\pi_3&= \gamma (t_1a_1\ldots a_{j_1}c_{m_1+1}\ldots c_{m_2}t_3 a_{j_2+1}\ldots a_{j}
c_1\ldots c_{m_1}
t_2a_{j_1+1}\ldots a_{j_2}c_{m_2+1}\ldots c_{m}) \, .
\end{align}
The lengths of the cycles are split as 
\be
n_1= 3+j+l~,\qquad p_1= 3+l+m~, \quad q_1= 3+j+m~.
\ee 
The length of the final conjugacy class can be written as 
\be 
q_1 = n_1+p_1-2l-3~.
\ee 
Consider the orbits of the various permutations in the space $O=\{ t, a, b, c\}$ of dimension $j+m+l+3$.
\begin{align}
\langle \pi_1 \rangle \setminus O
&= \{ \{t_1, t_2, t_3, a_i, b_r\}, c_s \}
\nonumber \\
\langle \pi_2 \rangle \setminus O
&= \{ \{t_1, t_2, t_3, b_r, c_s\}, a_i \}
\nonumber \\
\langle \pi_1 \pi_2 \rangle \setminus O
&= \{\{t_1, t_2, t_3, a_i, c_s\}, b_r \}
\nonumber \\
\langle \pi_1 , \pi_2 \rangle \setminus O
&= \{ t, a,b,c \} \, .
\end{align}
The graph defect can be computed as follows: 
\begin{align}
    g= \frac12\big(3+j+m+l+2 -(1+m)-(1+j)-(1+l) \big) = 1~.
\end{align}
The graph defect is one -- we have a genus one contribution.  We therefore  include the Euler character $e$ in the identification of the right hand side operator $\gamma$ -- see equation (\ref{Multiplication}) --:
\be 
\gamma = \Delta_\ast^{(l+1)}(\alpha\beta e)~.
\ee
 One can perform a consistency check by calculating and comparing the R-charges on both sides of the operator product. The new ingredient is the R-charge of the Euler class $q(e) = \frac{c}{3}$.  
%The R-charge of the $\Delta_\ast^{(l+1)}$ that has to appear is given by 
% %
% \be 
% Q(\Delta_\ast^{(l+1)}(e\cdot\phantom{e})) = n_1+p_1-2 - (q_1-1) = n_1+p_1-q_1-1 =2l+2~. 
% \ee 
% %
% This is consistent with the identification $Q(\Delta_\ast^{(l+1)})=2l$ and $Q(e) = 2$.
The calculation of the numerical coefficient appearing in the operator product proceeds along the same lines as before.
% \noindent
% Steps {\bf i.} and {\bf ii.} are identical to the other sub-cases in the $|T|=3$ sector. 
% % i. The number of elements in the first conjugacy class is 
% % \begin{equation}
% % |C_{n_1;n}|=\frac{n!}{n_1 (n-n_1)! } \, .
% % \end{equation}
% %
% % ii. We pick the colours $t_{1,2,3}$ in the following numbers of ways: 
% % \begin{equation}
% % \frac{1}{3!}n_1 (n_1-1)(n_1-2) \, .
% % \end{equation}
% % We need to divide by $3!$ because we don't yet assign an order to them. 
%
% \paragraph{ii'.} We pick the placement of the colours in $\pi_2$, respecting the ordering in $\pi_1$. We put $t_1$ first and then find
% \begin{equation}
% \frac{(p_1-1)(p_1-2)}{2}
% \end{equation}
% ways of placing the colours $t_{2,3}$. 
%
% \noindent
% Steps {\bf iii.} and {\bf iv.} are identical to the ones in the previous section.
%
% iii. We split the number $l$ into three parts:
% \begin{equation}
% \frac{(l+2)(l+1)}{2}
% \, .
% \end{equation}
%
% iv. The $m$ colours of type $c$ can be chosen in the following ways:
% \begin{equation}
% % \frac{(n-n_1)!}{(n-n_1-m_1)!} \frac{(n-n_1-m_1)!}{(n-n_1-m_2)!} \frac{(n-n_1-m_2)!}{(n-n_1-m_3)!} =  
% \frac{(n-n_1)!}{(n-n_1-m)!} \, .
% \end{equation}
%
Steps {\bf i.} and {\bf v.} are identical to the previous three sub-cases with $|T|=3$, leading to the factor
\be 
\frac{n!}{(n-n_1-m)!}\, \frac{1}{4}(j+1)(j+2)(l+1)(l+2)\, ,
\ee 
where we identify $j=n_1-l-3$ and $n_1+m=q_1+l$ from the expressions for the cycle lengths. 
\paragraph{ vi.} We divide by the number of elements in the last conjugacy class:
\begin{equation}
\frac{1}{|C_{q_1;n}|} =
\frac{q_1 (n-q_1)!}{n!}
\end{equation}

\paragraph{vii.} Finally we choose $l$ colours from among $n-q_1$, which gives the factor
\begin{equation}
\frac{l!(n-q_1-l)! }{(n-q_1)!}~.
\end{equation}
The structure constant is the product of all these factors:
\begin{align}
{\cal C}_{|T|=3} &=\frac14 (j+1)(j+2)(l+1)(l+2)\, l!\, q_1\nonumber \\ 
&= \frac14(n_1-l-1)(n_1-l-2)
(l+2)! q_1~.
\end{align}
Substituting the value of $q_1$ in terms of the cycle lengths and $l$, we find the operator product:
\begin{multline}
O_{[n_1]}(\alpha)\ast O_{[p_1]}(\beta)\Big|_{|T|=3} =\frac{1}{4}\sum_{l=0}^{n_1-3}(n_1-l-1)(n_1-l-2) (l+2)!\\
\times \sum_{q_1\ge 3} q_1~\delta_{q_1, n_1+p_1-2l-3}\, O_{[q_1, 1^l]}(\Delta_\ast^{(l+1)}(\alpha\beta e))~.
\end{multline}
This finishes the calculation of the operator product terms that arise from permutations with triple overlap equal to three. In particular, we identified a generic genus one contribution to the product of operators. 

\section{More Operator Products}
\label{novelOP}
\label{MoreOperatorProducts}

For most of our analysis in the bulk of the paper, we have worked with operators in which we associate non-trivial Frobenius algebra elements to active colours only. In this section we clarify a few basic properties of operators that have a non-trivial algebra element associated to inactive colours. 

\subsection{The Normalization of Operators}
\label{operatornorms}

% 
% In this subsection we explain the definition of various operators including their normalization. 
% Let us render explicit the details for the operators of the type:
% \begin{equation}
% C_{[p_1,\dots, p_k, q_j,1^l]} (\beta_1,\dots,{\beta}_k,\alpha_{(1}, \ldots \alpha_{l+1)} ) 
% \, ,
% \end{equation}
% where the symmetrization of the $\alpha_i$  associated to the cycles $q_j$ and $1^l$ has a coefficient $1/(l+1)!$ for the $(l+1)!$ terms. 

% Example:
% \begin{equation}
% C_{[p_1,\dots, \slashed{p}_j,\dots, q_j,1^l]} (\beta_1,\dots,\slashed{\beta}_j,\dots, \Delta_\ast^{(l+1)} (\alpha \beta_j))
% \rightarrow C_{[p_1,\dots, \slashed{p}_j,\dots, q_j,1^l]}
% \end{equation}
% where
% \begin{equation}
% C_{[p_1,\dots, \slashed{p}_j,\dots, q_j,1^l]}= \frac{|C_{[p_1,\dots, \slashed{p}_j,\dots, q_j,1^l]}|}{|C_{[p_1,\dots, \slashed{p}_j,\dots, q_j]}|} C_{[p_1,\dots, \slashed{p}_j,\dots, q_j]}
% \end{equation}
% where

We begin with a generic operator of the form 
\begin{equation}
O_{[p_1,\dots, p_k,1^l]} (\beta_1,\dots,\beta_k,\beta_{k+1},\dots,\beta_{k+l})
\end{equation}
in which we associate Frobenius algebra elements to both active and inactive colours. 
%and the operator
%\begin{equation}
%O_{[p_1,\dots, p_k, q_j]}
%\end{equation}
The number of terms in the operator is equal to:
\begin{equation}
|O_{[p_1,\dots,  p_k,1^l]}| = \frac{n!}{(p_1 \ldots p_k) (n-\sum_i p_i-l)!} \, .
\end{equation}
Let us exhibit these terms in detail. Consider a permutation $\pi=c(p_1)\ldots c(p_k)\in [p_1,p_2,\dots, p_k]$, where the $c(p_i)$ denotes a cycle of length $p_i$. We then define the operator as:
\begin{align}
O_{[p_1,\dots,  p_k,1^l]}(\beta_1, \dots \beta_{k+l}) &=
\sum_{\pi \in [p_1,p_2,\dots, p_k]}
 c(p_1)(\beta_1) \dots  c(p_k)(\beta_k)\nonumber\\  
&\hspace{2cm} \times\sum_{\rho \in S_{n-\sum p_j} \setminus S_{n-\sum p_j - l}} 
\rho ( \beta_{k+1} \otimes \dots 
\otimes \beta_{k+l} \otimes 1 \dots \otimes 1) 
\nonumber \\
&=
\sum_{\pi \in [p_1,p_2,\dots, p_k]}
c(p_1)(\beta_1) \dots  c(p_k)(\beta_k)\, ( \beta_{k+1} \otimes \dots 
\otimes \beta_{k+l} \otimes 1 \dots \otimes 1) \nonumber\\  
& 
\hspace{1cm}+ \text{terms with coefficient one symmetrized in the $l$ entries} \nonumber \\ 
& \hspace{2cm} \text{as well as the choice of inactive slots.} 
\end{align}
This should be contrasted with the number of terms in the operator
\begin{equation}
O_{[p_1,\dots, p_k]} (\beta_1,\dots,\beta_k)~,
\end{equation}
which equals the dimension of the conjugacy class $[p_1, \ldots p_k] $: 
\begin{equation}
|O_{[p_1,\dots,  p_k]}| = \frac{n!}{(p_1 \dots p_k) (n-\sum p_i)!} \, .
\end{equation}
Therefore we have the relation:
\begin{equation}
O_{[p_1,\dots, p_k,1^l]} (\beta_1,\dots,\beta_k,1,\dots,1)
= \frac{|O_{[p_1,\dots,  p_k,1^l]}|}{|O_{[p_1,\dots,  p_k]}|}
O_{[p_1,\dots,p_k]} (\beta_1,\dots,\beta_k) \, .
\end{equation}
This relation will prove to be useful when we perform some numerical checks on our master formula by setting the Frobenius algebra elements to the identity and counting the number of terms that appear in the operator product. 

% \subsubsection{The sub-subleading terms}

% \begin{equation}
% C_{[p_1, \ldots \slashed{p}_j\ldots q_j, q_k]}\big(\beta_1, \ldots \slashed{\beta}_j, \ldots \Delta_\ast^{(2)}(\alpha\beta_j )\big)
% \rightarrow C_{[p_1, \ldots \slashed{p}_j\ldots q_j, q_k]}
% \end{equation}
% at $l=0$.

% For higher $l$, we would need:
% \begin{equation}
% C_{[p_1, \ldots \slashed{p}_j\ldots q_j, q_k, 1^{l}]}\big(\beta_1, \ldots \slashed{\beta}_j, \ldots \Delta_\ast^{(l+2)}(\alpha\beta_j )\big)
% \rightarrow 
% \frac{|C_{[p_1,\dots, \slashed{p}_j,\dots, q_j,q_k,1^l]}|}{|C_{[p_1,\dots, \slashed{p}_j,\dots, q_j,q_k]}|} C_{[p_1,\dots, \slashed{p}_j,\dots, q_j, q_k]}
% \end{equation}

% We can try to find an example checking the third term at higher $l$ by computer. 

% For the fourth term

\subsection{A Non-trivial Inactive Colour Entry}
 
In this subsection we calculate the exact operator product
\begin{equation}
 O_{[2]}(\alpha) \ast O_{[n_1-1,1]} (e_l;e_k)~,
\end{equation}
in which we associate non-trivial basis vectors of the Frobenius algebra ($e_l$ and $e_k$) to both the non-trivial cycle $[n-1]$ and one inactive colour\footnote{
An operator product of this sort appears generically in the intermediate steps whenever higher point extremal correlators, or multiple single cycle fusions are carried out. See the example in section \ref{FourPointFunctions}.}. Such an operator product does not fall into the class of operator products encoded in the summary formula  \eqref{MasterFormula}, as in that calculation, we assumed that the unit operator was associated to the inactive colours. We therefore perform this calculation from first principles.
The colours in each of the operators can overlap in $0, 1$ or $ 2$ elements, and we consider these cases in turn. 
%denote these contributions by $P^{(i)}$. 

\paragraph{Case 1:} We first consider the term with zero overlap between the colours in the two operators. The contribution of such terms is  given by the fusion
\begin{align}
%P^{(0)}_{1} &= 
O_{[n_1-1,2,1]} (e_l;\alpha;e_k) \, .
\label{FusionContribution}
\end{align}
This can be checked by a  comparison of the number of terms in the product with zero overlap on the left hand side and checking that it is equal to the number of terms in the operator on the right hand side. 
% Our convention for that operator is that we symmetrize $e_l (1,\dots,n_1-1) \alpha (n_1,n_1+1) \otimes e_k \otimes 1 \otimes \dots$. We pick any element in $C_{[n_1-1,2]}$ and then symmetrize $e_k$ over the other $n-n_1-1$ entries, with weight one.
% This gives:
% \begin{equation}
% \frac{n!}{(2(n_1-1) (n-n_1-1)!} \times (n-n_1-1) =
% \frac{n!}{2(n_1-1) (n-n_1-2)!}\qquad \text{terms.}
% \end{equation}
% %
Let us first do a count on the left hand side. The number of terms in $ O_{[2]}(\alpha)$ is equal to the number of elements in the conjugacy class, equal to $\binom{n}{2}$.
% \begin{equation}
% |C_{[2]}| = \frac{n!}{2 (n-2)!} \, . 
% \end{equation}
The number of terms in the second operator is counted as follows.
%we start with a term $e_l (1, 2 ,\dots, n_1-1) \otimes e_k \otimes 1 \otimes \dots$. 
We first pick an element of the conjugacy class, and then we pick one out of $n-(n_1-1)$ elements so as to put $e_k$ in that entry. We sum with weight one over all these terms. The number of terms is:
\begin{align}
|O_{[n_1-1,1]}(\cdot~;~\cdot) | &=|O_{[n_1-1]}| \times (n-(n_1-1)) 
= 
%\frac{n!}{(n_1-1) (n-(n_1-1))!}\times (n-(n_1-1)) \nonumber \\= 
\frac{n!}{(n_1-1) (n-n_1)!} \, .
\end{align}
We know need to know how many of all these products contribute to the right hand side with zero overlap in all non-trivial entries. We need to pick two out of the remaining $n-n_1$ colours. There are therefore:
\begin{equation}
|O_{[n_1-1,1]}(\cdot~;~\cdot)| \times \frac{(n-n_1)!}{2 (n-n_1-2)!} = \frac{n!}{2(n_1-1) (n-n_1-2)!}~.
\end{equation}
such terms.
We need to divide this number by the number of terms in the right hand side operator $O_{[n_1-1,2,1]}(e_l,\alpha;e_k)$. Our convention for that operator is that 
%we symmetrize an example term $e_l (1,\dots,n_1-1) \alpha (n_1,n_1+1) \otimes e_k \otimes 1 \otimes \dots$ with weight one for each term. 
we pick any element in $O_{[n_1-1,2]}$ and then symmetrize $e_k$ over the other $n-n_1-1$ entries, with weight one.
This gives:
\begin{equation}
\frac{n!}{(2(n_1-1) (n-n_1-1)!} \times (n-n_1-1) =
\frac{n!}{2(n_1-1) (n-n_1-2)!}
\end{equation}
terms.
We need to divide the total number of terms (with weight one) by the number of terms on the right hand side and find the unit coefficient in the contribution (\ref{FusionContribution}).
% \begin{equation}
% \frac{n!}{(n_1-1) (n-n_1)!}\times \frac{(n-n_1)!}{2 (n-n_1-2)!}\times \frac{2(n_1-1) (n-n_1-2)!}{n!} = 1.
% \end{equation}

\paragraph{Case 2:} Next we consider the case in which there is zero overlap between the non-trivial cycles $[2]$ and $[n_1-1]$, but one overlap between the colours in $[2]$ and the colour associated to $e_k$. 
%Let's call these terms $P^{(0)}_{2}$. 
The operator on the right hand side is:
\begin{equation}
O_{[n_1-1,2]} (e_l,\alpha e_k) \, .
\end{equation}
The latter has
\begin{equation}
|O_{[n_1-1,2]}| = \frac{n!}{2(n_1-1) (n-n_1-1)!}
\end{equation}
terms. The number of terms contributing on the left hand side is:
\begin{equation}
|O_{[n_1-1,1]}(\cdot~;~\cdot)| \times (n-n_1) =\frac{n!}{(n_1-1)(n-n_1-1)!}\, .
\end{equation}
Therefore, the coefficient of the operator, which is  the ratio of these is given by two. 
% \begin{equation}
% \frac{n!}{(n_1-1)(n-n_1)!} (n-n_1) \frac{2 (n_1-1)(n-n_1-1)!}{n!}
% = 2 \, .
% \end{equation}
Thus we obtain a term
\be 
%P^{(0)}_2 = 
2\, O_{[n_1-1,2]} (e_l,\alpha e_k)~.
\ee 

\paragraph{Case 3:} Thirdly, we consider the case in which  we have overlap one between $[2]$ and $[n_1-1]$, and no overlap between the permutation in conjugacy class $[2]$ and the $e_k$ entry. The resulting operator is:
\begin{equation}
 O_{[n_1,1]} (\alpha e_l;e_k) \,.
\end{equation}
This has 
\begin{equation}
\frac{n!}{n_1 (n-n_1)!} (n-n_1) = \frac{n!}{n_1 (n-n_1-1)!} 
\end{equation}
terms. On the left hand side, we count all terms:
\begin{equation}
|O_{[n_1-1,1]}(\cdot~;~\cdot)| \times (n_1-1) (n-n_1) =\frac{n!}{(n-n_1-1)!}\, .
\end{equation}
that contribute to the single overlap. 
Therefore, the coefficient of that operator is  $n_1$:
% \begin{align}
% P_{1,1} &=  \frac{n!}{(n_1-1) (n-n_1)!}   \times (n_1-1) (n-n_1) \times \frac{n_1 (n-n_1-1)!} {n!} \nonumber \\
% &= n_1 \, .
% \end{align}
%Thus we have the term
\be 
%P^{(1)}_1 = 
n_1\,  O_{[n_1,1]} (\alpha e_l;e_k)~. 
\ee

\paragraph{Case 4:} Another possibility is a single overlap between $[2]$ and $[n-1]$ and a single overlap between $[2]$ and the colour associated to $e_k$.
The resulting operator is:
\begin{equation}
 O_{[n_1]}(\alpha e_l e_k) = {c_{lk}}^n \, O_{[n_1]}( \alpha\, e_n)~ \, ,
\end{equation}
where we have used the product in the Frobenius algebra. The operator  has
\begin{equation}
\frac{n!}{n_1 (n-n_1)!}
\end{equation}
terms. The number of terms contributing to the left hand side is:
\begin{equation}
|O_{[n_1-1,1]}(\cdot~;~\cdot)| \times (n_1-1) \, .
\end{equation}
Therefore the coefficient of the operator is $n_1$ again, leading to the term
\be 
%P^{(1)}_2 = 
n_1\,{c_{lk}}^n \, O_{[n_1]}( \alpha\, e_n)~.
\ee
%
% \begin{align}
% P_{1,2} &= \frac{n!}{(n_1-1) (n-n_1)!}   \times (n_1-1)\times \frac{n_1 (n-n_1)!}{n!} \nonumber \\
% &=n_1~.
% \end{align}

\paragraph{Case 5:} The last cases are where we have an overlap of two between $[2]$ and $[n_1-1]$. This splits the $n_1-1$ cycle into two (generically). We have to sum over the lengths with total length $n_1-1$.
%, unless one of the lengths is zero, then the length is $n_1-2$.  
The operator on the right hand side is given by 
\begin{equation}
\frac{1}{2} \sum_{ q_1=1}
^{n_1-2}
\, x(q_i)\, \delta_{q_1+q_2,n_1-1} O_{[q_1,q_2,1]} (\Delta_\ast^{(2)} (\alpha e_l); e_k) \, ,
\label{OperatorAndCoefficient}
\end{equation}
where $x(q_i)$ is a coefficient to be determined.
%and the $q_i$ take values from $1$ to $n_1-2$.  
Since each operator appears twice (if we assume $n_1$ even), we add a factor of 1/2.
These operators have
\begin{equation}
|O_{[q_1,q_2]}| \times (n-n_1+1)
\end{equation}
terms. The number of terms contributing at a given value of $q_1$ equals:
\begin{equation}
|O_{[n_1-1,1]}(\cdot~;~\cdot)| \times (n_1-1) \, .
\end{equation}
The resulting pre-factor is:
\begin{equation}
x(q_i) =\frac{n!}{(n_1-1) (n-n_1)!} \frac{q_1q_2(n-q_1-q_2)!}{n!  (n-n_1+1)}\times  (n_1-1) = q_1 q_2~.
\end{equation}
For the case $n_1$ odd and $q_1=q_2=(n_1-1)/2$, there is an extra factor of $2$ in the coefficient $x(q_i)$ which will cancel the factor of $1/2$ in equation (\ref{OperatorAndCoefficient}).

\paragraph{The OPE:} The final result for the operator product is: 
\begin{align}
 O_{[2]}(\alpha) \ast O_{[n_1-1,1]} (e_l;e_k) =&\, 
O_{[n_1-1,2,1]} (e_l;\alpha;e_k)
+2 O_{[n_1-1,2]} (e_l,\alpha e_k)\nonumber\\
& + n_1 (O_{[n_1,1]} (\alpha e_l;e_k)+{c_{lk}}^n \, O_{[n_1]}( \alpha\, e_n))\nonumber\\
& + %\frac{1}{2}
\sum_{\{ q_1,q_2 \}} q_1 q_2 \delta_{q_1+q_2,n_1-1} O_{[q_1,q_2,1]} (\Delta_\ast^{(2)} (\alpha e_l); e_k)~.
\label{extraOP}
\end{align}
A  check on the formula is to once again count the total number of terms on both sides and confirm equality. We use this formula in the bulk of the paper to compute extremal four-point functions in the untwisted theory. 

% Let's check whether the number of terms on the left and on the right are equal. On the left we have:
% \begin{equation}
% \frac{n! n!}{2 (n_1-1)(n-2)! (n-n_1)!}
% \end{equation}
% terms. 
% On the right, we have:
% \begin{align}
% & 
% 1 \times \frac{n!}{2 (n_1-1)(n-n_1-2)!}
% + n_1 ( \frac{n!}{n_1(n-n_1-1)!}+ \frac{n!}{n_1(n-n_1)!})\nonumber \\
% &\hspace{2cm} + \frac{1}{2} \sum_{q_1=1}^{n_1-2} q_1 q_2 (n-n_1+1) \frac{n!}{q_1 q_2 (n-n_1+1)!} + 2 \times \frac{n!}{2(n_1-1) (n-(n_1+1))!} \nonumber \\
% &= n! \Bigg( \frac{1}{2 (n_1-1)(n-n_1-2)!}
% + n_1 ( \frac{1}{n_1(n-n_1-1)!}+ \frac{1}{n_1(n-n_1)!})
% \nonumber\\
% &\hspace{2cm}+ \frac{1}{2} \sum_{q_1=1}^{n_1-2}  (n-n_1+1) \frac{1}{ (n-n_1+1)!}
% +  \frac{1}{(n_1-1) (n-n_1-1)!}\Bigg)
% \nonumber \\
% &= n! \Bigg( \frac{1}{2 (n_1-1)(n-n_1-2)!}
% +  ( \frac{1}{(n-n_1-1)!}+ \frac{1}{(n-n_1)!})
% \nonumber\\
% &\hspace{2cm}+ \frac{1}{2}(n_1-2)  \frac{1}{ (n-n_1)!}+  \frac{1}{(n_1-1) (n-n_1-1)!}\Bigg)
% \nonumber \\
% &= \frac{n!}{2(n_1-1)(n-n_1)!} \Big((n-n_1)(n-n_1-1)
% +   2(n_1-1)(n-n_1)+ 2(n_1-1)
% \nonumber \\
% &\hspace{2cm}+ (n_1-2)(n_1-1)+2  (n-n_1) \Big)\nonumber \\
% &= \text{ correct}. 
% \end{align}
% terms.

\section{Illustrations and Checks of the Product Formula}
\label{ConcreteChecks}

In this Appendix, we illustrate and check the master formula (\ref{MasterFormula}) using hands-on small cycle examples.  In particular we check the number of terms that appear on the right hand side against the ordinary multiplication of conjugacy classes. 

Our first comparison is for the multiplication of $S_n$ conjugacy class sums $C_{[3]}$ and $C_{[4]}$. These conjugacy classes multiply as follows (see e.g. \cite{IvanovKerov}):
\begin{align}
C_{[4]} C_{[3]} =& C_{[4,3]} 
+ 6 C_{[6]}+ 4 (n-4) C_{[4]}  + 2(n-2)(n-3) \, C_{[2]}
+ 12 C_{[3,2]} 
+ 4 C_{[4]} 
  \, . \label{ConjugacyClassProduct34}
\end{align}
Let us also write down the contributions from the master formula  \eqref{MasterFormula} where all seed operators are set equal to the identity. We have the fused contribution for $|T|=0$. We find non-vanishing contributions for  $|T|=1$ at $l=0,1,2$. For the case of double triple overlap $|T|=2$ only the $(1,1;2)$ part contributes and for the $|T|=3$ terms, only the $(1,1;1)$ contribution survives. 
Separating out the contributions for each triple overlap $|T|$, we find
\begin{align}
O_{[3]}(1) \ast O_{[4]}(1) =& \, \, O_{[3,4]}(1,1) 
\nonumber \\
& +6 O_{[6]}(1) +4 O_{[4,1]}(\Delta_\ast^{(2)} (1)) + 4 O_{[2,1,1]}(\Delta_\ast^{(3)} (1))
\nonumber \\
& + 12 O_{3,2} (\Delta_\ast^{(2)}(1)) 
\nonumber \\
&
+4 O_{[4]} (e)~.
\end{align}
The triple overlap zero term, $|T|=0$, is the leading, disconnected term, and it has coefficient one. 
At the next order, at triple overlap one, we wish to further split the total into the terms with differing double overlap numbers. We have for $|T|=1$ contributions from $l=0,1,2$.
The first case is the $O_{[6]}$ term which matches the second term in equation (\ref{ConjugacyClassProduct34}). The term with $l=1$ is the $O_{[4,1]}$ term. We have $n-4$ ways of identifying the inactive colour. Thus, the operator corresponds to the coefficient $4(n-4)$ in equation (\ref{ConjugacyClassProduct34}).
At $l=2$, we have the $O_{[2,1,1]}$ term. It has coefficient $4$. This is multiplied by $(n-2)(n-3)/2$ choices, which matches the expected number. 
At second order, 
for $|T|=2$, we have a contribution of a $l=0$ term.
The permutations have the form $[3] \times [4] \rightarrow [2,3]$. It has coefficient $12$ as expected.
For $|T|=3$, we again only have a $l=0$ term contribution. 
The permutations have the form $[3] \times [4] \rightarrow [4]$. This therefore contributes to the conjugacy class $[4]$, through the operator $O_{[4]}(e)$, with coefficient $4$ again matching equation (\ref{ConjugacyClassProduct34}). This ends our rudimentary illustration and check. %

\paragraph{A Few Computer Checks:} 
We  mention two additional computer generated checks on two non-trivial aspects of our master formula (\ref{MasterFormula}). Using a symbolic manipulation program, we studied the conjugacy class product
$C_{[4]} C_{[5]}$ in the symmetric group $S_9$ and identified a right hand side conjugacy class corresponding to a triple overlap $|T|=3$ and where the total double overlap is also two (such that the quantum number $l$ measuring the extra double overlap in the master formula equals zero).
We found the coefficient:
\begin{equation}
C_{[4]} C_{[5]} = 48 \, C_{[2,2,2]} + \dots \, . 
\end{equation}
We can compare this to the $|T|=3$ and $l=0$ contribution in the master formula and find that the overall coefficient agrees. A second check is that we studied another term on the right hand side corresponding to $|T|=1, l=3$:
\begin{equation}
C_{[4]} C_{[5]} = 420 \, C_{[2]} + \dots \, . 
\end{equation}
This can be checked against the corresponding term in the master formula which includes a factor of $l!$ -- this  confirms the recurring factorial feature in the master formula. 
% If we can check a term for $l=3$ (which needs higher numbers, e.g. $C_5 C_4$ giving $C_2$), then we can check the $l!$ (i.e. the factorial). 

% New example, at $n=9$ (which may well be in the book as well !? No. Only goes to $n=8$.): at triple overlap $3$ and double overlap also $3$, we have $C_4 C_5 = 60480/ (9!/ 2^3 3! 3!) C_{[2,2,2]}= 48 C_{[2,2,2]}$. We have coefficient: $1/2 3.2.2. 2^3=48$. These agree (if one deletes the $1/3!$ as seems appropriate for equal $q_i$ on the right hand side). 

\bibliographystyle{JHEP}

\end{document}